\documentclass[reprint,aps,prx,groupedaddress,amsmath,amssymb,superscriptaddress]{revtex4-2}
\usepackage{graphicx}
\usepackage{dcolumn}
\usepackage{bm}
\usepackage{xcolor}
\usepackage[normalem]{ulem}
\usepackage[USenglish]{babel}
\usepackage{float}
\usepackage{times}
\usepackage{physics}
\usepackage{comment}
\usepackage{xr}
\usepackage[colorlinks]{hyperref}

\global\long\def\bkt#1{\left(#1\right)}

\makeatletter
\let\ORIbbl@fixname\bbl@fixname
\def\bbl@fixname#1{%
  \@ifundefined{languagealias@\expandafter\string#1}
    {\ORIbbl@fixname#1}
    {\edef\languagename{\@nameuse{languagealias@#1}}}%
}
\newcommand{\definelanguagealias}[2]{%
  \@namedef{languagealias@#1}{#2}%
}
\makeatother

\definelanguagealias{en}{english}
\definelanguagealias{EN}{english}

\begin{document}



\title[]{Spectral Theory for Non-linear Superconducting Microwave Systems: Extracting Relaxation Rates and Mode Hybridization}

\makeatletter

\date{\today}

\author{Dung N. Pham}
\affiliation{Department of Electrical and Computer Engineering, Princeton University, Princeton, NJ 08544, USA}
\author{Richard D. Li}
\affiliation{Department of Physics, Yale University, New Haven, CT 06511, USA}
\author{Hakan E. T\"ureci}
\affiliation{Department of Electrical and Computer Engineering, Princeton University, Princeton, NJ 08544, USA}
\begin{abstract}
The accurate modeling of mode hybridization and calculation of radiative relaxation rates have been crucial to the design and optimization of superconducting quantum devices. In this work, we introduce a spectral theory for the electrohydrodynamics of superconductors that enables the extraction of the relaxation rates of excitations in a general three-dimensional distribution of superconducting bodies. Our approach addresses the long-standing problem of formulating a modal description of open systems that is both efficient and allows for second quantization of the radiative hybridized fields. This is achieved through the implementation of finite but transparent boundaries through which radiation can propagate into and out of the computational domain. The resulting spectral problem is defined within a coarse-grained formulation of the electrohydrodynamical equations that is suitable for the analysis of the non-equilibrium dynamics of multiscale superconducting quantum systems. 
\end{abstract}

\maketitle


\section{Introduction}

Engineered superconducting systems employed in analog quantum simulation~\cite{houck2012chip, carusotto2020photonic}, quantum sensing~\cite{hatridge2011dispersive,macklin2015near} and quantum computing~\cite{blais2004cqed, devoret2013superconducting} are described by the electrohydrodynamic theory of superconducting systems (EHDS), first proposed without justification by R.P. Feynman in a special lecture he delivered in the 1960s~\cite{feynman}. Few attempts have been made to either justify~\cite{ao1995nonlinear,aitchison1995effective} or solve~\cite{dec-qed} this model's equations of motion exactly, but simplified versions~\cite{greiter1989hydrodynamic, PhysRevLett.64.587, salasnich2009hydrodynamics, PhysRevLett.64.587} have been considered via various modifications to minimal coupling. This low-energy theory can be credited as the underpinning of Josephson phenomena, flux quantization, and, generally, the physics of vortices. The second-quantized version of it constitutes the basis of the formulation known as circuit quantum electrodynamics (cQED)~\cite{blais_rmp}, which has been the workhorse behind the current understanding of the physics governing superconducting quantum computers. 
As systems grow increasingly sophisticated and their electromagnetic environments become more complex, efficient computational strategies in both classical and quantum regimes are much needed to produce accurate reduced models containing the degrees of freedom of interest. This has given rise to an active research area at the intersection of computational electromagnetism and quantum electrodynamics of superconducting devices~\cite{BBQ_2012,solgun2014blackbox,quantize_shunted_sc_2016, cuttoff_free_cqed_2017, parra2019canonical,EPR,TCGtoolbox}, to analyze and model the physics of these circuits.

An indispensable component in the electrodynamic modeling of superconducting circuits is the extraction of relaxation rates. Purcell modification of radiative lifetimes of qubits is today an essential mechanism for the protection of qubits and plays an important role in the electromagnetic design of the entire processor~\cite{houck2008spontaneous, reed2010fast, jeffrey2014fast, bronn2015broadband}. Additionally, in applications that require strong RF or microwave excitations of individual oscillators, new dissipative processes and inter-level transitions can be activated~\cite{Blais_etal_dispersive_cQED,Slichter2012StateMixing,Devoret_March_Meeting,Minev_Nature,PetrescuReadout2020,Hanai2021,Pop_et_al_readout_with_large_photon_number,Blais_etal_transmon_ionization,Blais_et_al_chaos_in_transmon}. These processes, in turn, depend strongly on the spectral characteristics of the electromagnetic environment in which the non-linear elements are embedded~\cite{PetrescuReadout2020}. Resource-intensive numerical simulations are needed to extract such information for complex quantum-electrodynamic systems. Such simulations are not only computationally demanding but also face conceptual difficulties arising from the infinite degrees of freedom inherent in quantum electrodynamics (QED)~\cite{Caldeira_Leggett}. These conceptual issues~\cite{houck2008spontaneous,Filipp2011, cuttoff_free_cqed_2017,Gely2017multimode} can be effectively addressed by the introduction of a rigorous spectral theory~\cite{kanupaper}. This approach allows the accurate quantification of limitations brought about by noise~\cite{yoshihara2014flux,kumar2016origin,wilen2021correlated}, dissipation~\cite{pop2014,frattini2018optimizing} and undesired interactions~\cite{Crosstalk1_2011,huang2021microwavepackage} within the system and its electromagnetic environment. 

In the present work, we introduce a spectral theory of EHDS from which the relaxation rates of general three-dimensional superconducting circuits can be obtained. This spectral theory is adapted to a coarse-grained description of the EHDS equations, known as DEC-QED, previously derived and analyzed in Ref.~\cite{dec-qed}. DEC-QED provides the following advantages: (1) through its structure-preserving geometric discretization procedure, known as discrete exterior calculus (DEC), this formulation enables stable long-time simulations. (2) By virtue of the fundamental fields being hybridized gauge-invariant fields rather than the standard electromagnetic potentials, different materials can be handled in a uniform fashion. (3) At a superconducting-normal-superconducting (SNS) junction, the hybridized field is identical to the gauge-invariant Josephson phase across the junction. (4) For transmission lines and lumped-element circuits and in the limit of a perfect superconductor ($\lambda_L$ = 0), DEC-QED equations reduce to the standard cQED equations. 

The difficulty with developing a spectral theory stems from the desire to solve the EHDS equations, a set of non-linear PDEs describing the evolution of the order parameter of a charged fluid coupled to Maxwell's equations, in a finite spatial domain. Ideally, one would like to keep the computational domain as small as possible, just as large as the superconducting system itself. However, this is not possible, because we also need to keep the volume as large as possible to allow for radiative relaxation of the excitations in the superconductor. The optimal solution to avoid this trade-off is to use a modal description for a finite but open domain subject to transparent boundary conditions for electromagnetic (EM) radiation. This allows us to keep the computational domain small enough to be computationally feasible, while still allowing for the radiative relaxation of the excitations. While boundary conditions, such as the perfectly matched layer (PML)~\cite{berenger1994pml, berenger1996_3Dpml} or absorbing boundary conditions~\cite{clayton1977absorbing,absorbingBCs} are known and very useful in classical EM problems, it is also desirable to implement a transparent boundary so that solving the corresponding spectral problem would provide a modal decomposition which can serve as the basis for second quantizing the EHDS equations. This, however, has been a difficult problem that remains unaddressed  since the full formulation of quantum electrodynamics~\cite{Gell-Mann_qed} because such boundary conditions result in non-Hermitian modes~\cite{qi2018defect, el2018non, gigli2020quasinormal}. Recently, it was shown that by using the Heisenberg equations of motion for quantum field operators, this issue can be circumvented in the context of one-dimensional transmission line systems through the use of singular function expansion and a suitable spectral problem~\cite{kanupaper}. Here we generalize the statement of this spectral problem to the solution of the non-linear EHDS equations (the original spectral problem has been defined~\cite{kanupaper} in the context of cQED which assumes fields do not penetrate the superconductors), and to general three dimensional domains. We do not address the issue of quantization but focus on the spectral problem that arises through the linearization of the EHDS equations, the resulting non-Hermitian modes, and the calculation of dissipation rates for all modes. The dissipation rates for ``qubit-like" modes are precisely their Purcell radiative lifetimes~\cite{purcell1946resonance}, while for the more spatially extended modes, they represent their losses due to their hybridization with radiative channels in the system~\cite{malekakhlagh2016non,bosman2017multi,puertas2019tunable}. 

With reference to the prior work on modeling of superconducting devices, our approach is within the class of full-wave methods~\cite{roth2021_fullwaveCQED, chew2016quantum}. With respect to existing full-wave methods for superconducting systems, DEC-QED has two important distinctions: {\it 1. Calculation of Radiative Losses:} The boundary conditions implemented in existing approaches to capture radiative losses often implement absorbing boundary conditions. Most frequently, radiative losses at a 3D cavity’s ports are modeled by adapting from the lumped-element models and terminating the ports with $50\,\Omega$ matched resistors \cite{solgun2014blackbox} or by covering the holes on the cavity walls with resistive disks \cite{EPR}. Although this approach is effective in obtaining the frequencies of the eigenemodes, it is known to fail to accurately capture the spatial structure of the open modes. More importantly, no known method exists to second quantize using such modes, and therefore such EM full wave simulations can not rigorously be used to synthesize a quantum mechanical Hamiltonian or Liouvillian. Often the quantum Hamiltonian is derived from simulations ignoring radiative losses, and the relaxation rates are introduced through a perturbative approach. In contrast, DEC-QED builds on prior work in classical electromagnetic systems for the implementation of correct open (transparent) boundary conditions~\cite{tureci2008randomlaser,wiersig2002boundary}, which recently has been shown to serve as a mathematically consistent basis for second quantization of the modes of a transmission line cavity~\cite{kanupaper}. \textit{2. Time-dependent Evolution}: DEC-QED aims at numerically solving the set of non-linear PDEs describing the electrohydrodynamics of the superconducting condensate coupled to fields described by Maxwell equations. This allows DEC-QED to describe the evolution of gauge-invariant hybridized fields living throughout the free space and the superconductors, forgoing issues that arise in defining boundary conditions at superconductor/vacuum boundaries whose accuracy is not controlled. Compared to existing full-wave methods, this is a unique feature of our approach. The linearized spectral solutions of DEC-QED then serves as a basis for expanding this non-linear dynamics. In the current work, we do not discuss the non-linear aspects of evolution, which is discussed at length in Ref.~\,\cite{dec-qed}.

The main results presented in this paper are organized as follows: in Section \ref{sec:EHDSformulation}, we discuss the electrohydrodynamic model used to describe the dynamics of the superconducting condensate coupled to the EM environment and derive the spectral problem to be solved. Then in Section \ref{sec:DEC_background} we briefly discuss the formulation of DEC-QED used in the rest of the article for coarse-grained calculations. Next, we demonstrate the convergence in the calculations for the spectral profile of a sample superconducting cavity. Both simplicial and cubical meshing strategies are benchmarked in Section \ref{sec:simplicial_vs_cubical}. In Section \ref{sec:accidental_degeneracy} we then show how mode hybridization due to accidental degeneracies can be encountered in the calculations of the modes for a symmetric cavity and that our method can distinguish the degenerate modes through the inclusion of a small perturbation to the cavity shape. Next, in Section \ref{sec:multiscale_sys} we demonstrate the coarse-grained calculations of hybridized modes for multiscale systems containing multiple components. 
Finally, we present two main approaches for implementing open boundary conditions based on (1) Green's boundary integrals in Sections \ref{sec:scalargreen_BIM}-\ref{sec:vectorgreen_BIM} and (2) vector spherical harmonic expansions in Section \ref{sec:VSH_openbc}, along with numerical examples, to demonstrate the accuracy and effectiveness of these methods. We have also made the implementation of the formulation discussed here available as part of an open-source DEC-QED repository~\cite{decqed_repo}.  

\section{Electrohydrodynamic formulation of superconducting materials}\label{sec:EHDSformulation}
Consider a multiply connected region composed of different dielectric and superconducting materials. Well below the critical temperature, where most superconducting devices operate, the superconducting material can be modeled as a charged condensate~\cite{ao1995nonlinear,aitchison1995effective} trapped by a background of positive charge ($\rho_{\text{src}}$). The dynamics of the condensate are wholly determined through minimal coupling to the dynamical electromagnetic potentials ($\bm{A}$,  $V$) and the Coulomb attraction to the static positive background ($U$) defining the superconductor. We refer to the resulting equations as the ElectroHydroDynamics of Superconductors (EHDS)~\cite{feynman}
\begin{equation}\label{eq:generalSE}
    i\hbar\frac{\partial \Psi({\bf r},t)}{\partial t} 
    = \bigg[\frac{1}{2m}\Big(\!\!-i\hbar{\bf \nabla}-q{\bf A} \Big)^2 + qV({\bf r},t) + U({\bf r}) \bigg]\Psi\bkt{{\bf r},t}
\end{equation}
and Maxwell's equations
\begin{align}
    {\bf\nabla}\times{\bf\nabla}\times{\bf A} + \mu_0\epsilon\ddot{{\bf A}} &= \mu_0({\bf J}_s + {\bf J}_{\text{src}}) - \mu_0\epsilon\nabla\dot{V}, \label{eq:Ampere} \\
    \nabla^2V + \frac{\partial}{\partial t}({\mathbf \nabla}\cdot {\mathbf A})&= -\frac{q}{\epsilon}(\rho +\rho_{\text{src}}). \label{eq:Gauss}
\end{align}
where $\rho$ and ${\bf J}_s$ are the condensate density and the supercurrent, respectively, and $\rho_{\text{src}}$ and ${\bf J}_{\text{src}}$ are the external charge and current sources. Here, $q=2e$ and $m=2m_e$ are the charge and mass of a cooper pair, respectively, which are twice the charge $e$ and mass $m_e$ of an electron. Generally, we include the positive background defining the superconducting regions in $\rho_{\text{src}}$. Dielectric regions are defined by $\epsilon(\bm{r}) = \Tilde{\epsilon}(\bm{r})\epsilon_0$, where $\Tilde{\epsilon}$ is the step-wise constant relative permittivity function that is unity where dielectric is not present and is greater than one otherwise. Using the Madelung representation for the condensate wavefunction, $\Psi({\bf r},t) = \sqrt{\rho({\bf r},t)}e^{i\theta({\bf r},t)}$, and introducing the gauge-invariant hybridized field \mbox{$\bm{\mathcal{A}} = {\bf A} - \frac{\hbar}{q}\nabla\theta$}, Eqs.\,(\ref{eq:generalSE})-(\ref{eq:Gauss}) can be written in the form~\cite{dec-qed}
\begin{align}\label{eq:Aprime_waveeq_rho}
 {\bf\nabla}\!\times\!{\bf\nabla}\!\times\!\bm{\mathcal{A}} + \mu_0 &\epsilon\frac{\partial^2\bm{\mathcal{A}}}{\partial t^2} + \frac{\mu_0 q^2}{m}\rho\bm{\mathcal{A}}  -\frac{\mu_0\epsilon q}{2m}\frac{\partial}{\partial t}\nabla\big|\bm{\mathcal{A}}\big|^2 \nonumber \\
  & + \frac{\mu_0\epsilon\hbar^2}{2mq}\frac{\partial}{\partial t}\nabla\bigg[\frac{\nabla^2(\sqrt{\rho})}{\sqrt{\rho}}\bigg] =  \mu_0{\bf J}_{src},
\end{align}
and 
\begin{equation}\label{eq:chargeconserve2_rho}
    \frac{\partial\rho}{\partial t}  = {\bf \nabla} \cdot \Bigg[\frac{q}{m}\rho\bm{\mathcal{A}} - \frac{{\bf J}_{src}}{q}\Bigg] - \frac{\partial\rho_{src}}{\partial t}.
\end{equation}
These equations were derived and the resulting real-time dynamics under specific conditions were analyzed in Ref.~\cite{dec-qed}. Here, we are interested in the spectral problem associated with these non-linear equations. Linearization can be done by splitting the condensate density $\rho$ into the mean value $\rho_0$ that exactly balances the positively charged ionic background and the fluctuation $\delta\rho$ arising from interactions with the external EM field. The linear sector of Eq.\,(\ref{eq:Aprime_waveeq_rho}) that corresponds to the transverse excitations then reproduces London theory~\cite{londontheory} and is sourced by the fluctuation in the source current. A derivation of the linearization is provided in Appendix \ref{append:linearization}. The resulting inhomogeneous source-field equations can be solved through the spectral problem of a vector Helmholtz equation for the gauge-invariant hybridized field $\bm{{\mathcal A}}$:
\begin{align}\label{eq:vectorHelmholtz}
    {\mathbf \nabla}\times{\mathbf \nabla}\times\bm{{\mathcal A}} + \bigg (\frac{1}{\lambda_L^2({\mathbf r})} - n^2({\mathbf r})k^2\bigg )\bm{{\mathcal A}} = 0,
\end{align}
where the distribution of superconducting and dielectric materials is given by the local London penetration depth function $\lambda_L({\mathbf r}) = \sqrt{\frac{m}{\mu_0 q^2\rho_0({\mathbf r})}}$ and the refractive index function $n({\mathbf r})=\sqrt{\Tilde{\epsilon}({\mathbf r})}$, respectively, and $k$ is the wavenumber.

\section{DEC-QED formulation}\label{sec:DEC_background}
DEC-QED provides a spatially coarse-grained description of a physical system governed by nonlinear PDEs. The fundamental field variables are small integrals of the original microscopic continuous fields over finite spatial intervals. This results in a discretized model that is computationally efficient and still accurate within the resolution of a given measurement apparatus. In this section, we provide a minimal introduction to the geometric constructions in DEC that are needed for computing electromagnetic modes in systems that may contain an arbitrary distribution of superconducting and dielectric materials, all within a finite computational domain. For a more comprehensive discussion of the formulation, we refer to Ref.~\cite{dec-qed}.

\begin{figure}[t]
    \centering
    \includegraphics[scale=0.07]{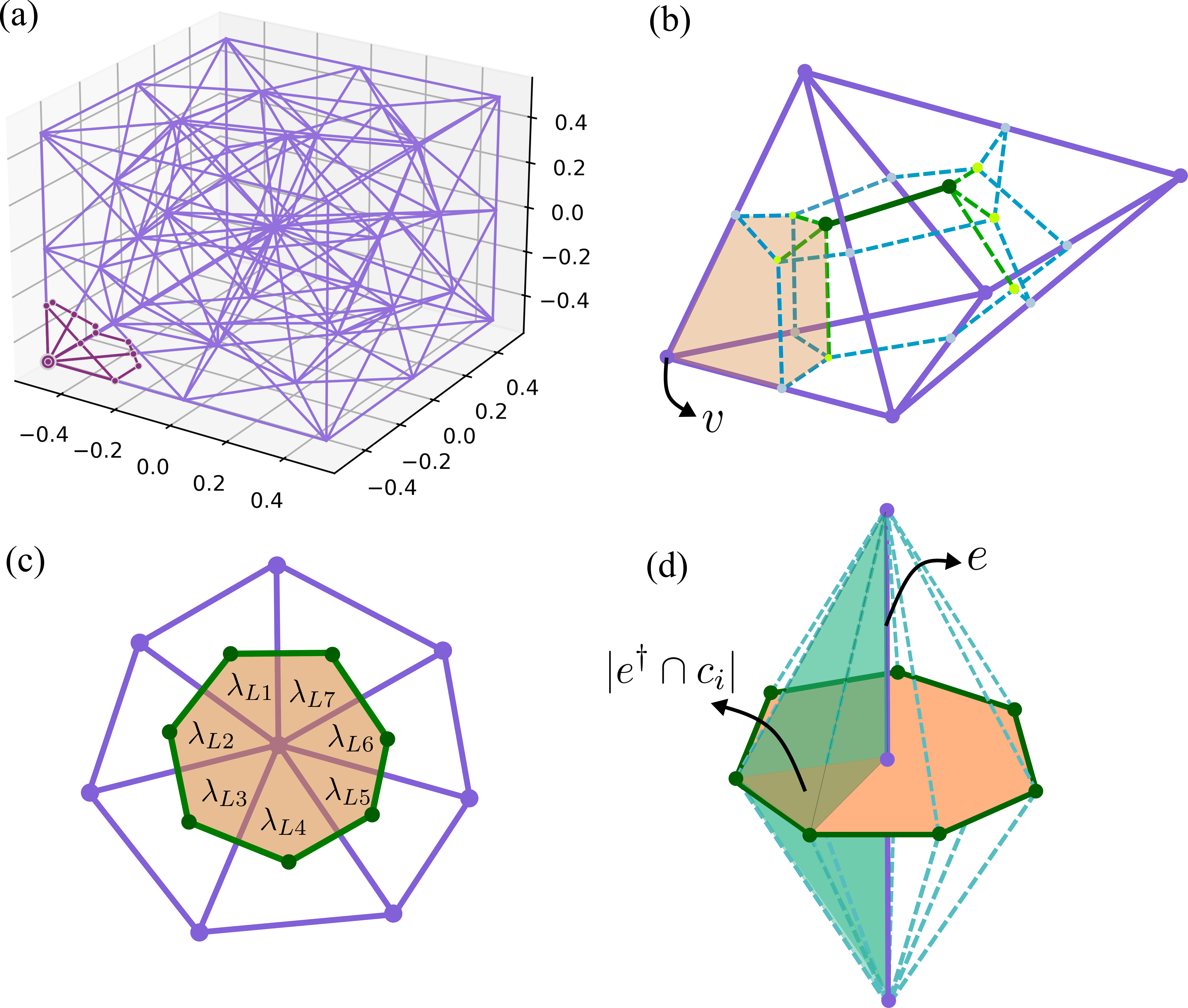}
    \caption{(a) An example primal mesh. The highlighted edges at the lower-left corner form the boundary of an auxiliary dual volume. (b) Close-up view of a node $v$ that lies on the boundary of the primal mesh. The neighboring dual nodes are labeled in dark green, while the auxiliary boundary dual nodes are labeled in bright green. The shaded volume is the truncated volume $v^\dagger$ dual to the node $v$. (c) 2D example of a vertex that lies at the interface among multiple material regions. The material properties assigned to this node is the weighted average of the values in the surrounding regions. (d) 3D example of an edge lying at a material interface composed of multiple cells ${c_i}$. The darkened triangle is the intersection of the dual face $e^\dagger$ with the cell $c_i$, and the shaded volume is the portion of the support volume of $e$ that lies inside $c_i$.}
    \label{fig:mesh_schematics}
\end{figure}

In DEC, the discretization of PDEs requires a dual-mesh construction, within which the $d$-dimensional computational space is discretized by a {\it primal mesh} $M$ that conforms to the boundaries of the enclosed physical domain and the interfaces between materials. The fundamental building blocks of the primal mesh can be simplices (i.e. triangles in 2D and tetrahedra in 3D) or cubical elements. The vertices of the {\it dual mesh} $M^\dagger$ are then circumcenters of the primal $d-$cells, and the edges of $M^\dagger$ are generated by connecting the neighboring circumcenters. Throughout this paper, we will use $\dagger$ to denote a dual quantity. For elements strictly inside the computational space, the mappings from vertices ($v$), edges ($e$), faces ($f$), and cells ($c$) in $M$ to cells ($v^\dagger$), faces ($e^\dagger$), edges ($f^\dagger$), vertices ($c^\dagger$) respectively in $M^{\dagger}$ are one-to-one. This bijective correspondence is not applicable at the computational boundary, where the dual elements are truncated, which leads to auxiliary dual nodes lying on the boundary of $M$ (See Figs.\,(\ref{fig:mesh_schematics}a) and (\ref{fig:mesh_schematics}b)). During numerical calculations, this truncation of the dual mesh is taken care of by the appropriate application of boundary conditions. 

In this DEC framework, scalar fields live on either primal vertices $v$ or their dual volume $v^\dagger$, while vectorial quantities are projected onto the discrete primal edges $e$ or assigned to the dual faces $e^\dagger$. For example, given a vector field $\bm{{\mathcal A}}$ we construct the coarse-grained edge field
\begin{equation}
    \Phi(e) = \int_{e}{\mathbf d\ell}\cdot\bm{{\mathcal A}},
\end{equation}
where the integral is done along the primal edge $e$, and given a scalar function $\rho$ we define the coarse-grained variable
\begin{equation}
    Q(v^\dagger) = \int_{v^\dagger}dV\rho,
\end{equation}
where the integral is done over the dual volume $v^\dagger$. By solving the equations governing these coarse-grained fields, we can probe the properties of the system. 

To properly account for the distribution of different materials within the computational domain, material properties such as dielectric function $n^2$ or $1/\lambda_L^2$ (which is proportional to the bulk condensate density) can be assigned to objects living on the dual mesh. For example, for every edge $e$ a value for $1/\lambda_L^2$ is assigned to its dual $e^\dagger$. This procedure is particularly convenient if $e$ lies at the interface between multiple materials. The dual $e^\dagger$ in such cases would intersect with all the different material domains that share this edge, and the effective value $\overline{1/\lambda_L^2}(e^\dagger)$ there will be the weighted average of the values in the surrounding domains (See Fig.\,\ref{fig:mesh_schematics}d). In a three-dimensional setting, this is formally defined as followed: let $\{c_i\}$ be the list of all the cells that share an edge $e$, then
\begin{equation}\label{eq:eff_lambda_L}
    \overline{\bigg(\frac{1}{\lambda_L^2}\bigg)}(e^\dagger) = \sum_{c_i} \frac{|e^\dagger\cap c_i|}{\Delta A(e^\dagger)}\frac{1}{\lambda_L^2(c_i)},
\end{equation}
where $|e^\dagger\cap c_i|$ in the area of the intersection between $e^\dagger$ and $c_i$, and $\Delta A(e^\dagger)$ is the area of $e^\dagger$.

This DEC formulation of the electrohydrodynamics of coarse-grained quantities offers several benefits: (1) it provides a natural framework for describing gauge-invariant fields that are consistent with the standard flux-based description of Josephson dynamics. (2) Since the gauge-invariant fields $(\bm{{\mathcal A}},\rho)$ permeate all of space, the issues related to boundary conditions at material interfaces are simplified by introducing effective material properties. (3) Because the Josephson phase is, by definition, a gauge-invariant and coarse-grained generalized flux variable, our approach is a rigorous generalization of the coarse-grained formulation beyond the junction and apply to the entire discretized 3D(2D) computational space. It is therefore different from other fullwave approaches because here the fundamental variables of interest are the coarse-grained fields that hybridize light and matter. The quantization of this electromagnetic theory therefore follows a similar procedure as is typically done for 1D transmission line cQED, but now with fluxes living on the edges and charges living on the nodes of the fully discretized 3D(2D) system. This approach, in which all of the classical and quantum analysis is done at the EM field level, is distinct from circuit diagram-based methods such as blackbox quantization \cite{BBQ_2012, solgun2014blackbox} or energy-participation ratio \cite{EPR}. We sacrifice the simplicity brought about by the reduction to the circuit picture to obtain accurate mode structure and the full transparency in the ensuing quantization.

That being said, before the theory can be quantization-ready, a rigorous spectral theory for the coarse-grained quantities that can correctly account for radiative losses and mode hybridization, both in the eigenfrequencies and in the spatial profile of the modes (including and particularly on the open boundaries), is needed and is the focus of this paper. In the following sections, we demonstrate with specific examples how this is achieved in our approach.

\section{Modes of closed superconducting systems}\label{sec:closedsystem}
\subsection{Performance benchmarks for simplicial- and cubical-DEC}\label{sec:simplicial_vs_cubical}
\begin{figure}[t]
    \centering
    \includegraphics[scale=0.42]{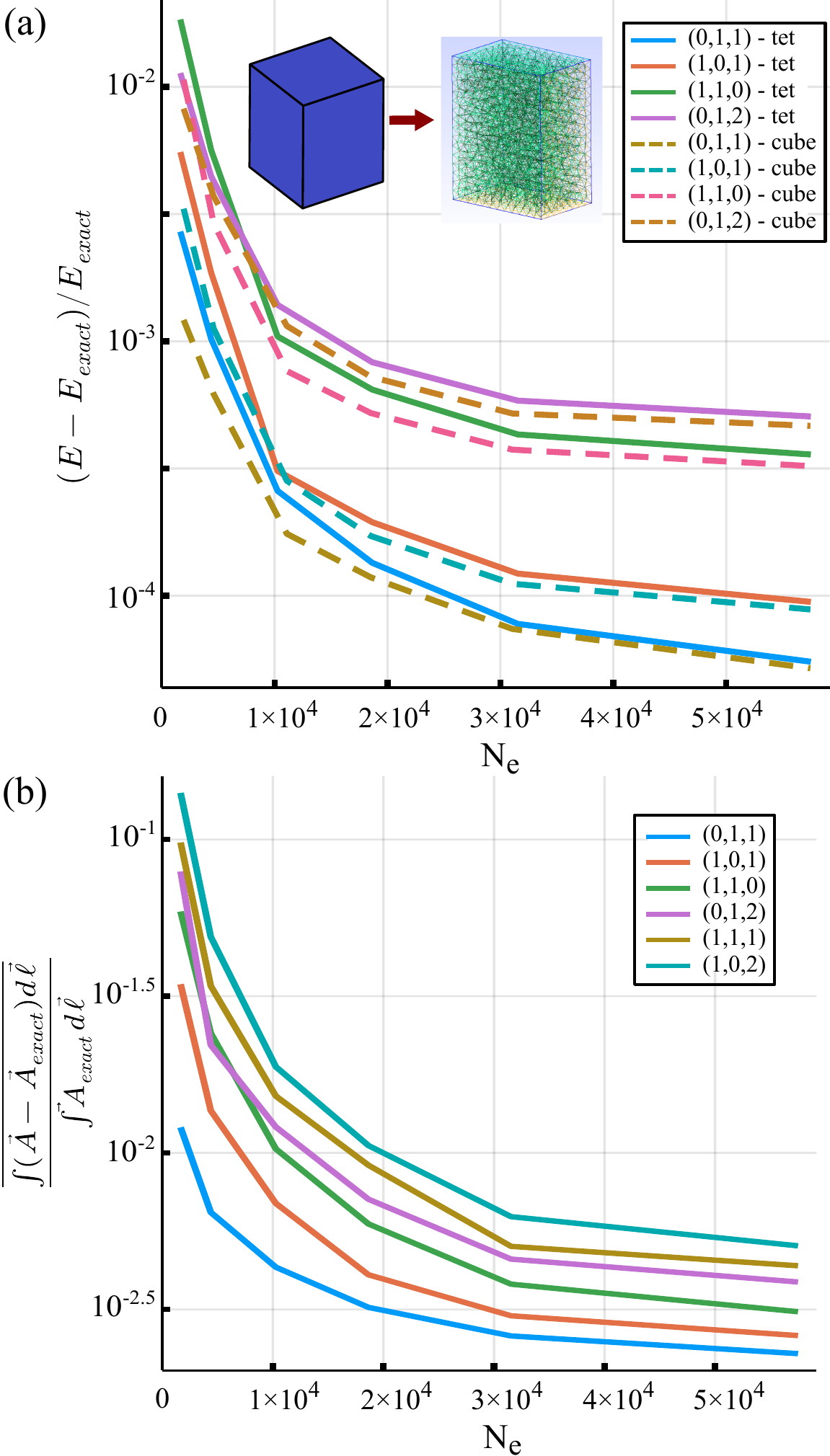}
    \caption{(a) Convergence in the eigenvalues $E=k^2$ of the vector Helmholtz equation applied to a perfect superconducting cavity as a function of the number of edges $N_e$ used in the discretization. The solid lines correspond to results obtained using simplicial meshing, while the dashed lines are results obtained using cubical meshing. The relative error is calculated with respect to the exact analytical solution. (b) The relative error in the edge fields of the eigenmodes with respect to analytical solutions. The errors are averaged over all edges and plotted as a function of $N_e$. The cavity dimensions are  \mbox{$L_x =1 \text{\,cm}, L_y=1.5 \text{\,cm}$, and $L_z = 2 \text{\,cm}$}. }
    \label{fig:rectangular_cav_conv}
\end{figure}

We are interested in solving Eq.\,(\ref{eq:vectorHelmholtz}) for the electromagnetic modes of systems composed of possibly spatially disjoint superconducting structures. To numerically compute the modes of such systems, we derive the DEC equations corresponding to Eq.\,(\ref{eq:vectorHelmholtz})
\begin{align}\label{eq:discrete_vectorhelmholtz}
\sum_{e_0\in\partial(e^{\dagger})}&\sum_{e_1\in\partial(e_0^\dagger)} \frac{\Delta\ell(e_0)}{\Delta A(e_0^{\dagger})}\Phi(e_1) \\
&+ \bigg(\overline{\frac{1}{\lambda_L^2}}(e^\dagger) -  \overline{n^2} (e^\dagger)k^2\!\bigg)\frac{\Delta A(e^\dagger)}{\Delta\ell(e)}\Phi(e) = 0, \nonumber
\end{align}
where $\partial(e^\dagger)$ is the boundary of the dual face $e^\dagger$, and $\{\overline{1/\lambda^2_L},\overline{n^2}\}$ are defined as in Eq.\,(\ref{eq:eff_lambda_L}). This form is universal and is independent of the mesh elements used. 

First, we investigate the numerical convergence of DEC equations using two types of elements: simplicial and cubical. Simplicial elements are generally the preferred choice for a meshing scheme that conforms to arbitrary superconducting domains. To analyze the convergence of simplicial meshing, we consider shapes of superconducting cavities for which analytical solutions are available. A good choice is a rectangular cavity, which conforms well to cubical elements. A comparison of the convergence of error of simplicial and cubical meshing would illustrate the efficacy of simplicial meshing.

Consider a three-dimensional rectangular cavity with dimensions \mbox{$L_x =1 \text{\,cm}, L_y=1.5 \text{\,cm}$, and $L_z = 2 \text{\,cm}$}, where $L_x, L_y,$ and $L_z$ are the sizes of the cavity along $x,y,$ and $z$ respectively. To allow for direct comparisons with analytical solutions, we first assume the cavity walls are perfect superconductors $(\lambda_L\rightarrow 0)$, so that the tangential component of $\bm{{\mathcal A}}$ vanishes at the boundary. This is equivalent to the boundary condition
\begin{equation}\label{eq:hardwall_bc}
    \Phi(e_t) = 0,
\end{equation}
where $e_t$ is an edge lying tangentially on the cavity boundary. The eigenmodes of the  cavity are computed with two choices of meshing: tetrahedral and cubical. The convergence with respect to analytical solutions is shown in Fig.\,\ref{fig:rectangular_cav_conv}. As shown in Fig.\,\ref{fig:rectangular_cav_conv}a, the error in the eigenvalues $E=k^2$ of both mesh choices converge at the same rate as the number of edges $N_e$ in the discretized domain increases. The effectiveness of simplical-DEC is therefore shown to be comparable to cubical-DEC, even when it is applied to a geometry where cubical meshing holds an advantage due to its elements sharing the same symmetry as the cavity. In Fig.\,\ref{fig:rectangular_cav_conv}b, we also study how simplicial-DEC produces accurate solutions to the edge fields of the eigenmodes. The convergence  to analytical solutions of the edge fields is achieved. 

These results help justify our shift to using simplicial meshing onwards, as its flexibility allows us to apply DEC to complicated, realistic structures where cubical symmetry is rarely present. Moreover, simplicial meshing allows for the implementation of different spatial resolutions in different regions, an important feature needed for the efficient application of DEC to systems comprised of multiple spatial scales.

\begin{figure}[t]
    \centering
    \includegraphics[scale=0.5]{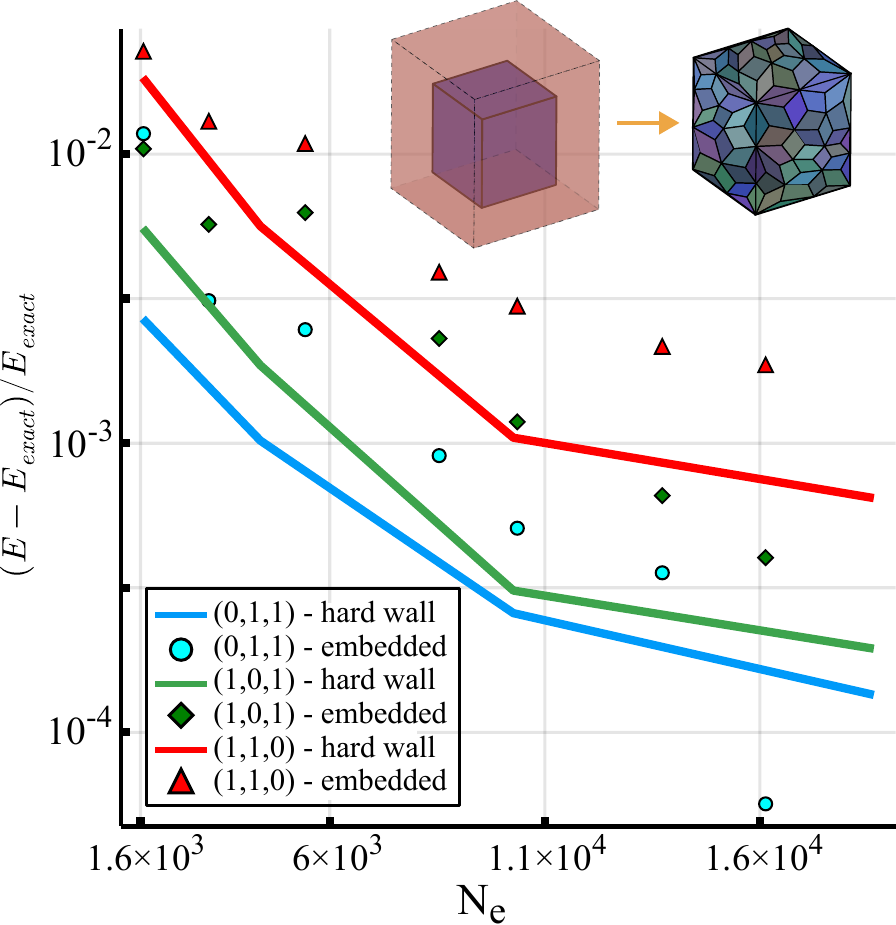}
    \caption{Comparison between the convergence of computed eigenvalues for a cavity embedded in a superconducting shell and those for a perfect cavity with the hard-wall boundary condition imposed. $N_e$ is the number of edges used in discretizing the cavity. The dimensions of the vacuum regions within both cavities are  \mbox{$L_x =1 \text{\,cm}, L_y=1.5 \text{\,cm}$, and $L_z = 2 \text{\,cm}$}. A schematic of the embedded cavity is shown in the inset at the center top, while the inset at the top-right corner shows schematically how the effective penetration depth is assigned on the inner boundary of this cavity via coarse-graining.}
    \label{fig:embedded_cav_conv}
\end{figure}

Next, we consider a rectangular cavity enclosed by a superconducting shell that has a finite thickness. The penetration depth of this shell is set to be short enough so that the field inside the cavity decays immediately at the material interface, i.e. the inner walls of the cavity. Instead of directly imposing the Dirichlet boundary condition at the cavity inner walls as in the previous case, the values of the field are left floating there, and we only impose a ``hard-wall" boundary condition (Eq.\,(\ref{eq:hardwall_bc})) at the outer boundary of the shell, which is also the boundary of the computational domain. The purpose of this numerical experiment is to investigate the validity of DEC when there are sharp interfaces, such as the vacuum-superconductor boundary here, where the field undergoes abrupt variations. The procedure for applying material properties to the edges lying on the interface is given in Eq.\,(\ref{eq:eff_lambda_L}). In a non-uniform tetrahedral mesh, the partitioning of the dual faces $e^\dagger$ that lie on multi-material interfaces to the neighboring material domains is also highly non-uniform. This leads to the effective penetration depth of each edge on the same boundary being different from one another. One can imagine the vacuum-material interface made of many small patches, each with a slightly different material property, as shown in the top-right inset in Fig.\,\ref{fig:embedded_cav_conv}. We compute the eigenmodes of Eq.\,(\ref{eq:discrete_vectorhelmholtz}) for an embedded cavity that has the same dimensional ratios as in the perfect cavity case, and the convergence is shown in Fig.\,\ref{fig:embedded_cav_conv}. The embedded cavity calculation achieves a similar order of accuracy as the perfect cavity case, and the two converge at the same rate as mesh density increases.

\subsection{Detection and removal of hybridization between degenerate modes}\label{sec:accidental_degeneracy}
\begin{figure}[t]
    \centering
    \includegraphics[scale=0.3]{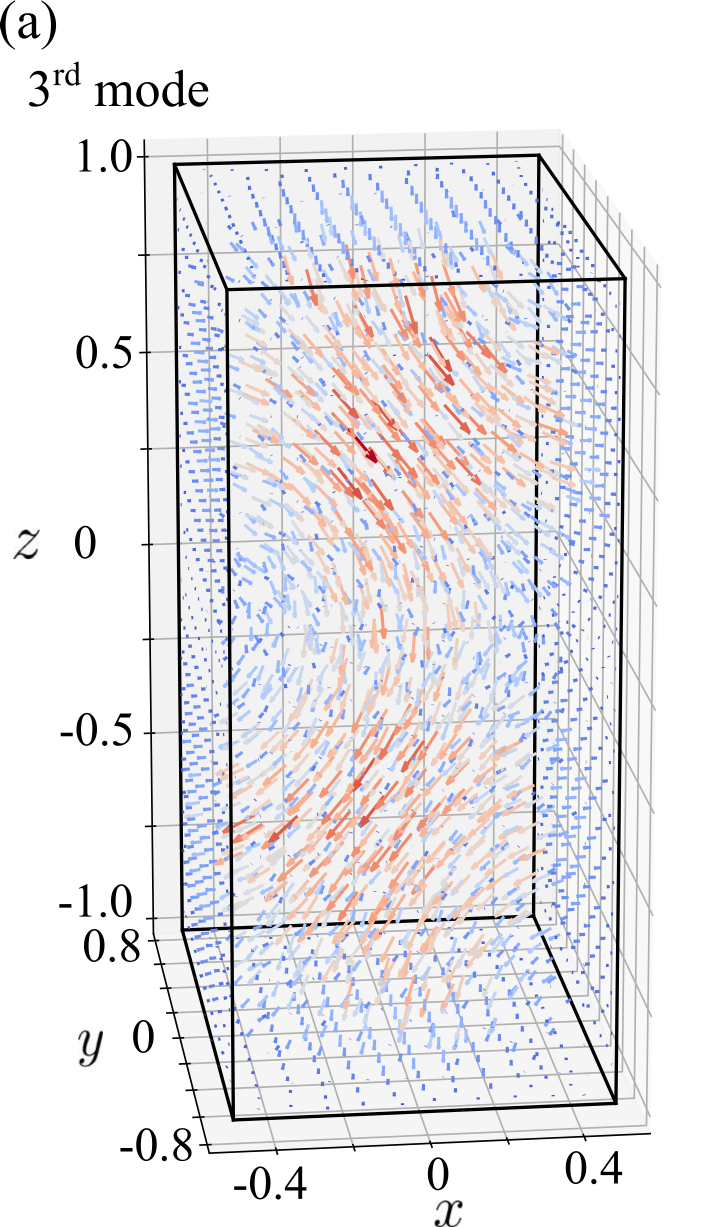}\hspace{0.5in}
    \includegraphics[scale=0.3]{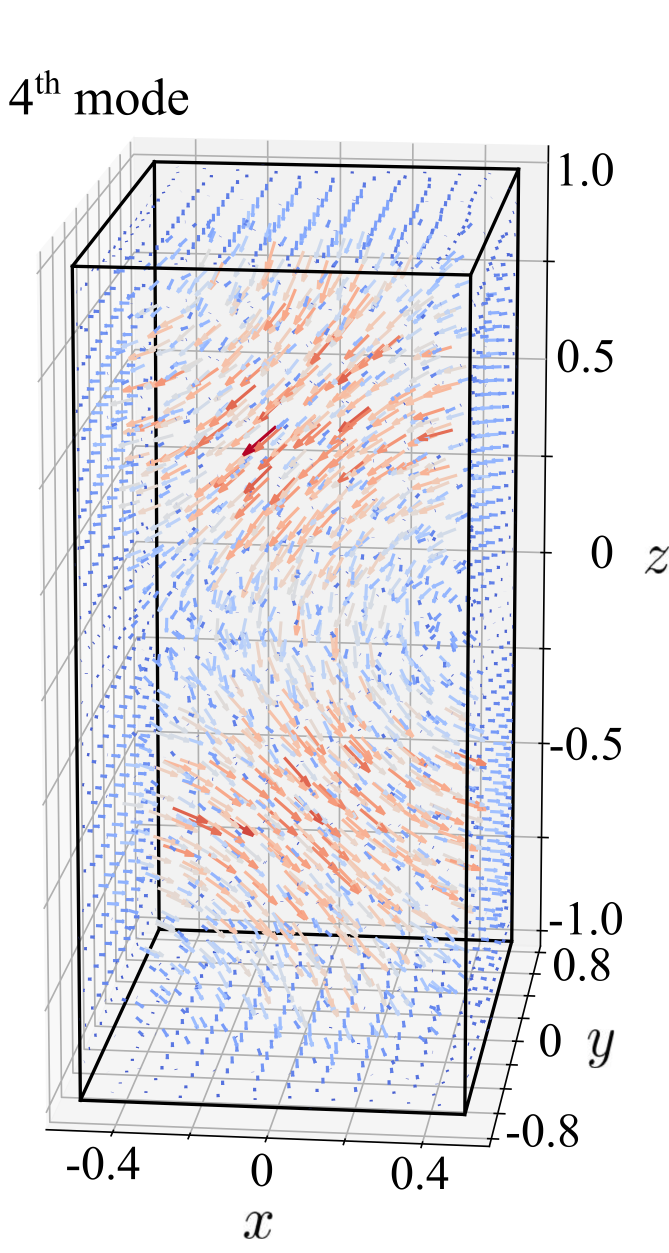}\\
    \includegraphics[scale=0.3]{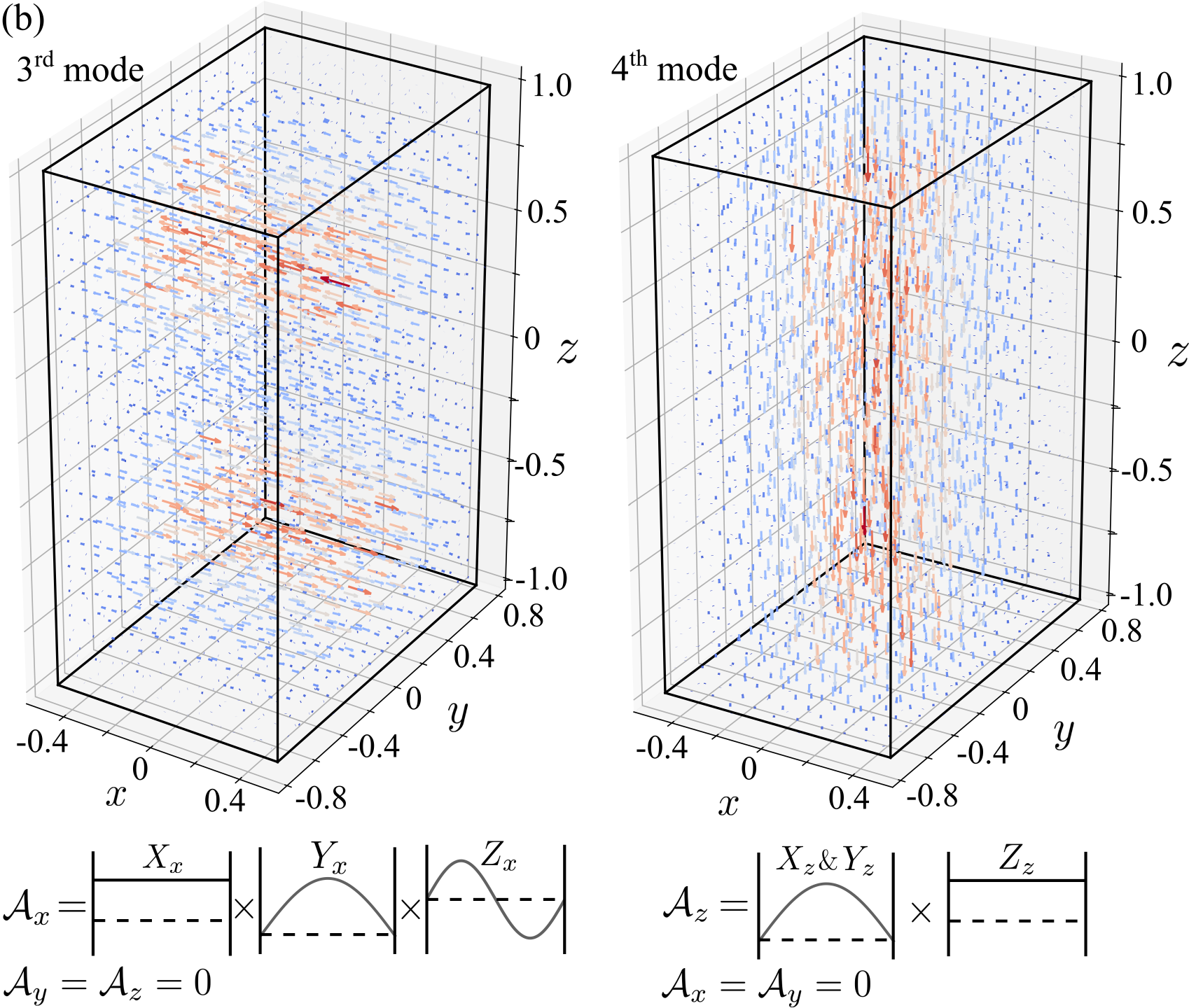}
    \caption{Demonstration of the appearances and removals of hybridization between degenerate modes in a system with hidden symmetries. (a) The degenerate $3^{\text{rd}}$ and $4^{\text{th}}$ modes of a rectangular cavity with perfect superconducting boundaries and dimension ratios of $1\!:\!1.5\!:\!2$. (b) The $3^{\text{rd}}$ and $4^{\text{th}}$ modes of a perturbed cavity, when the dimension ratios are now $1\!:\!1.5\!:\!2.01$. The two modes are now well-characterized by their respective set of quantum numbers, which are schematically shown below the field distributions.}
    \label{fig:acc_degeneracy}
\end{figure}

In symmetric structures such as the rectangular cavity discussed earlier in this article, there is a possibility of hybridization of degenerate modes in the numerically obtained eigenspectrum. These degeneracies are sometimes called ``accidental" because they are not predicted by the symmetry group of the Hamiltonian but originate from a hidden symmetry of the system~\cite{Shaw_1974}. In such cases, numerical solvers tend to face difficulties in distinguishing these degenerate modes. For a rectangular superconducting cavity, the eigenvalues corresponding to each field component ${\mathcal A_i}$ of the field $\bm{{\mathcal A}}$ reads
\begin{equation}
    E_i \sim \bigg(\frac{n_x^2}{L_x^2} + \frac{n_y^2}{L_y^2} + \frac{n_z^2}{L_z^2} \bigg),
\end{equation}
with the corresponding eigenfunction that is separable in cartesian coordinates
\begin{equation}
    {\mathcal A_i} = X_i(x)Y_i(y)Z_i(z).
\end{equation}
If the ratio between any two of the three lengths $L_x, L_y$ or $L_z$ are integers, then there can be degeneracies. To numerically lift these degeneracies, the hidden symmetry needs to be eliminated. In this specific case, we can do so by either extending the cavity boundary to have a finite thickness, as was done in our earlier discussion on the embedded cavity, or by introducing a small perturbation to the cavity shape. 
To demonstrate this, we consider again the perfect cavity with dimensions $L_x\!\!:\!\!L_y\!\!:\!\!L_z=1\!:\!1.5\!:\!2$.
In Fig.\,\ref{fig:acc_degeneracy}a, where the $3^{\text{rd}}$ and $4^{\text{th}}$ modes of the unperturbed cavity are plotted, we can see that each field component of the two modes is not characterized by any single set of quantum numbers $(n_x, n_y, n_z)$, but rather a linear combination of the two accidentally degenerate modes. In Fig.\,\ref{fig:acc_degeneracy}b, we plot the same modes, but the cavity is now slightly perturbed by an amount $\Delta L_z=0.01$ along $z$, so that now  $L'_z= L_z + \Delta L_z = 2.01$. The degeneracy is then lifted, and each of the two modes is now well-described by a distinct set of quantum numbers, which helps demonstrate the sensitivity of our numerical scheme to small changes in the geometry of the system simulated.

\subsection{Calculations for closed multiscale systems}\label{sec:multiscale_sys}
\begin{figure*}[t]
    \centering
    \includegraphics[scale=0.13]{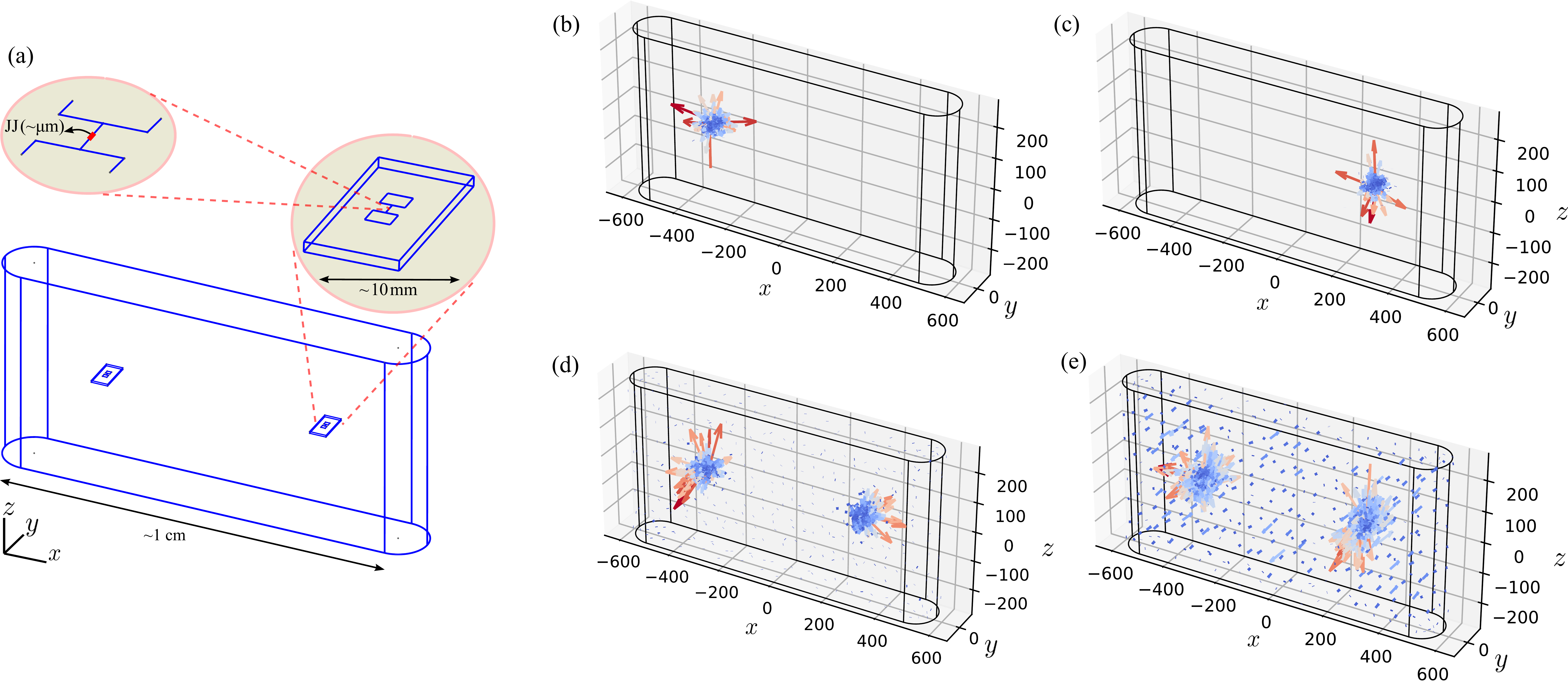}
    \caption{(a) Schematic of a system consisting of a three-dimensional superconducting cavity containing two spatially separated chips. Each chip is a dielectric substrate on top of which a Josephson qubit is mounted. The qubit is made of two superconducting capacitor pads connected together by a bridge containing a Josephson junction. A few exemplary modes of the system are plotted, such as (b) the mode where only qubit 1 participates, (c) only qubit 2 participates, (d) both qubits are activated, and (e) a hybridization of the two qubits and the cavity field. The units for all the axes in plots (b)-(e) are in millimeters.}
    \label{fig:3dcav_twoqbits}
\end{figure*}

So far, in this article, we have validated the accuracy of DEC by applying the method to simple systems with a large degree of symmetry and comparing the results to analytical solutions. The strength of DEC, however, lies in its ability to correctly capture the properties of systems containing multiple spatial scales through coarse-graining. 
The ability to simulate the full multiscale system is especially important for analyzing package modes and how they affect the on-chip operations \cite{huang2021microwavepackage} - a study that requires a computational mesh  capable of encompassing multiple spatial scales to cover the volume of the entire package while simultaneously resolve the fine details on the quantum chip.
Another example of a multiscale system is a three-dimensional superconducting cavity containing one or a few superconducting chips. The dimensions of a 3D cavity are typically on the order of $\sim\! 1$\,cm, while a dielectric substrate holding the qubits is a few mm in size, and a qubit itself can have its smallest components in the range from micrometers down to tens of nanometers. 
To model such systems, one possible workaround to avoid computational bottlenecks is dividing the problem into individual simulations of separate parts. However, due to hybridization, the spectral characteristics of the entire system composed of these devices being in the vicinity of each other can be vastly different from that of the individual components. Hence, it is imperative to be able to model the entire structure and directly extract its modes.

A schematic of the system we shall consider is shown in Fig.\,\ref{fig:3dcav_twoqbits}a. It consists of a 3D cavity containing two dielectric substrates. A transmon qubit is mounted on each substrate. Each transmon qubit is composed of two superconducting capacitor pads connected by a Josephson junction. It is a well-known property of Josephson junctions that their dynamics are well characterized by the coarse-grained phase across it~\cite{JosephsonReview_1965}. In other words, to an observer outside and away from the junction and who can only make measurements with a limited precision, the detailed dynamics inside the junction is unimportant to the dynamics that results from its interaction with the surrounding electromagnetic environment. This property can be utilized in coarse-grained calculations using DEC to reduce the complexity of the mesh needed; this is done by modeling a junction by a single edge instead of finely meshing the junction geometry. In the calculations of electromagnetic modes, this is equivalent to the linearization of the junction across the edge. To obtain modes whose wavelengths are orders of magnitude larger than the longitudinal size of the junction, the fine details of the material distribution within the junction are irrelevant and do not need to be resolved during meshing. In Fig.\,\ref{fig:3dcav_twoqbits}b-\ref{fig:3dcav_twoqbits}e, a few example modes of this multiscale system are shown. In this calculation, we have set the dimensions of the cavity to be \mbox{$1160\!\times\!160\!\times\!550$}\,(mm), while the sizes of the two identical substrates are both $25\!\times\!50\!\times\!3$\,(mm) and are separated by a distance of $600$\,mm. The size of the capacitor pads of the qubits is $6\!\times\!3\!\times\!0.1$\,(mm), and the two pads belonging to the same qubit are connected by a $0.2$\,mm-long bridge. We show the results for different types of modes that the system supports; in Figs.\,\ref{fig:3dcav_twoqbits}b and \ref{fig:3dcav_twoqbits}c, single qubit modes are shown, where the hybridization with the other qubit and the cavity field is suppressed. On the other hand, there can also be modes in which both qubits participate, such as the one shown in Fig.\,\ref{fig:3dcav_twoqbits}d. Finally, in Fig.\,\ref{fig:3dcav_twoqbits}e, we demonstrate  a hybridized mode in which both the qubits and the cavity participate.

\section{Calculations of open modes}
So far in this article, we have considered only closed systems, i.e. systems that only support modes that decay exponentially beyond their boundaries (e.g. a system surrounded by a superconducting enclosure much thicker than the relevant penetration depths). Here, we would like to extend DEC-QED to allow for a mathematically rigorous consideration of radiative losses. The correct modeling of open systems, where fields can propagate into and out of a confined domain, has been of great interest since it is directly related to the quantification of qubit lifetimes and radiative losses. A superconducting qubit placed in an open cavity acquires a relaxation rate through hybridization of the qubit with the electromagnetic modes of the cavity. Therefore its relaxation rate is given by the imaginary part of the complex-valued eigenfrequency of the qubit-like mode, while the difference between the real part in the frequency of a hybridized mode with the internal natural frequency of the qubit gives the Lamb shift~\cite{malekakhlagh2016non}. In a similar vein, radiative loss of cavity modes is given by the imaginary parts of the cavity-like modes and is modified through hybridization with the qubit.
Beyond the realm of superconducting microwave circuits, the problem has also been important to the broader scope of electromagnetic devices. We are particularly interested in the implementation of finite, open boundaries that are transparent so that fields can propagate through without any reflection. Moreover, the formulation needs to be compatible with the eventual second quantization of the electromagnetic field everywhere within the system, including the boundary itself. Here, we provide the derivations of two such approaches and present numerical demonstrations.

\subsection{Green's boundary integral formulation for scalar fields}\label{sec:scalargreen_BIM}
\begin{figure}[t]
    \centering
    \includegraphics[scale=0.35]{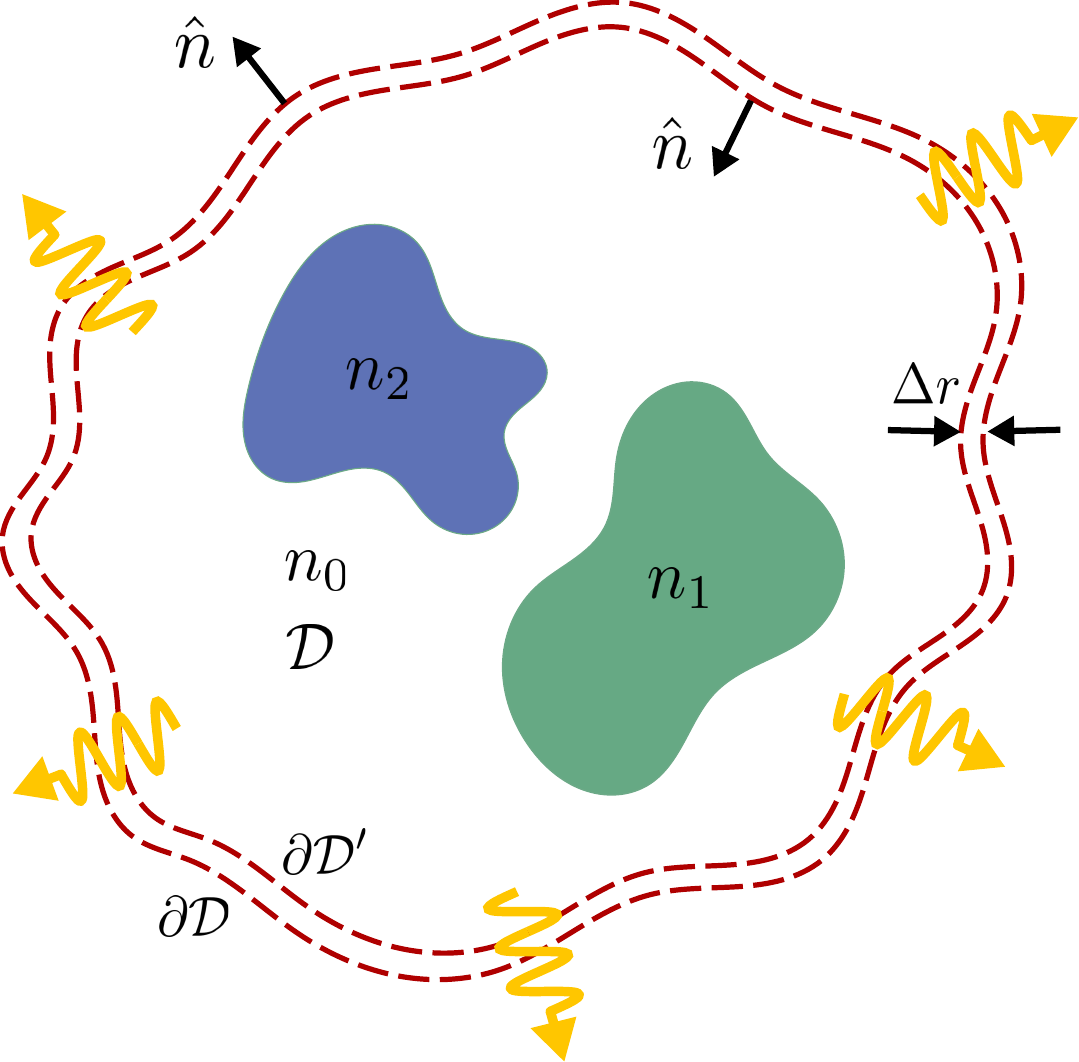}
    \caption{Schematic of an open system $D$ consisting of multiple material regions with refractive indices $n_i$. The {\it Surface of Last Scattering} (SoLS), the imaginary (transparent) boundary $\partial\mathcal{D}$ (the dashed contour) containing all material systems can have an arbitrary shape as long as all regions of interest are enclosed within it. The secondary boundary $\partial\mathcal{D}'$ is created by contracting everywhere the original boundary $\partial\mathcal{D}$ by an amount $\Delta r$.}
    \label{fig:schematic_openbc}
\end{figure}
First, we discuss an implementation of open boundaries using Green's boundary integral formalism. The definition of radiative boundary conditions is analytically stated only at spatial infinity $(r\!\rightarrow\!\infty)$. However, our goal is to state a mathematically equivalent condition at a finite distance from the cavity to keep the computationally volume as small as possible. We refer to this imaginary boundary that includes all material systems within its volume as the ``Surface of Last Scattering" (SoLS). We then employ the analytically known free-space frequency-domain Green’s function to ‘propagate’ back the boundary condition at infinity to the chosen SoLS.

Consider a domain $\mathcal{D}$ consisting of possibly multiple disjoint regions all enclosed by an imaginary boundary surface. Before considering the spectral problem associated with the vector Helmholtz equation \ref{eq:vectorHelmholtz}, we consider the warm-up problem of the  Helmholtz equation for a scalar field $\phi({\mathbf r})$: 
\begin{equation}\label{eq:scalarHelmholtz}
    \nabla^2\phi({\mathbf r}) + n_i^2({\mathbf r})k^2\phi({\mathbf r}) = 0,
\end{equation}
where $n({\mathbf r})$ is the dielectric function of the $i^\text{th}$ region $\Gamma_i$, and $k$ is again the wavenumber. 
The Green's function of the Helmholtz operator reads
\begin{align}\label{eq:green_function_def}
    G({\bf r, r'}, k) = \begin{cases}
      -\frac{i}{4}H_0^{(1)}(n_ik|{\mathbf r}-{\mathbf r'}|) & \text{in 2D,}\\
      -\frac{e^{in_ik|{\mathbf r}-{\mathbf r'}|}}{4\pi|{\mathbf r}-{\mathbf r'}|} & \text{in 3D,}
    \end{cases} 
\end{align}
where $H_0^{(1)}$ is the Hankel function of the first kind. At first glance, the Green's functions in Eq.\,(\ref{eq:green_function_def}) and their derivatives diverging at ${\mathbf r}\!=\!{\mathbf r'}$ might seem like a problem. This difficulty can be circumvented by casting the boundary condition in an integral form to regularize the Green's function singularity~\cite{kosztin1997boundary, pham2023singularfields}. 
Upon applying Green's second identity and taking the limit as \mbox{${\mathbf r'}\rightarrow \partial\Gamma_i$}, the field value at a point ${\mathbf r'}$ that lies on the boundary $\partial \Gamma_i$ of a domain is~\cite{wiersig2002boundary}
\begin{equation}\label{eq:scalar_green_boundaryInt}
\phi({\bf r'}) = 2\mathcal{P}^{d-1}\!\!\int_{\partial\Gamma_i}\!\!\big[\phi({\bf r})\nabla G({\bf r, r'}, k) - G({\bf r, r'}, k)\nabla\phi({\bf r}) \big]\cdot {\mathbf ds},
\end{equation}
where $\mathcal{P}^{d-1}$ indicates the $(d\!-\!1)$-dimensional Cauchy principal value integral, with $d$ being the dimension of the domain. A detailed derivation of Eq.\,(\ref{eq:scalar_green_boundaryInt}) is given in Appendix \ref{append:green_integral} for completeness. In Eq.\,(\ref{eq:scalar_green_boundaryInt}), the field at a point on the boundary is determined by the field and its gradient everywhere else on the same boundary. Eq.\,(\ref{eq:scalar_green_boundaryInt}) is applicable to smooth boundaries of any form and shape. They can also be made up of different disjoint but closed segments. 

To use Eq.\,(\ref{eq:scalar_green_boundaryInt}) for applying boundary condition at the domain boundary $\partial\mathcal{D}$, we need to numerically evaluate the normal gradient of the boundary field in a way that is self-contained within the field living on the boundary itself. To do so, consider a secondary boundary $\partial\mathcal{D}'$ that is formed by contracting $\partial\mathcal{D}$ by an amount $\Delta r$ along the normal direction everywhere on the surface, as shown schematically in Fig.\,\ref{fig:schematic_openbc}. 
There is now a thin layer bounded by $\partial\mathcal{D}\cup \partial\mathcal{D}'$ over which Green's identity can be performed to obtain Eq.\,(\ref{eq:scalar_green_boundaryInt}) that determines $\phi({\bf r'})$ for every point ${\mathbf r'}\in \partial\mathcal{D}$. The surface integral is now done over \mbox{$\partial\mathcal{D}\cup\partial\mathcal{D}'$}. Note that during the evaluation of Eq.\,(\ref{eq:scalar_green_boundaryInt}) the unit normal vector $\hat{n}$ on $\partial\mathcal{D}'$ points into $\mathcal{D}$.

The application of open boundaries using Eq.\,(\ref{eq:scalar_green_boundaryInt}) results in a boundary condition that depends parametrically on $k$. To solve for the eigenmodes of Eq.\,(\ref{eq:scalarHelmholtz}), we rewrite it into the following matrix representation
\begin{equation}\label{eq:Helmholtz_matrixform}
    \big[\mathbb{H}_s(k)+ k^2\big]\Phi = 0, 
\end{equation}
where $\mathbb{H}_s(k)$ is the scalar Helmholtz operator, and the vector $\Phi$ contains the scalar field evaluated at all vertices in the computational mesh. 
The task of finding the eigenvalues of $\mathbb{H}_s(k)$ can be cast into a singular value decomposition (SVD) problem~\cite{Backer_2003}. The eigenvalues $k_\alpha$ correspond to locations of local minima in the lowest singular value of \mbox{$\mathbb{M}_s(k) = \mathbb{H}_s(k)+ k^2$}.

\begin{figure}[t]
    \centering
    \includegraphics[scale=0.8]{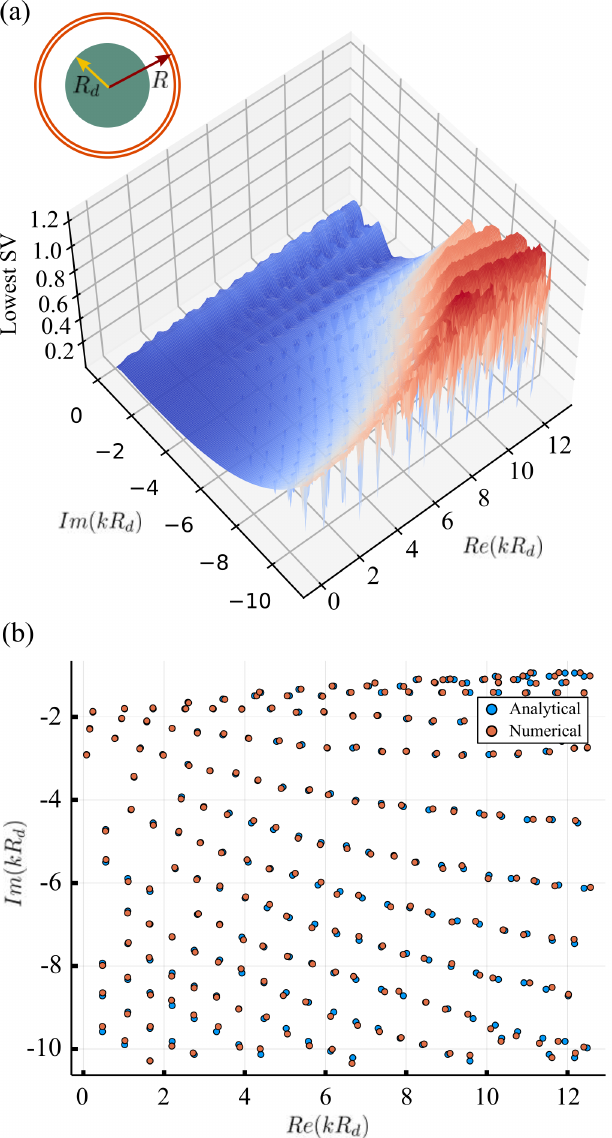}
    \caption{(a) The lowest singular value of $\mathbb{M}_s(k)$ computed for a sample range of $k$. The physical system is a dielectric disk with radius $R_d =5$\,mm and $n=1.5$. A schematic of the system is given in the top-left corner. (b) The eigenvalues $k_n$ of $\mathbb{H}_s(k)$, are indicated by the locations of the local minima of the lowest singular value of $\mathbb{M}_s(k)$. The semi-analytical results are plotted in blue, while the numerically computed values are plotted in orange.}
    \label{fig:lowest_SVs_2dscalar}
\end{figure}

\begin{figure*}[t]
    \centering
    \includegraphics[scale=0.26]{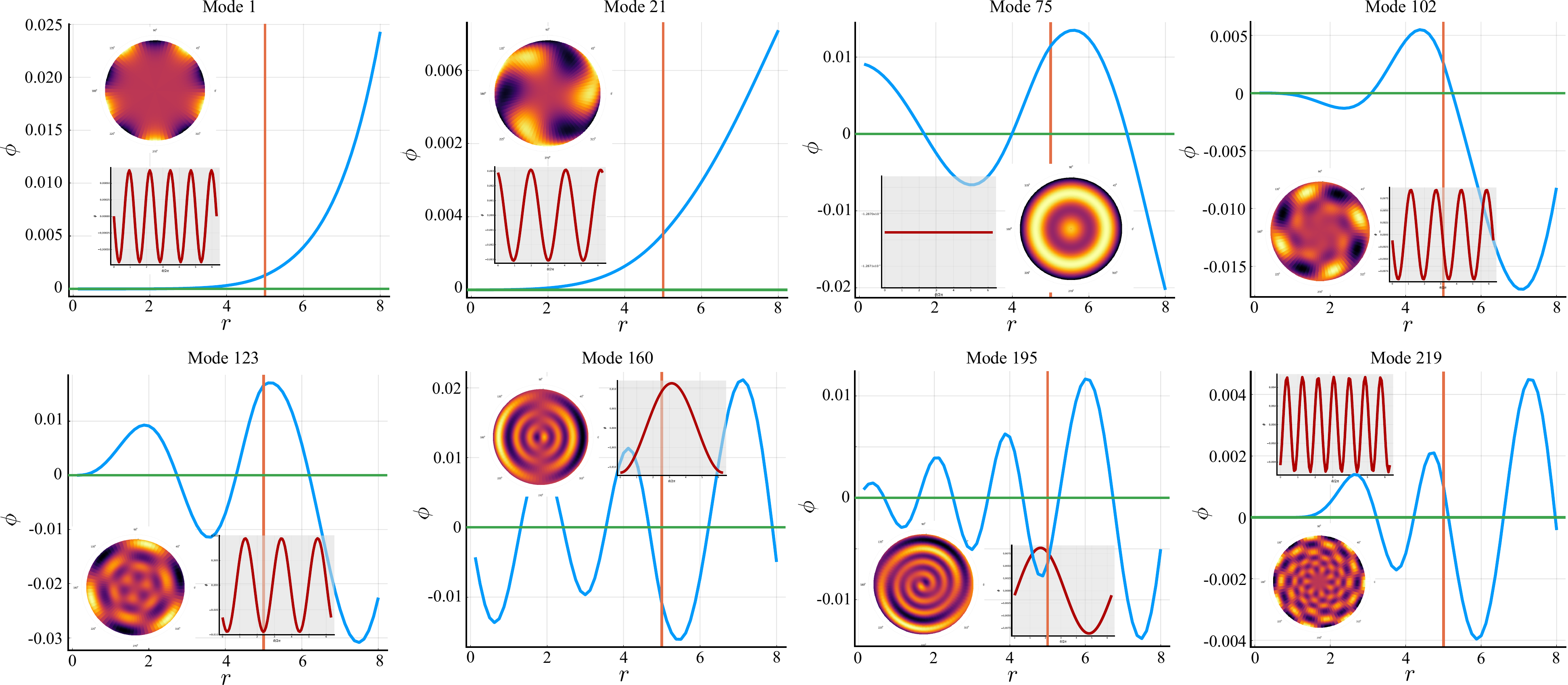}
    \caption{A selection of open modes of the scalar Helmholtz equation on a dielectric disk whose radius is $R_d=5$\,mm and refractive index is $n=1.5$. In each panel: the surface plot shows the distribution of $Re(\phi)$ in the entire domain inside the boundary radius $R$. The blue curve shows the value of $Re(\phi)$ at a fixed angle $\theta = 4\pi/5$ as a function of the radial coordinate $r$. The units for the horizontal axes are in mm. The orange vertical line indicates the edge of the disk. The red curve shows the field at a fixed radius $r=4$ as a function of the angular coordinate $\theta$.}
    \label{fig:2dscalar_modes}
\end{figure*}

We demonstrate the formulation presented here through the calculations of the open modes of Eq.\,(\ref{eq:scalarHelmholtz}) applied to a 2D dielectric disk placed in a vacuum. The disk has radius $R_d\!=\!5$\,mm and $n=1.5$, while the boundary is chosen to be a concentric circle with radius $R\!=\!8$\,mm. The results for the lowest singular value of $\mathbb{M}_s(k)$ is shown in Fig.\,\ref{fig:lowest_SVs_2dscalar}a for $k$ within the range $0\leq Re(kR_d)\leq 4\pi$ and $-3.5\pi\leq Im(kR_d)< 0$. Since the system supports incoming and outgoing modes equally, we only need to focus on the outgoing modes, whose real part of $k$ is positive. The distribution of the eigenvalues has a mirror symmetry across $Re(k)=0$, and the incoming modes are only different from the outgoing ones by the sign of $Re(k)$. The local minima in Fig.\,\ref{fig:lowest_SVs_2dscalar}a, whose locations correspond to the eigenvalues $k_\alpha$, are collected and plotted in Fig.\,\ref{fig:lowest_SVs_2dscalar}b. Note that since all modes are radiative, their eigenvalues are complex with the imaginary parts giving the corresponding relaxation rates.
We also compare these numerically computed eigenvalues with the semi-analytical solutions obtained by solving the transcendental equation arising from matching the continuity condition of the field and its normal derivative at the edge of the disk. As seen in Fig.\,\ref{fig:lowest_SVs_2dscalar}b, the numerically obtained eigenvalues (orange) agree with those obtained through the analytical equation (blue) over a wide range of $k$.

The field distributions of a few eigenmodes are presented in Fig.\,\ref{fig:2dscalar_modes}, where the real part of the fields, $Re(\phi)$, are plotted.
The numerical labeling of the modes are done first in ascending order of $Im(k_{\alpha}R_d)$ within the range $\{-3.5\pi,0\}$, then in ascending order of $Re(k_{\alpha}R_d)$ in the range $\{0,4\pi\}$. The features exhibited by these distributions can be understood through analytical considerations; in polar coordinates, one can write the fundamental solution to the scalar Helmholtz equation as $\phi(r,\theta)\!=\!R(r)\Theta(\theta)$. The angular dependence of the fields is of the form \mbox{$\Theta(\theta) \sim e^{\pm im\theta}$}, which is confirmed by the curves in red in Fig.\,\ref{fig:2dscalar_modes}, where the angular dependence of the fields are plotted at a fixed radius.
The radial component, on the other hand, is given by \mbox{$R(r)\sim Z_m(nkr)$}, where $Z_m$ is either the Bessel function of the first kind $J_m$, Bessel function of the second kind $Y_m$, or the Hankel functions $H^{(1,2)}_m$. Inside the dielectric disk, the radial component of the field is given by $J_m$, while the field outside the disk is described by $H_m^{(1)}$ \big($H_m^{(2)}\big)$ for an outgoing (incoming) wave. This radial dependence is shown in the blue curves in Fig.\,\ref{fig:2dscalar_modes}.

\subsection{Green's boundary integral formulation for vector fields}\label{sec:vectorgreen_BIM}
A problem of greater interest is, however, the implementation of open boundaries for vector fields. For this, we have extended Green's boundary integral method for scalar fields to address the vector Helmholtz problem as well. Consider a divergence-free vector field $\bm{\mathcal A}({\mathbf r'})$ that satisfies Eq.\,(\ref{eq:vectorHelmholtz}) defined over the domain $\mathcal{D}$. The dependence of the field value at a location ${\mathbf r'}\in\partial D$ on the field everywhere else on the same boundary is given by
\begin{align}\label{eq:vector_green_boundaryInt}
\bm{\mathcal A}({\mathbf r'}) = -2&\mathcal{P}^{d-1}\int_{\partial D}\bigg\{G({\mathbf r},{\mathbf r'},k)\big[({\mathbf \nabla}\times \bm{\mathcal A})\times \hat{n}\big] \\
&- \nabla G({\mathbf r},{\mathbf r'},k)(\bm{\mathcal A}\cdot\hat{n}) -\nabla G\times({\bm{\mathcal A} \times\hat{n}}) \bigg\}ds, \nonumber
\end{align}
where $G({\mathbf r},{\mathbf r'},k)$ is again the scalar Green's function as defined in Eq.\,(\ref{eq:green_function_def}), and the principal value integral is generalized to the vector case. A detailed derivation of Eq.\,(\ref{eq:vector_green_boundaryInt}) is given in Appendix \ref{append:green_integral}.

\begin{figure*}[t]
    \centering
    \includegraphics[scale=0.35]{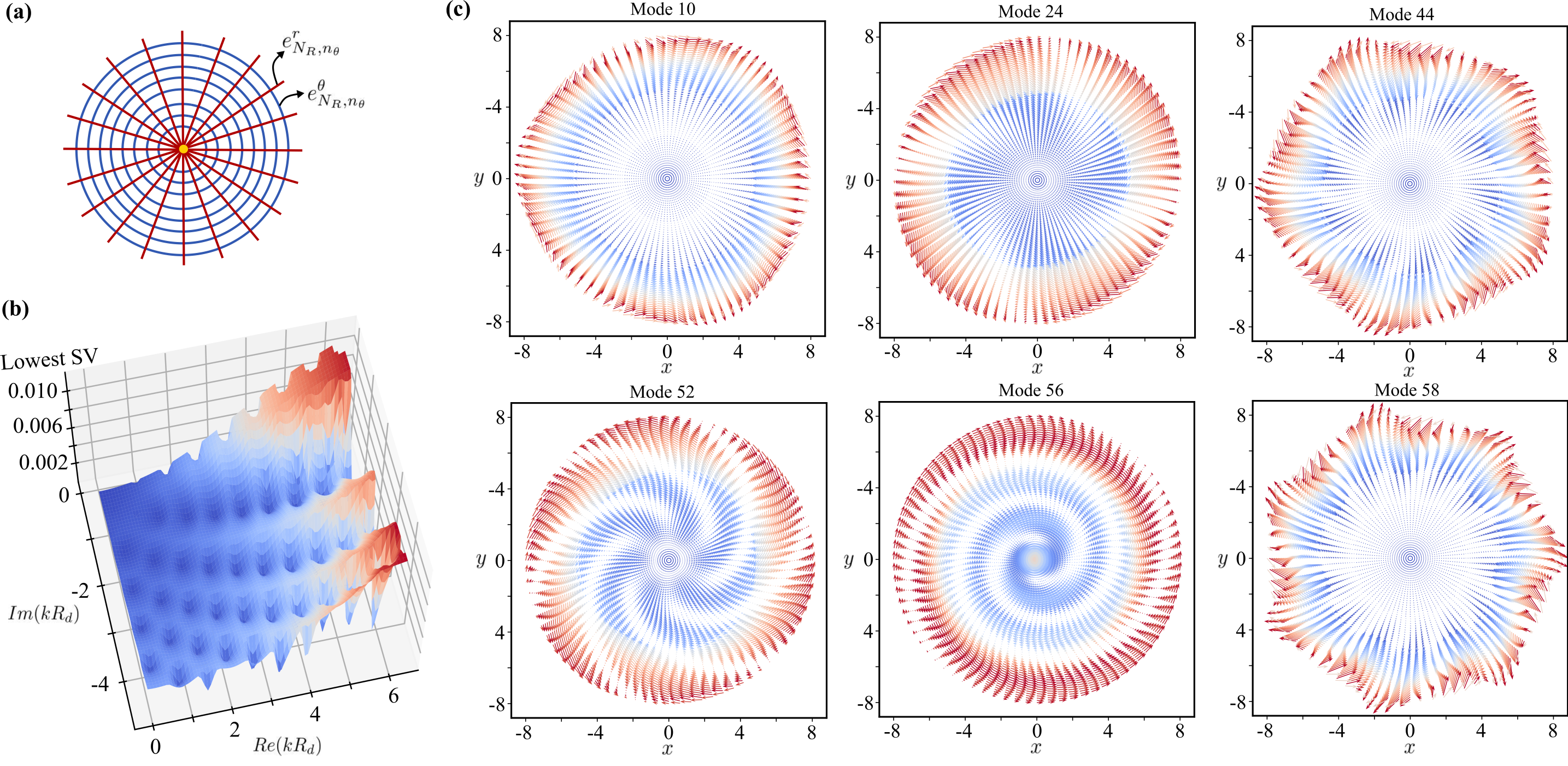}
    \caption{Open-mode calculations of the vector Helmholtz problem for a dielectric disk placed in a vacuum. The disk has radius $R_d=5$\,mm and $n=1.5$. (a) A schematic of the polar grid used in the calculation. (b) The surface plot of the lowest singular value of the operator $\mathbb{M}_v(k)$ as a function of $k$. The local minima indicate the eigenvalues of the vector Helmholtz operator $\mathbb{H}_v(k)$. (c) The distributions of $Re(\bm{\mathcal A})$ in a number of radiative modes (the units for all axes are in mm).}
    \label{fig:2dvector_modes}
\end{figure*}

We demonstrate the implementation of the boundary condition in Eq.\,(\ref{eq:vector_green_boundaryInt}) in DEC by calculating the modes of Eq.\,(\ref{eq:vectorHelmholtz}) applied to the same 2D dielectric disk studied in the scalar case. Consider a discretization of the computational domain into a polar grid that has $N_R$ and $N_\theta$ vertices along each radial ray and circle, respectively. In the case of open BC, it is natural to allow the mesh to also be ``open" such that there are additional edges on the physical boundary of the system that protrude outwards in the direction normal to the surface, as shown in Fig.\,\ref{fig:2dvector_modes}a.
A primal edge $e$ in this setup is identified by $n_r$ and $n_\theta$, its radial and angular indices, accompanied by the superscript $\{r,\theta\}$ that indicates the direction of the edge. In DEC, applying BC to a vector field translates to the application of BC for the normal and tangential edge fields $\Phi(e^r_{N_R,n_\theta})$ and $\Phi(e^\theta_{N_R,n_\theta})$ living on the boundary edges. The condition in Eq.\,(\ref{eq:vector_green_boundaryInt}), when applied to a 2D circular boundary, then becomes
\begin{widetext}
\begin{align}
\Phi(e^r_{N_R,n_\theta}) =  -\Phi(e^r_{N_R-1,n_\theta}) + \sum_{\theta} 2\bigg\{\frac{\partial G}{\partial r}\Big[\Phi^r_{N_R+1}(\theta) + \Phi_{N_R}^r(\theta) \Big]R\Delta\theta - \frac{\partial G}{\partial\theta}\Big[\Phi^\theta_{n_\theta-1}(R) + \Phi^\theta_{n_\theta}(R) \Big]\frac{\Delta R}{R} \bigg\}, 
\end{align}
for the radial boundary edge, and
\begin{align}
\Phi(e^\theta_{N_R,n_\theta}) =  -\Phi(e^\theta_{N_R,n_\theta-1}) - \sum_{\theta}&\bigg\{ 2\Delta\theta\bigg[G({\mathbf r},{\mathbf r'},k)\Big(1 + \frac{R}{\Delta R}\Big) - R\frac{\partial G}{\partial r} \bigg]\Big(\Phi_{n_\theta}^{\theta}(R) + \Phi_{n_\theta-1}^\theta(R) \Big) \\
 &- 2G\frac{R\Delta\theta}{\Delta R}\Big(\Phi_{n_\theta}^{\theta}(R-\Delta R) + \Phi_{n_\theta-1}^\theta(R-\Delta R) \Big) - 2R\frac{\Delta\theta^2}{\Delta R}\frac{\partial G}{\partial\theta}\Big[\Phi_{N_R+1}^r(\theta) + \Phi_{N_R}^r(\theta) \Big] \nonumber\\
&- G\frac{R\Delta\theta}{\Delta R}\Big[\Phi_{N_R+1}^r(\theta+\Delta\theta) + \Phi_{N_R}^r(\theta + \Delta\theta) - \Phi_{N_R+1}^r(\theta-\Delta\theta) - \Phi_{N_R}^r(\theta-\Delta\theta) \Big] \bigg\} \nonumber
\end{align}
for the angular boundary edges, where $\Delta R$ and $\Delta \theta$ are the spacing between consecutive grid points on a radial ray and circle, respectively.
\end{widetext}
Similar to the scalar problem, to search for the eigenvalues, we compute the lowest singular values of the operator $\mathbb{M}_v(k) =\mathbb{H}_v(k)-k^2$, where $\mathbb{H}_v(k)$ is the vector Helmholtz operator. The distribution of the local minima of  $\mathbb{M}_v(k)$ in $k$-space is shown in Fig.\,\ref{fig:2dvector_modes}b. The field distributions of a few randomly selected eigenmodes are presented in Fig.\,\ref{fig:2dvector_modes}c, where the plotted vector fields exhibit the correct behavior of how radiations permeate and escape a dielectric disk.

\begin{figure*}[t]
    \centering
    \includegraphics[scale=0.38]{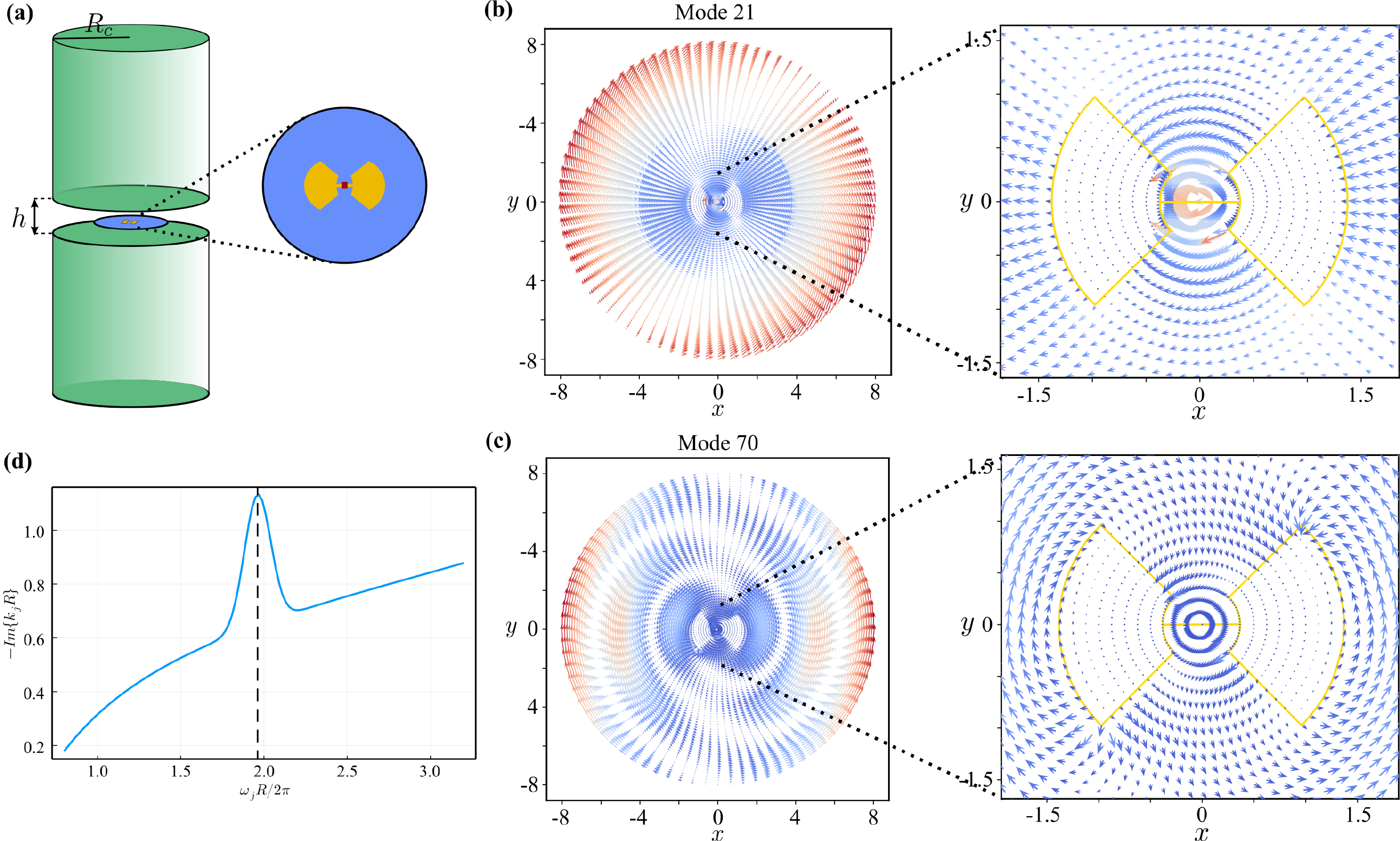}
    \caption{Calculations of radiative modes of a transmon qubit sandwiched between two cylindrical superconducting capacitors. (a) The schematic of the system considered; the qubit is mounted on a circular dielectric disk that has radius $R_d=4$\,mm, refractive index $n=1.5$. The qubit is composed of two superconducting islands that are $0.75$\,mm away from each other and connected by a bridge containing a junction in the middle. The width of both the bridge and the junction is 50$\mu$m. (b) An example of a ``dipole" mode; the qubit acts like a dipole with the field lines starting from the surface of one island and ending on the surface of another island. (c) An example of a mode where the field lines flow around the individual capacitor islands with no normal component at each island's surface. The boundary of the qubit is painted in yellow in the enlarged insets, and all the axes of the plots are in units of mm. (d) Spontaneous emission rate of a qubit-like mode as a function of the qubit frequency.}
    \label{fig:cylindrical_cap_qbit}
\end{figure*}

To demonstrate the versatility of DEC in computing radiative modes of arbitrarily shaped systems, we apply the formulation to the modeling of a superconducting qubit chip placed in the region between two cylindrical capacitors, as schematically shown in Fig.\,\ref{fig:cylindrical_cap_qbit}a. 
The qubit is made of two superconducting islands that are separated by a distance of $0.75$\,mm and connected by a bridge that contains a Josephson junction. The width of both the bridge and the junction is $50\mu$m, and the qubit is mounted on a circular dielectric disk that has a radius $R_d=4$\,mm with refractive index $n=1.5$. We assume the two ideal superconducting cylindrical capacitors extend indefinitely towards both ends, and the qubit is placed in the middle of the slit separating them. The width $h$ of the vacuum spacing between the two cylinders is taken to be very small compared to their radius $R_c$. This allows us to effectively discretize the distance between the surfaces of the two cylinders by two edges. Due to mirror symmetry at $z=0$, the amplitude of the coarse-grained field on one $z$-edge is identical to that of its mirroring edge. This allows us to decouple the in-plane field component from the out-of-plane component, and we can focus on solving for the modes of Eq.\,(\ref{eq:discrete_vectorhelmholtz}) applied to the 2D in-plane field on the $z=0$ slice where the substrate containing the qubit is located. 
We choose the computational boundary to be a circle whose radius $R=8$\,mm is half that of the cylindrical capacitors $R_c$. There are two notable types of modes, whose examples are shown in Figs.\,\ref{fig:cylindrical_cap_qbit}b-\ref{fig:cylindrical_cap_qbit}c. Due to the geometry of the qubit being two islands connected by a bridge, it can act as a dipole. A mode that behaves this way is shown in Fig.\,\ref{fig:cylindrical_cap_qbit}b, where the field lines exit in the normal direction from one island and enter the other island. The field also decays in the bulks of the islands at the rate determined by their penetration depths. In another scenario, due to their separation, the two islands can behave as individual superconducting objects around which the field lines flow with no normal component at the surfaces of the islands. A demonstration of such a mode is shown in Fig.\,\ref{fig:cylindrical_cap_qbit}c. Finally, in Fig.\,\ref{fig:cylindrical_cap_qbit}d, we show how one can tune the internal resonance frequency of the qubit through a cavity resonance, exposing the Purcell enhancement of the qubit relaxation rate \cite{purcell1946resonance}. The relaxation rate due to radiative loss of a qubitlike mode, i.e. the spontaneous emission rate, is plotted as a function of the qubit frequency. We observe a finite peak as the qubit freqency is on-resonant with a cavitylike mode, signifiying an enhanced emission rate due to the hybridization of this qubit mode with the cavity field \cite{cuttoff_free_cqed_2017}. In this case, the Purcell enhancement factor with respect to its value away from resonance is not large, because we intentionally chose a very (radiatively) lossy cavity represented by placing two cylinders facing each other, a situation that corresponds to the regime of overlapping resonances (\textit{finesse} $\ll 1$). 

As was mentioned earlier, the formulation for calculating radiative fields using Green's function method is flexible in terms of the shape and topology of boundary surfaces it can be applied. The physical boundary of the system can be made of multiple, possibly disjoint, but closed segments that bound a non-simply connected structure. However, the vector field being calculated has to be divergent-free as a prerequisite. Although this restriction is not a concern in many useful cases, such as the calculation of the electric field in a source-less region or of the magnetic vector potential in the Coulomb gauge, it is sometimes helpful to have the freedom of not necessarily choosing a divergence-less field. In the following section, we introduce an alternative formulation that achieves this. 

\subsection{Vector spherical harmonics expansion}\label{sec:VSH_openbc}
In this section, we discuss the implementation of open boundaries using vector spherical harmonics (VSH) expansions. The VSH~\cite{barrera1985vector} is an extension for vector fields of the perhaps more well-known scalar spherical harmonics. There are three sub-classes of these vectors, defined in spherical coordinates $(r,\theta,\varphi)$ as
\begin{align}
    {\mathbf Y_{lm}} &= Y_{lm}(\theta,\varphi)\hat{r} \\
    {\mathbf \Psi_{lm}} &= r\nabla Y_{lm}(\theta,\varphi) \\
    {\mathbf \Phi_{lm}} &= \mathbf{r}\times\nabla Y_{lm}(\theta,\varphi)
\end{align}
that altogether form an orthonormal and complete basis. Any vector field can then be expanded as follows
\begin{align}\label{eq:vsh_expand}
    \bm{\mathcal A}({\mathbf r}) = \sum_{l=0}^{\infty}\sum_{m=-l}^l \mathcal{A}^r_{lm}{\mathbf Y_{lm}} + \mathcal{A}_{lm}^{(1)}{\mathbf \Psi_{lm}} + \mathcal{A}_{lm}^{(2)}{\mathbf \Phi_{lm}},
\end{align}
where coefficients $\mathcal{A}^r_{lm}, \mathcal{A}_{lm}^{(1)}$ and $\mathcal{A}_{lm}^{(2)}$ are functions of the radial coordinate $r$. In Eq.\,(\ref{eq:vsh_expand}) above, of the three mutually orthogonal terms, the first term on the right-hand side is the radial component of the field, while the remaining two terms are angular contributions. Note that neither of the two angular terms aligns with azimuthal or polar directions but is a linear combination of both. Using Eq.\,(\ref{eq:vsh_expand}) as an ansatz for the field, Eq.\,(\ref{eq:vectorHelmholtz}) can then be split into three decoupled ODEs, each for one of the three coefficients. Their solutions are
\begin{align}
    \mathcal{A}_{lm}^r =& \frac{Z_{l+1/2}(kr)}{r^{3/2}}, \\
    \mathcal{A}_{lm}^{(1)} =& \frac{1}{\l(l+1)}\bigg\{\frac{Z_{l+1/2}(kr)}{r^{3/2}} + \frac{k}{\sqrt{r}}\Big[ Z_{l-1/2}(kr)\nonumber\\
     &\hspace{1in} -Z_{l+3/2}(kr) \Big] \bigg\}, \\
    \mathcal{A}_{lm}^{(2)} =& \frac{\pi}{2kr}Z_{l+1/2}(kr)
\end{align}
where $Z$ in general can be the Bessel function of the first kind $J$, the Bessel function of the second kind $Y$, or the Hankel functions $H^{(1,2)}$. For the outgoing(incoming) waves, the Hankel functions of the first(second) kind provide the appropriate description. After the VSH expansion in Eq.\,(\ref{eq:vsh_expand}), the implementation of the open boundary condition on the field $\bm{\mathcal A}$ now reduces to the BC on each of the three field components $\bm{\mathcal{A}^r}, \bm{\mathcal{A}^{(1)}}$ and $\bm{\mathcal{A}^{(2)}}$. Here we discuss the BC for $\mathcal{A}^r$ as an example, while the details about BCs for $\mathcal{A}^{(1)}$ and $\mathcal{A}^{(2)}$ are given in Appendix \ref{append:VSH}. Consider a spherical boundary of radius $R$. The series expansion for the radial component of the radiating field on the sphere is given by
\begin{align}\label{eq:VSH_fieldexpand}
    \bm{\mathcal A}^r(R,\theta,\varphi) &= \sum_{l,m}\Tilde{a}_{lm}^rA_{lm}^r(R){\mathbf Y_{lm}}(R, \theta, \varphi) \nonumber\\
    &= \sum_{l,m} \tilde{a}_{lm}^r\frac{H_{l+1/2}(kR)}{R^{-3/2}} Y_{lm}\hat{r},
\end{align}
where $\Tilde{a}_{lm}^r$ are the expansion coefficients. Performing an integral over the surface of the sphere and utilizing the orthonormality condition of VSH, we arrive at an expression for the expansion coefficients as follows
\begin{align}\label{eq:ar_coef}
    a^r_{lm} &\equiv \tilde{a}_{lm}^r H_{l+1/2}(kR)\\
     &= R^{3/2}\int d\Omega\bm{\mathcal A}^r\cdot{\mathbf Y^*_{lm}} \nonumber\\
     &= R^{-1/2}\sum_{i,j}\mathcal{A}^r(R,\theta_i,\varphi_j)Y_{lm}^*(\theta_i,\varphi_j)\Delta A(f), \nonumber
\end{align}
where $\int d\Omega$ is the solid angle integral that covers the entire sphere, and the last line of Eq.\,(\ref{eq:ar_coef}) above is the discretized version of the integral written as a sum over all the triangular patches $f$ whose centers are located at $(\theta_i,\varphi_j)$ on the spherical boundary. To apply the BCs, consider the sphere beneath the boundary that has radius $R\!-\!\Delta r$ (i.e. the second-to-last layer). The radial derivative of $\mathcal{A}^r$ is given by
\begin{align}\label{eq:Ar_deriv}
    &\frac{\mathcal{A}^r(R,\theta,\varphi)-\mathcal{A}^r(R-\Delta R,\theta,\varphi)}{\Delta R} = \\
    &\sum_{l,m}\tilde{a}^r_{lm}\bigg[\frac{-3}{2R^{5/2}}H_{l+1/2}(kR) + \frac{k}{R^{3/2}}H'_{l+1/2}(kR)\bigg]Y_{l,m}, \nonumber
\end{align}
which allows us to write the field at the boundary $\mathcal{A}^r(R,\theta, \varphi)$ in terms of the expansion coefficients $\tilde{a}^r_{lm}$ and the field at the layer with radius \mbox{$R\!-\!\Delta R$}. Using Eq.\,(\ref{eq:Ar_deriv}) and the expression for $\tilde{a}^r_{lm}$ derived in Eq.\,(\ref{eq:ar_coef}), the boundary condition for $\mathcal{A}^r$ is then
\begin{align}\label{eq:Ar_BC_VSH}
    \mathcal{A}^r&(R,\theta,\varphi) \!=\!\sum_{l.m}\!\sum_{i,j}\!\mathcal{A}^r(R\!-\!\Delta R, \theta_i,\varphi_j)Y^*_{lm}(\theta_i,\varphi_j)\frac{\Delta A(f)}{R^2}\nonumber\\
    &\times\bigg\{1 + \Delta R\Big[-\frac{3}{2R} + k\frac{H_{l-1/2}(kR)-H_{l+3/2}(kR)}{2H_{l+1/2}(kR)} \Big] \bigg\} \nonumber\\
    &\hspace{1.2in} \times Y_{lm}(\theta,\varphi).
\end{align}
The expression in Eq.\,(\ref{eq:Ar_BC_VSH}) is for the radial component of the field, which in DEC is a scalar living on vertices of the primal mesh. To arrive at  the appropriate BC for the edge field $\Phi(e)$ governed by the discrete vector Helmholtz equation given in Eq.\,(\ref{eq:discrete_vectorhelmholtz}), we need to rewrite $\mathcal{A}^r(R,\theta,\varphi)$ and $\mathcal{A}^r(R\!-\!\Delta R, \theta_i,\varphi_j)$ in terms of the edges belonging to the vertex located at $(R,\theta,\varphi)$. This means two consecutive spherical layers beneath the boundary are needed to compute $\mathcal{A}^r(R\!-\!\Delta R, \theta_i,\varphi_j)$. The procedure as a whole can be substantially simplified by designing the mesh so that the boundary surface and the two layers beneath it have identical triangulation patterns. This ensures that for any vertex $v_b$ located at $(R, \theta,\varphi)$ on the boundary surface, there is an edge $e^r$ oriented radially that connects $v_b$ with the vertex $v_b'$ at $(R\!-\!\Delta R, \theta,\varphi)$ that has the same angular coordinates as $v_b$. Similarly, there is always an edge connecting $v_b'$ with a vertex $v_b''$ at $(R\!-\!2\Delta R, \theta,\varphi)$. Note that we only need the three outmost layers of the mesh to be spherical and have identical triangulation patterns, while the rest of the internal domain can be freely and randomly discretized with tetrahedra. Similar to the discussion on Green's function approach in the previous section, here, for open boundaries, it is convenient to use an open mesh such that there are radial edges that protrude out of the boundary surface. The BC applied to these boundary radial edges $e^r_b$ is then
\begin{widetext}
\begin{align}\label{eq:Phi_r_BC_VSH}
    \Phi(e^r_b) = -\Phi(e^{r'}_b) + \sum_{e^r}\sum_{f\supset v|v\subset e^r}\sum_{l,m}\frac{\Delta A(f)}{3R^2}\Phi(e^r)\Bigg\{1 + \Delta R\bigg[-\frac{3}{2R} + k \frac{H_{l-1/2}(kR)-H_{l+3/2}(kR)}{2H_{l+1/2}(kR)} \bigg]Y^*_{lm}(f)Y_{lm}(\theta,\varphi) \Bigg\},
\end{align}
\end{widetext}
where $e^{r'}_b$ is the other radial edge that shares the boundary vertex with $e_b^r$. In Eq.\,(\ref{eq:Phi_r_BC_VSH}), the first sum is done over all radial edges $e^r$ living in the two layers beneath the boundary surface, while the second sum is done over all faces $f$ that share a vertex with $e^r$. This concludes our derivation for the open BC applied to a radial edge in the DEC framework. The boundary conditions for edges lying tangential to the surface are discussed in Appendix \ref{append:VSH}.

\begin{figure}[t]
    \centering
    \includegraphics[scale=0.35]{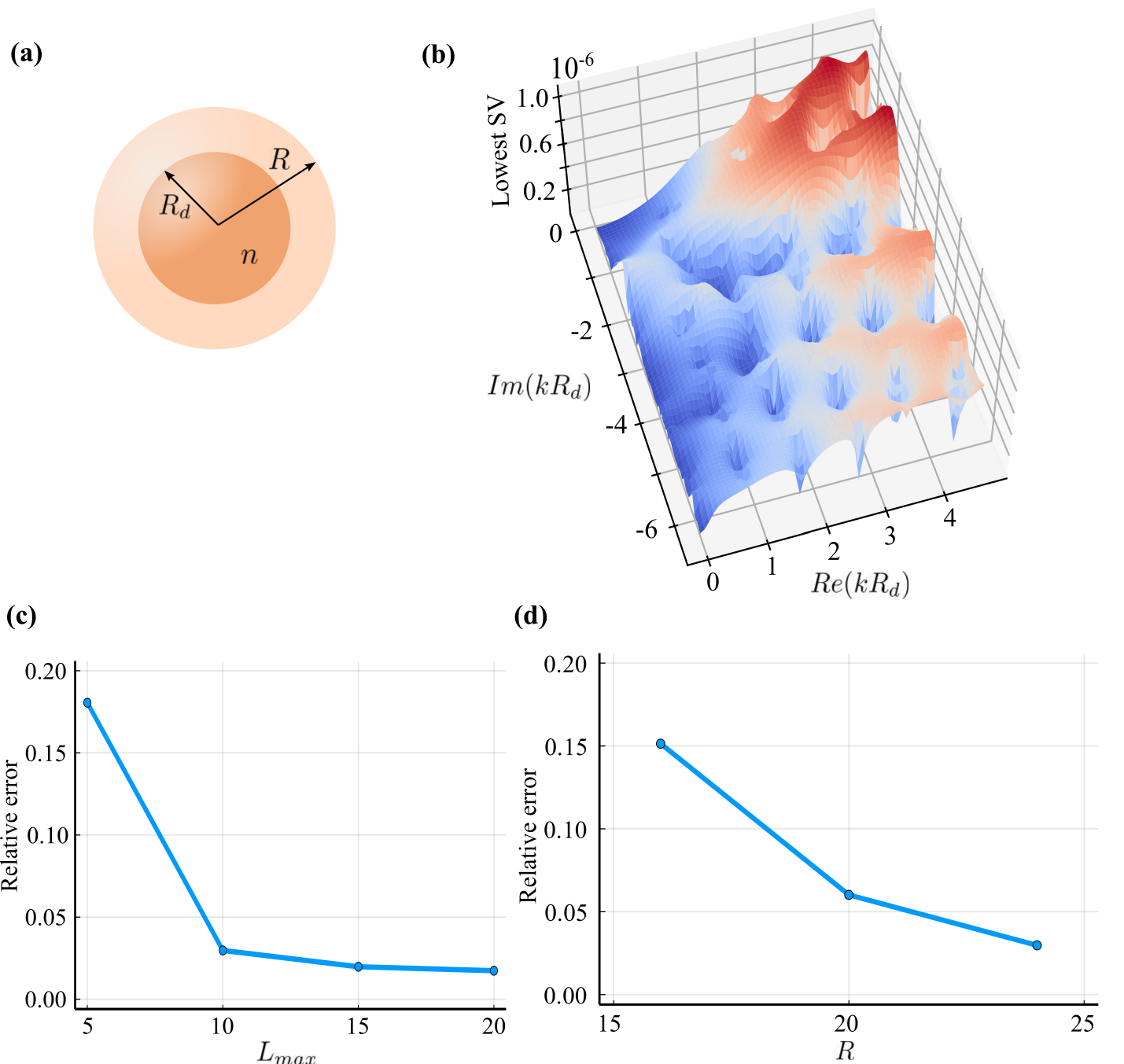}
    \caption{Results for the open-boundary modes of a dielectric microsphere. (a) Schematic of the system studied: a sphere of radius $R_d=12 \mu$m. (b) The lowest singular value of $\mathbb{M}_v(k)$ as a function of $k$. (c) Error of the lowest eigenvalue as a function of the largest ordering of Bessel terms. (d) Error of the lowest eigenvalue as a function of the boundary radius $R$.}
    \label{fig:3Dsphere_VSH}
\end{figure}

We apply the formulation developed here to calculate the modes for a three-dimensional dielectric microsphere surrounded by a vacuum. The radius of this dielectric sphere is kept fixed at \mbox{$R_d=12\mu$m} throughout all calculations discussed in this section. The computational boundary is an ``imaginary" spherical surface at a finite radius $R>R_d$, as schematically shown in Fig.\,\ref{fig:3Dsphere_VSH}a. The results of the eigenvalue search found through singular value calculations are shown in Fig.\,\ref{fig:3Dsphere_VSH}b, where we have chosen $R=24 \mu$m. With a primal mesh of $\approx 2900$ vertices and $\approx 18000$ edges, the local minima of the smallest singular value are of order $10^{-7}$, proving them to be reliable indicators of the eigenvalues of the system. The dielectric microsphere is chosen to demonstrate our formulation because there exist semi-analytical solutions that can be obtained by solving the transcendental equation arising from enforcing the continuity of the field and its normal derivative at the dielectric-vacuum interface. We utilize these solutions to investigate the efficacy of our method in a number of ways. First, the numerical implementation of the series expansion in Eq.\,(\ref{eq:VSH_fieldexpand}) requires a truncation in the number of terms considered. We, therefore, investigate the dependence of convergence rate on the number of Bessel functions included while keeping the mesh size fixed. The results can be seen in Fig.\,\ref{fig:3Dsphere_VSH}c, where we compute the error of the numerically obtained lowest eigenvalue with the semi-analytical solution. With a manageable number of $L_{max}=10$, where $L_{max}$ is the largest ordering of Bessel terms in the expansion, the error reduces to below $4\%$. Another important factor is how large the computational boundary needs to be in order to achieve precise solutions since accuracy is expected to improve as the boundary is moved further away from where the devices are concentrated, but doing so also increases mesh complexity, and hence the computational burden is exacerbated. In Fig.\,\ref{fig:3Dsphere_VSH}d, the convergence as a function of boundary radius $R$ of the lowest eigenvalue is plotted. We see that even with a tight boundary when $R=16\mu$m (while the dielectric sphere is kept at $12\mu$m) the error is relatively low at $15\%$, and for $R=2R_d$ it is reduced to $<4\%$. These results verify that our coarse-grained implementation of open boundaries using VSH is able to produce accurate solutions using a boundary of reasonable size.

The formulation based on VSH expansions introduced here allows for effective calculations of open modes while also removing the requirement of divergence-less fields. However, unlike Green's boundary integral formulation, which is adaptable to arbitrarily-shaped boundaries, this approach needs to be implemented on a spherical surface. This makes the formulation most ideal when applied to three-dimensional structures that have comparable lengths along three directions, such as 3D cavities. In some other specialized cases, this requirement may limit us from drawing the tightest boundary possible, which in turn might cause the computational domain to be larger than it needs to be. An example could be a coplanar waveguide whose length is much larger than its planar width, which in turn is much larger than the thickness along the cross-section of the waveguide. 
In such a case, replacing a spherical boundary with an ellipsoidal one is preferable since an ellipsoid has three tuning parameters that allow squeezing and stretching in three directions. Fortunately, the series expansion-based approach discussed here is extendable to ellipsoidal coordinates. This relies on the fact that the Helmholtz equation is separable in ellipsoidal coordinates as well, which allows for the expansion of functions, now in terms of ellipsoidal waves. Although the mathematically intensive discussion on how to generalize our method to ellipsoidal coordinates is beyond the scope of this article, in Appendix \ref{append:ellipsoidal_coords}, we provide a preamble by discussing a brief proof of the separability of the Helmholtz equation.

\section{Conclusion}
In this article, we have introduced a spectral theory for the modeling of mode hybridization and relaxation rates in general systems composed of superconducting and dielectric materials. In doing so, we have demonstrated the following important points:

(1) We show that the coarse-grained formulation of electrodynamics using discrete exterior calculus is adaptable to both simplicial and complex meshing strategies with equal precision. Although based on the calculations of coarse-grained quantities, the numerical scheme can still display the high sensitivity to small changes in the geometry and topology of the system that is warranted for instance in the categorization of degenerate modes arising from hidden symmetries when a small perturbation is introduced. 

(2) The method, which maps the problem of modeling vector fields to the calculation of averaged projections living on discrete edges, is particularly effective when applied to systems composed of objects of multiple spatial scales, such as those used in superconducting electronic circuits. Since the method keeps track of the average variables, it allows for more efficient meshing strategies that take into account the sizes of the Josephson junctions as well as the finite resolution imposed by a given measurement apparatus. This is an important advantage that this formulation offers because, here, the mesh does not need to be finer than the size of JJs or the measurement apparatus. Future work will focus on adaptive meshers that adapt to the finite spatio-temporal resolution provided by the measurement apparatus interrogating a particular dynamics, allowing for the efficient and accurate simulation of only what is measurable.  

(3) We introduce two implementations of open boundaries for the vector wave equation that are applicable to a wide range of electromagnetic systems. By drawing ``imaginary,'' transparent boundaries that are reasonably sized, we are able to faithfully produce the open modes that agree well with analytical solutions. We also demonstrate the flexibility of the formulation by applying it to the calculation of radiative modes of a superconducting qubit.  

The spectral theory presented here is suitable for the second quantization of the electromagnetic field and the time-dependent numerical simulation of the non-linear dynamics of open superconducting systems with arbitrary complexity. We expect the ability to accurately and efficiently extract relaxation rates, as well as the quantification of mode hybridization demonstrated here to be useful in the modeling and optimization of superconducting circuits.

\section{Acknowledgements}
We gratefully acknowledge discussions with Nicholas Bronn, Thomas G. McConkey, Anil N. Hirani, Thomas Maldonado, Zoe Zager, and Haley M. Cole. 
We are grateful to Benjamin Lienhard and Wentao Fan for giving the paper their critical read and for the insightful comments. 
We acknowledge support from the US Department of Energy, Office of Basic Energy Sciences, Division of Materials Sciences and Engineering, under Award No. DESC0016011. The simulations presented in this article were performed on computational resources managed and supported by Princeton Research Computing, a consortium of groups including the Princeton Institute for Computational Science and Engineering (PICSciE) and the Office of Information Technology's High-Performance Computing Center and Visualization Laboratory at Princeton University.

\appendix
\section{Linearization of the EHDS equations}\label{append:linearization}
\noindent In this Section, we provide a derivation for the linearization of the EHDS equations which leads to the spectral problem analyzed in this article. We start with the gauge-invariant form of the EHDS equations (Eq.\,(\ref{eq:Aprime_waveeq_rho}) and (\ref{eq:chargeconserve2_rho}) in the main text) and imposing that charge conservation is individually respected by both the charged condensate $\rho$ and the source $\rho_{src}$, meaning that the last two terms in Eq.\,(\ref{eq:chargeconserve2_rho}) cancel each other. Consider a time-harmonic source current ${\mathbf J}_{src}={\mathbf J_0}e^{i\omega t}$ that, to first order in frequency, results in fluctuations in the gauge-invariant field $\bm{\mathcal{A}} = \bm{\mathcal{A}_0}e^{i\omega t}$. According to Eq.\,(\ref{eq:chargeconserve2_rho}), this in turns leads to a fluctuating condensate 
\begin{equation}\label{eq:rho_harmonic}
    \rho =\rho_0 +\delta\rho_0e^{i\omega t}
\end{equation}
where $\delta\rho_0\ll\rho_0$. While the quadratic term $\partial_t\nabla|\bm{\mathcal{A}}|^2$ does not have a first-order contribution and can be neglected, the last term on the lhs of Eq.\,(\ref{eq:Aprime_waveeq_rho}), which is related to the rate of change of the quantum pressure, needs careful consideration
\begin{widetext}

\begin{align}\label{eq:qpress_lin}
    \frac{\mu_0\epsilon_0\hbar^2}{2mq}\frac{\partial}{\partial t}\nabla\bigg[\frac{\nabla^2(\sqrt{\rho})}{\sqrt{\rho}}\bigg] &= \frac{\mu_0\epsilon_0\hbar^2}{4mq}\nabla\left[\left(\frac{1}{\rho}\nabla^2 + \frac{|\nabla\rho|^2}{\rho^3} - \frac{\nabla^2\rho}{\rho^2} - \frac{\nabla\rho\cdot\nabla}{\rho}\right)\frac{\partial\rho}{\partial t}\right] \nonumber\\
    &= \frac{\mu_0\epsilon_0\hbar^2}{4m^2}\nabla\nabla^2\nabla\!\cdot\!\bm{\mathcal{A}},
\end{align}
\end{widetext}

where in the last line of Eq.\,(\ref{eq:qpress_lin}) we have used the ansatz (\ref{eq:rho_harmonic}) for $\rho$ with $\rho_0$ being a step-wise constant function, used Eq.\,(\ref{eq:chargeconserve2_rho}) to replace the time derivative $\partial_t\rho$, and only kept the leading-order terms. For transverse excitations, the operation on $\bm{\mathcal{A}}$ in Eq.\,(\ref{eq:qpress_lin}) vanishes, resulting in the inhomogeneous source-field equation 
\begin{align}\label{eq:inhomo_linear}
    {\mathbf \nabla}\times{\mathbf \nabla}\times\bm{{\mathcal A}_0} + \bigg (\frac{1}{\lambda_L^2({\mathbf r})} - n^2({\mathbf r})k^2\bigg )\bm{{\mathcal A}_0} = \mathbf{J_0},
\end{align}
where $\lambda_L({\mathbf r}) = \sqrt{\frac{m}{\mu_0 q^2\rho_0({\mathbf r})}}$ is the London penetration depth, \mbox{$n({\mathbf r})=\sqrt{\Tilde{\epsilon}({\mathbf r})}$, and $k=\mu_0\epsilon_0\omega^2$}. 

\section{Derivation of Green's boundary integrals}\label{append:green_integral}
In this Section, we show the derivations for the integral forms of the boundary fields that obey the scalar and vector Helmholtz equations as were shown in Eqs.\,(\ref{eq:scalar_green_boundaryInt}) and (\ref{eq:vector_green_boundaryInt}), respectively.  

\begin{figure}[t]
    \centering
    \includegraphics[scale=0.38]{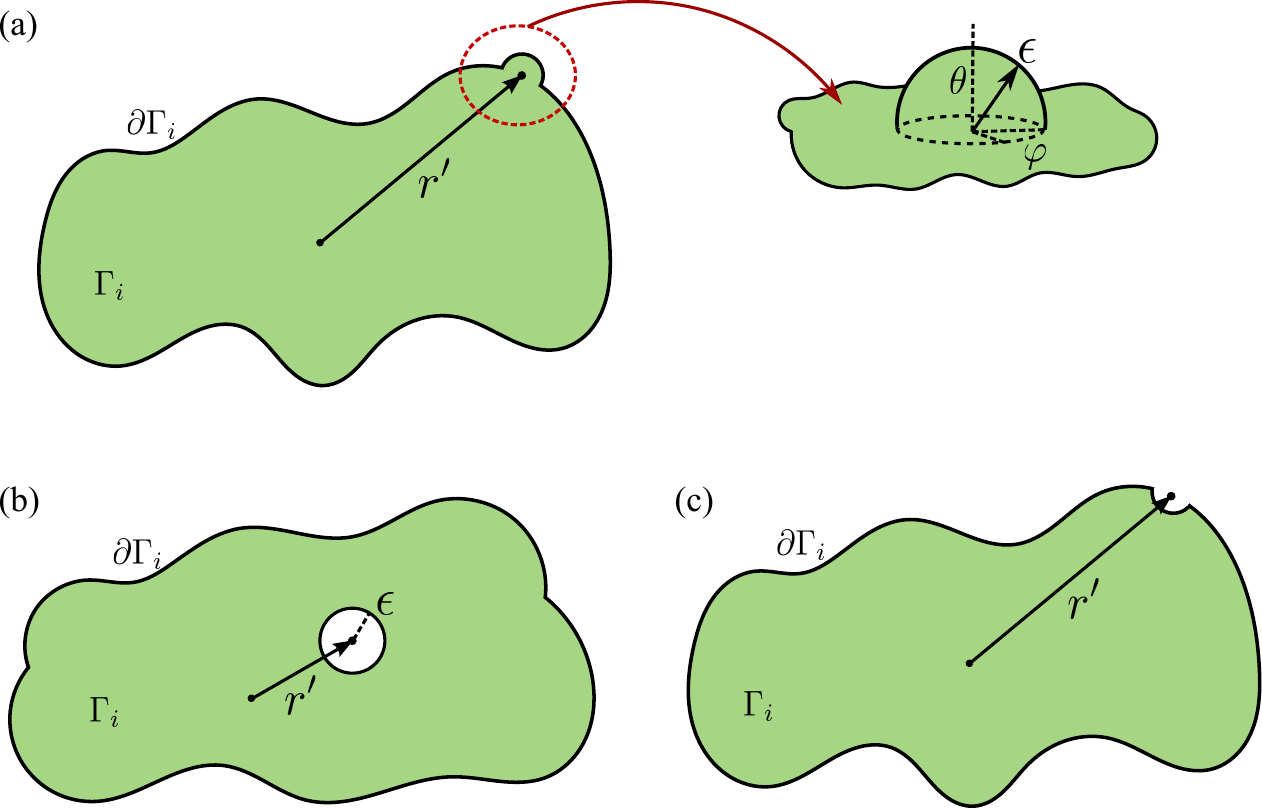}
    \caption{Sketches of how deformations are done to evaluate diverging integrals containing the Green's function. (a) A deformation on the boundary to include the boundary point. (b) The insertion of a hollow region internal to $\Gamma_i$ surrounding the point of interest. (c) Deformation of the boundary to exclude the boundary point of interest.}
    \label{fig:boundary_int_schematic}
\end{figure}

Consider a homogeneous domain $\Gamma_i$ bounded by the boundary $\partial\Gamma_i$, the Green's function is the solution of the impulse response of scalar Helmholtz operator and is defined by
\begin{equation}\label{eq:scalargreen_def}
    \big[\nabla^2 + n_i^2k^2\big]G({\mathbf r}, {\mathbf r'},k) = \delta({\mathbf r}-{\mathbf r'}),
\end{equation}
where $\delta({\mathbf r}\!-\!{\mathbf r'})$ is a $d-$dimensional Dirac $\delta-$function. 
Multiplying 
Eq.\,(13)
by $G({\mathbf r}, {\mathbf r'},k)$ and Eq.\,(\ref{eq:scalargreen_def}) by $\phi({\mathbf r})$, then applying Green's theorem gives
\begin{equation}\label{eq:green_scalaridentity}
\phi({\bf r'}) = \int_{\partial\Gamma_i}\!\!\big[\phi({\bf r})\nabla G({\bf r, r'}, k) - G({\bf r, r'}, k)\nabla\phi({\bf r}) \big]\cdot \mathbf{ds},
\end{equation}
for ${\mathbf r'}$ strictly inside the domain boundary $\partial\Gamma_i$. As we take the limit ${\mathbf r'}\rightarrow \partial\Gamma_i$, both $G({\bf r, r'}, k)$ and its normal gradient diverges. However, we can show for the 3D case that Eq.\,(\ref{eq:green_scalaridentity}) is still integrable. A proof for 2D can be found in Ref.\,\citenum{kosztin1997boundary}. We first consider an infinitesimal half-spherical deformation $C$ of the boundary at ${\mathbf r'}$ such that the surface now avoids and encloses the singular point inside the domain $\Gamma_i$ as seen Fig.\,(\ref{fig:boundary_int_schematic}a). Let the radius of the deformation be $\epsilon$, then at the limit $\epsilon\rightarrow 0$ the integral of second term in  Eq.\,(\ref{eq:green_scalaridentity}) over $C$ vanishes
\begin{align}\label{eq:lim_2nd_term}
    \lim_{\epsilon\rightarrow 0}& \int_{C} \!\!G({\bf r, r'}, k)\nabla\phi({\bf r})\cdot \mathbf{ds} \nonumber\\
    &= \lim_{\epsilon\rightarrow 0} -\frac{\partial\phi}{\partial n}\bigg\rvert_{\mathbf r'}\frac{1}{4\pi}\bigg(\frac{1}{\epsilon}+ in_ik + ... \bigg)2\pi\epsilon^2 = 0,
\end{align}
while the integral of the first term over $C$ gives
\begin{align}\label{eq:lim_1st_term}
    \lim_{\epsilon\rightarrow 0} \int_{C} \!\!\phi\frac{\partial G}{\partial n}ds &= \lim_{\epsilon\rightarrow 0} -\phi({\mathbf r'}) \frac{e^{in_ik\epsilon}}{4\pi}\bigg(\frac{in_ik}{\epsilon} - \frac{1}{\epsilon^2}\bigg)2\pi\epsilon^2\nonumber\\
    &= \frac{1}{2}\phi({\mathbf r'}),
\end{align}
where in the maneuvers above we have used the series expansion of the Green's function, which reads
\begin{equation}
    G({\bf r, r'}, k) = -\frac{e^{in_ik|{\mathbf r}-{\mathbf r'}|}}{4\pi|{\mathbf r}-{\mathbf r'}|} = -\bigg(\frac{1}{|{\mathbf r}-{\mathbf r'}|} + in_ik + ...\bigg).
\end{equation}
From the results in (\ref{eq:lim_2nd_term}) and (\ref{eq:lim_1st_term}) it follows that the integral in Eq.\,(\ref{eq:green_scalaridentity}) when ${\mathbf r'}\in \partial\Gamma_i$ evaluated over the entire boundary surface is indeed the form shown in Eq.\,(\ref{eq:scalar_green_boundaryInt}).

For the vector case, we start by introducing the vector analog of Green's identity
\begin{widetext}
\begin{align}\label{eq:vectorgreen_identity}
    \int_{\Gamma_i}\big({\mathbf u}\cdot{\mathbf\nabla}\!\times\!{\mathbf\nabla}\!\times\!{\mathbf v} - {\mathbf v}\cdot{\mathbf\nabla}\!\times\!{\mathbf\nabla}\!\times\!{\mathbf u} \big)dV = \int_{\partial\Gamma_i}\big({\mathbf v}\!\times\!\nabla\!\times\!{\mathbf u}- {\mathbf u}\!\times\!\nabla\!\times\!{\mathbf v}\big)\cdot\mathbf{ds},
\end{align}
where $\mathbf{u}$ and $\mathbf{v}$ are vector fields. We now choose ${\mathbf v}= \bm{\mathcal{A}}(\mathbf{r})$, our field of interest, and ${\mathbf u} = {\mathbf G}\equiv G(\mathbf{r},\mathbf{r'},k)\hat{\nu}$, where $\hat{\nu}$ is a unit vector that has a randomly selected, but constant, orientation. For a point ${\mathbf r'}$ outside of $\Gamma_i$, Eq.\,(\ref{eq:vectorgreen_identity}) reduces to
\begin{align}
    \int_{\Gamma_i}\!(\hat{\nu}\!\cdot\!\nabla G)(\nabla\!\cdot\!\bm{\mathcal{A}})dV &= \int_{\partial\Gamma_i}\!\!\Big\{(\hat{\nu}\!\cdot\!\nabla G)(\bm{\mathcal{A}}\!\cdot\!\hat{n}) + \bm{\mathcal{A}}\!\times\!(\nabla G\!\times\!\hat{\nu}) - \mathbf{G}\!\times\!{\mathbf \nabla}\!\times\!\bm{\mathcal{A}} \Big\}\cdot\mathbf{ds} \\
    &= \int_{\partial\Gamma_i}\!\!\Big\{(\hat{\nu}\!\cdot\!\nabla G)(\bm{\mathcal{A}}\!\cdot\!\hat{n}) + (\hat{\nu}\!\cdot\!\bm{\mathcal{A}})(\nabla G\!\cdot\!\hat{n}) - (\hat{\nu}\!\cdot\!\hat{n})(\nabla G\!\cdot\!\bm{\mathcal{A}})  -\mathbf{G}\!\cdot\!\big[({\mathbf\nabla}\!\times\!\bm{\mathcal{A}})\!\times\!\hat{n} \big]\Big\}ds\nonumber,
\end{align}
where $\hat{n}$ is the unit normal vector on $\partial\Gamma_i$. Since $\hat{\nu}$ is a common factor in all the terms, it can be dropped and we arrive at
\begin{align}\label{eq:vec_helm_intermid}
    \int_{\Gamma_i}\!\nabla G(\nabla\!\cdot\!\bm{\mathcal{A}})dV = \int_{\partial\Gamma_i}\!\!\Big\{\nabla G(\bm{\mathcal{A}}\cdot\hat{n}) + \nabla G\!\times\!(\bm{\mathcal{A}}\!\times\!\hat{n})  -G\big[({\mathbf\nabla}\!\times\!\bm{\mathcal{A}})\!\times\!\hat{n} \big] \Big\}ds. 
\end{align}
We are interested in, however, the evaluation of the field inside $\Gamma_i$.
To proceed, consider when ${\mathbf r'}$ is inside the material region enclosed by $\partial\Gamma_i$, but there exists an infinitesimally small spherical hole $S$ of radius $\epsilon$ centered around ${\mathbf r'}$. The boundary of $\Gamma_i$ now consists of the original outer surface $\partial\Gamma_i$ and the boundary of the sphere, as seen in Fig.\,(\ref{fig:boundary_int_schematic}b). At the limit $\epsilon\rightarrow 0$, each term in Eq.\,(\ref{eq:vec_helm_intermid}) evaluated on the sphere gives
\begin{align}
    \lim_{\epsilon\rightarrow 0}\int_{S}G\big[({\mathbf\nabla}\!\times\!\bm{\mathcal{A}})\!\times\!\hat{n} \big]ds &= \lim_{\epsilon\rightarrow 0}-\frac{e^{in_jk\epsilon}}{4\pi\epsilon}\big[(\nabla\!\times\!\bm{\mathcal{A}})\!\times\!\hat{n}\big]\bigg\rvert_{\mathbf r'}\!4\pi\epsilon^2 \nonumber\\
    &= 0,
\end{align}
also
\begin{align}
    \lim_{\epsilon\rightarrow 0}\int_{S}\!\!\Big\{\nabla G(\bm{\mathcal{A}}\cdot\hat{n}) + \nabla G\!\times\!(\bm{\mathcal{A}}\!\times\!\hat{n})\Big\}ds &= \lim_{\epsilon\rightarrow 0} \int_S\Big(\frac{1}{\epsilon} - in_ik \Big)\frac{e^{in_ik\epsilon}}{4\pi\epsilon}\bm{\mathcal{A}}ds \nonumber\\
    &= \bm{\mathcal{A}}{(\mathbf r'}).
\end{align}
Therefore, the integral form of the field at ${\mathbf r'}$ inside $\Gamma_i$ is given by
\begin{align}
    \bm{\mathcal{A}}({\mathbf r'}) = \int_{\partial\Gamma_i}\!\!\bigg\{G\big[({\mathbf \nabla}\times \bm{\mathcal{A}})\times \hat{n}\big] - \nabla G(\bm{\mathcal{A}}\cdot\hat{n}) -\nabla G\times({\mathbf A \times\hat{n}}) \bigg\} + \int_{\Gamma_i}\!\nabla G({\mathbf \nabla}\cdot\bm{\mathcal{A}})dV,
\end{align}
where the volume term on the right hand side is an improper integral. However, this term vanishes if the field is divergence-free, in which case the field at anywhere strictly inside $\Gamma_i$ is determined by the field on the boundary $\partial\Gamma_i$. 
\end{widetext}

Finally, to find the integral form of the field on the boundary, at ${\mathbf r'}\in\partial\Gamma_i$, we consider a half-spherical deformation of $\partial\Gamma_i$ such that the boundary avoids ${\mathbf r'}$ and leaves it outside of the enclosed domain $\Gamma_i$, as shown in Fig.\,(\ref{fig:boundary_int_schematic}c). Note that this deformation is in the opposite direction as in the scalar case discussed earlier, where ${\mathbf r'}$ was included into $\Gamma_i$. By following a similar limiting procedure as was laid out above for ${\mathbf r'}$ strictly inside $\Gamma_i$ but now applied to a half-sphere, we arrive at the self-consistent integral form for the boundary field shown in Eq.\,(\ref{eq:vector_green_boundaryInt}).

\section{Coarse-grained open boundary conditions using VSH}\label{append:VSH}
In this Section, we derive the boundary conditions for the angular components of the field $\bm{\mathcal{A}}({\mathbf r})$ satisfying the vector Helmholtz equation. We also discuss the DEC implementation of such conditions on the tangential boundary edges.

The series expansion of the $\bm{\mathcal{A}}^{(1)}$ component is 
\begin{widetext}
\begin{align}
    \bm{\mathcal{A}}^{(1)}(r,\theta,\varphi)
    &= \sum_{l,m}\Tilde{a}_{lm}^1\mathcal{A}_{lm}^{(1)}{\mathbf \Psi}_{lm}(r,\theta,\varphi) \nonumber\\
    &= \Tilde{a}_{lm}\frac{1}{2l(l+1)}\bigg[\frac{H_{l+1/2}(kr)}{r^{3/2}} + \frac{k}{\sqrt{r}}\big(H_{l-1/2}(kr) -H_{l+3/2}(kr) \big)\bigg]{\mathbf \Psi}_{lm}(r,\theta,\varphi). \nonumber
\end{align}
The coefficients $a^1_{lm}(R)$ in the expansion of the field on the boundary surface are then given by
\begin{align}\label{eq:a1_coeff}
    a^1_{lm}(R) &= \Tilde{a}_{lm}^1\mathcal{A}_{lm}^{(1)}(R)\frac{1}{2l(l+1)}\bigg[\frac{H_{l+1/2}(kr)}{r^{3/2}} + \frac{k}{\sqrt{r}}\big(H_{l-1/2}(kr) -H_{l+3/2}(kr) \big)\bigg] \nonumber\\
    &= \frac{1}{l(l+1)}\int d\Omega\bm{\mathcal{A}}^{(1)}\cdot{\mathbf \Psi}^*_{lm} \nonumber\\
    &= \frac{1}{l(l+1)}\sum_f\frac{\Delta A(f)}{R^2}\bm{\mathcal{A}}_t(f)\cdot{\mathbf\Psi}_{lm}^*(f),
\end{align}
where $\bm{\mathcal{A}}_t$ is the component of $\bm{\mathcal{A}}({\mathbf r})$ tangential to the boundary surface, and the sum is perform over all the triangular faces $f$ on the boundary. In the last line of Eq.\,(\ref{eq:a1_coeff}) we have utilized the orthogonality between $\bm{\mathcal{A}}^{(1)}$ and ${\mathbf \Psi}_{lm}^*$ to switch to using $\bm{\mathcal{A}}_t$ instead of $\bm{\mathcal{A}}^{(1)}$ in the computation of $a^1_{lm}(R)$, since the former is more readily accessible during numerical implementation. The field $\bm{\mathcal{A}}^{(1)}$ can be decomposed into azimuthal and polar contributions, for each of which we derive the boundary condition. The radial derivative of the polar component of $\bm{\mathcal{A}}^{(1)}$ at the boundary is given by 
\begin{align}
    \frac{\mathcal{A}_{\theta}^{(1)}(R,\theta,\varphi)-\mathcal{A}_{\theta}^{(1)}(R\!-\!\Delta R,\theta,\varphi)}{\Delta R} = \sum_{l,m}\Tilde{a}_{lm}^{(1)}\frac{1}{2l(l+1)}\bigg[\frac{H_{l+1/2}(kr)}{r^{1/2}} + k\sqrt{r}\big(H_{l-1/2}(kr) -H_{l+3/2}(kr) \big)\bigg]'\bigg\rvert_{R} (\nabla Y_{lm})_{\theta}
\end{align}
from which the expression for $\bm{\mathcal{A}}_\theta^{(1)}$ on the boundary surface can be extracted
\begin{align}\label{eq:A1_theta_bc}
    \mathcal{A}_{\theta}^{(1)}(R,\theta,\varphi) = \sum_{l,m} a^{(1)}_{lm}(R)\left(R + \Delta R\left\{1 + R\frac{\Big[\frac{H_{l+1/2}(kr)}{r^{3/2}} + \frac{k}{\sqrt{r}}\big(H_{l-1/2}(kr) -H_{l+3/2}(kr) \big)\Big]'\Big\rvert_R }{\Big[\frac{H_{l+1/2}(kR)}{R^{3/2}} + \frac{k}{\sqrt{R}}\big(H_{l-1/2}(kR) -H_{l+3/2}(kR) \big)\Big]} \right\} \right)(\nabla Y_{lm})_\theta,
\end{align}
where $(\nabla Y_{lm})_\theta$ is the polar component of $\nabla Y_{lm}$, and $a^1_{lm}(R)$ given as in Eq.\,(\ref{eq:a1_coeff}). The expression for the azimuthal component $\mathcal{A}_{\varphi}^{(1)}(R,\theta,\varphi)$ is exactly similar, with a replacement of $\theta\rightarrow\varphi$ in the subscripts in Eq.\,(\ref{eq:A1_theta_bc}).
\end{widetext}
Similar to the steps for $\mathcal{A}^{(1)}$, the derivation of BC for $\mathcal{A}^{(2)}$ starts with the expansion
\begin{align}
    \bm{\mathcal{A}}^{(2)} = \sum_{l.m} \tilde{a}^{(2)}_{lm}\frac{\pi}{2kR}H_{l+1/2}(kR){\mathbf\Phi}_{l.m}(r,\theta,\varphi)
\end{align}
from which the coefficients $a^{(2)}_{lm}$ can be computed 
\begin{align}\label{eq:a2_coeff}
    a_{lm}^{(2)}(R) &= \tilde{a}^{(2)}_{lm}\frac{\pi}{2kR}H_{l+1/2}(kR) \nonumber\\
    &= \frac{1}{l(l+1)}\int d\Omega\bm{\mathcal{A}}^{(2)}\cdot{\mathbf\Phi}_{lm}^* \nonumber\\
    &= \frac{1}{l(l+1)}\sum_f\frac{\Delta A(f)}{R^2}\bm{\mathcal{A}}_t(f)\cdot{\mathbf\Phi}_{lm}^*(f).
\end{align}
We also write the boundary conditions for $\mathcal{A}_{\theta}^{(2)}$ and $\mathcal{A}_{\varphi}^{(2)}$ separately using their discrete form of radial derivative at the boundary. This leads to
\begin{widetext}
    \begin{align}\label{eq:A2_theta_bc}
    \mathcal{A}_{\theta}^{(2)}(R,\theta,\varphi) = \sum_{l,m}a_{lm}^{(2)}(R)\left\{R + \frac{\Delta R}{2}\left[1 + kR\frac{H_{l-1/2}(kR)-H_{l+3/2}(kR)}{H_{l+1/2}(kR)}\right] \right\} (\hat{r}\times\nabla Y_{lm})_{\theta},
\end{align}
and by replacing $\theta$ with $\varphi$ in the subscripts, we obtain the expression for $\mathcal{A}_{\varphi}^{(2)}(R,\theta,\varphi)$.
\end{widetext}

The boundary conditions given in Eqs.\,(\ref{eq:A1_theta_bc}) and (\ref{eq:A2_theta_bc}) are for vector individual components, which are scalars living on vertices of the primal computational mesh. We would like to translate these BCs to fields living the edges that lie on the boundary surface. First of all, the practical computation of the expansion coefficients in Eqs.\,(\ref{eq:a1_coeff}) and (\ref{eq:a2_coeff}) requires knowledge about the tangential field $\bm{\mathcal{A}}_t$ defined on the centers of the triangular faces $f$ that discretize the boundary surface. In DEC, however, vectors are replaced by their projections onto the discrete edges, which are quantities that we have access to. Therefore, at each triangular face, given the edge fields that are on the three sides of the triangle, we need to determine the value of the vector $\bm{\mathcal{A}}_t$ at the circumcenter $f^{\dagger}$. This mapping from primal edge field to dual vector field is done through a sharp (\#) operator, whose formal definition is given in Ref.\,\citenum{DEC_HiraniThesis}. Now given that the coefficient expansions are obtained, for every vertex on the boundary, the polar(azimuthal) component of $\mathcal{A}_t$ is the sum of projections from $\mathcal{A}^{(1)}$ and $\mathcal{A}^{(2)}$
\begin{equation}\label{eq:angular_field_sum}
    \mathcal{A}_{\theta(\varphi)}(R,\theta,\varphi) =  \mathcal{A}_{\theta(\varphi)}^{(1)}(R,\theta,\varphi) + \mathcal{A}_{\theta(\varphi)}^{(2)}(R,\theta,\varphi).
\end{equation}
For an edge $e[v_1,v_2]$ lying on the boundary, where $v_1$ and $v_2$ are the starting and ending vertices of the edge, the edge field is given by
\begin{equation}\label{eq:angular_edge_bc}
    \Phi(e[v_1,v_2]) = \frac{\big[\bm{\mathcal{A}}_t(v_1) + \bm{\mathcal{A}}_t(v_2)\big]}{2}\cdot({\mathbf v}_2 - {\mathbf v}_1),
\end{equation}
where ${\mathbf v}_{1}$ and ${\mathbf v}_2$ are the locations of the vertices $v_1$ and $v_2$, respectively, with the components of $\bm{\mathcal{A}}_t$ on the boundary given in Eq.\,(\ref{eq:angular_field_sum}). Eq.\,(\ref{eq:angular_edge_bc}) is the boundary condition to be imposed on the edges lying tangentially to the boundary surface.

\section{Separability of Helmholtz equation in ellipsoidal coordinates}\label{append:ellipsoidal_coords}
The ellipsoidal coordinate system is based on the equations
\begin{align}\label{eq:ellipse}
    \frac{x^2}{\xi^2_i-a^2} + \frac{y^2}{\xi^2-b^2} + \frac{z^2}{\xi^2-c^2} = 1,
\end{align}
where $a\geq b\geq c$, with $i=1,2,3$ such that
\begin{equation}
    \xi_1>a>\xi_2>b>\xi_3>c.
\end{equation}
Eqs.\,(\ref{eq:ellipse}) represent three families of confocal quadric surfaces sharing the same foci. In the discussion here, to simplify the algebraic manipulations we consider the case where $c=0$. The relationship between the ellipsoidal coordinates $(\xi_1,\xi_2,\xi_3)$ and the Cartesian coordinates are given by
\begin{align}
    x &= \sqrt{\frac{(\xi_1^2-a^2)(\xi_2^2-a^2)(\xi_3^2-a^2)}{a^2(a^2-b^2)}}, \nonumber\\
    y &= \sqrt{\frac{(\xi_1^2-b^2)(\xi_2^2-b^2)(\xi_3^2-b^2)}{b^2(b^2-a^2)}}, \\
    z &= \frac{\xi_1\xi_2\xi_3}{ab}, \nonumber
\end{align}
with the scale factors being
\begin{align}
    h_1 &= \sqrt{\frac{(\xi_1^2-\xi_2^2)(\xi_1^2-\xi_3^2)}{(\xi_1^2-a^2)(\xi_1^2-b^2)}} \nonumber\\
    h_2 &= \sqrt{\frac{(\xi_2^2-\xi_1^2)(\xi_2^2-\xi_3^2)}{(\xi_2^2-a^2)(\xi_2^2-b^2)}}.
\end{align}
It is known that the Helmholtz equation 
(Eq.\,(13) in the main text)
is separable in eleven three-dimensional coordinate systems, with the ellipsoidal coordinates being the most general of them and the remaining ten - including the spherical coordinates used in the main text - are derived from it through limiting processes \cite{morse1954methods}. The separability in ellipsoidal coordinates can be shown by using the ansatz
\begin{equation}
    \phi(\xi_1,\xi_2,\xi_3) = \psi_1(\xi_1)\psi_2(\xi_2)\psi_3(\xi_3)
\end{equation}
for the field in 
Eq.\,(13), resulting in the separated ODEs of the form
\begin{align}\label{eq:ellipsoidal_ode}
    4\sqrt{f(\kappa_i)}\frac{d}{d\kappa_i}\!\left(\!\sqrt{f(\kappa_i)}\frac{d\psi_i}{d\kappa_i} + (\lambda_1 + \lambda_2\kappa_i + k^2\kappa_i^2)\! \right)\psi_i = 0,
\end{align}
where \mbox{$f(\kappa_i) = \sqrt{(\kappa_i-a^2)(\kappa_i^2-b^2)\kappa_i}$}, with $\kappa_i=\xi^2$. Upon performing a change of variables
\begin{equation}
    \kappa_i = b^2sn^2\left(\alpha, \frac{b^2}{a^2} \right),
\end{equation}
where $sn(u,m)$ is a Jacobi elliptic function, we can write 
\begin{align}
    f(\kappa_i) = b^4a^2sn^2\left(\alpha,\frac{b^2}{a^2} \right)dn^2\left(\alpha,\frac{b^2}{a^2}\right)cn^2\left(\alpha,\frac{b^2}{a^2}\right),
\end{align}
with $dn(u,m)$ and $cn(u,m)$ also Jacobi elliptic functions. Eq.\,(\ref{eq:ellipsoidal_ode}) can then be written in terms of the new variable as
\begin{equation}\label{eq:ellipsoidal_wave_eq}
    \frac{d^2\psi_i}{d\alpha^2} + \left[\frac{\lambda_1}{a^2} + \frac{\lambda_2}{a^2}sn^2\left(\alpha,\frac{b^2}{a^2}\right) + \frac{k^2}{a^2}sn^4\left(\alpha,\frac{b^2}{a^2}\right)\right]\psi_i = 0.
\end{equation}
Eq.\,(\ref{eq:ellipsoidal_wave_eq}) is the ellipsoidal wave equation whose solutions are known as ellipsoidal wave functions. With the Helmholtz equation being separable in ellipsoidal coordinates with known solutions, the generalization to this coordinate system of our spherical harmonics method for open systems is therefore straightforward.  

\bibliography{refs}

\begin{thebibliography}{71}%
\makeatletter
\providecommand \@ifxundefined [1]{%
 \@ifx{#1\undefined}
}%
\providecommand \@ifnum [1]{%
 \ifnum #1\expandafter \@firstoftwo
 \else \expandafter \@secondoftwo
 \fi
}%
\providecommand \@ifx [1]{%
 \ifx #1\expandafter \@firstoftwo
 \else \expandafter \@secondoftwo
 \fi
}%
\providecommand \natexlab [1]{#1}%
\providecommand \enquote  [1]{``#1''}%
\providecommand \bibnamefont  [1]{#1}%
\providecommand \bibfnamefont [1]{#1}%
\providecommand \citenamefont [1]{#1}%
\providecommand \href@noop [0]{\@secondoftwo}%
\providecommand \href [0]{\begingroup \@sanitize@url \@href}%
\providecommand \@href[1]{\@@startlink{#1}\@@href}%
\providecommand \@@href[1]{\endgroup#1\@@endlink}%
\providecommand \@sanitize@url [0]{\catcode `\\12\catcode `\$12\catcode `\&12\catcode `\#12\catcode `\^12\catcode `\_12\catcode `\%12\relax}%
\providecommand \@@startlink[1]{}%
\providecommand \@@endlink[0]{}%
\providecommand \url  [0]{\begingroup\@sanitize@url \@url }%
\providecommand \@url [1]{\endgroup\@href {#1}{\urlprefix }}%
\providecommand \urlprefix  [0]{URL }%
\providecommand \Eprint [0]{\href }%
\providecommand \doibase [0]{https://doi.org/}%
\providecommand \selectlanguage [0]{\@gobble}%
\providecommand \bibinfo  [0]{\@secondoftwo}%
\providecommand \bibfield  [0]{\@secondoftwo}%
\providecommand \translation [1]{[#1]}%
\providecommand \BibitemOpen [0]{}%
\providecommand \bibitemStop [0]{}%
\providecommand \bibitemNoStop [0]{.\EOS\space}%
\providecommand \EOS [0]{\spacefactor3000\relax}%
\providecommand \BibitemShut  [1]{\csname bibitem#1\endcsname}%
\let\auto@bib@innerbib\@empty
\bibitem [{\citenamefont {Houck}\ \emph {et~al.}(2012)\citenamefont {Houck}, \citenamefont {T{\"u}reci},\ and\ \citenamefont {Koch}}]{houck2012chip}%
  \BibitemOpen
  \bibfield  {author} {\bibinfo {author} {\bibfnamefont {A.~A.}\ \bibnamefont {Houck}}, \bibinfo {author} {\bibfnamefont {H.~E.}\ \bibnamefont {T{\"u}reci}},\ and\ \bibinfo {author} {\bibfnamefont {J.}~\bibnamefont {Koch}},\ }\bibfield  {title} {\bibinfo {title} {On-chip quantum simulation with superconducting circuits},\ }\href@noop {} {\bibfield  {journal} {\bibinfo  {journal} {Nature Physics}\ }\textbf {\bibinfo {volume} {8}},\ \bibinfo {pages} {292} (\bibinfo {year} {2012})}\BibitemShut {NoStop}%
\bibitem [{\citenamefont {Carusotto}\ \emph {et~al.}(2020)\citenamefont {Carusotto}, \citenamefont {Houck}, \citenamefont {Koll{\'a}r}, \citenamefont {Roushan}, \citenamefont {Schuster},\ and\ \citenamefont {Simon}}]{carusotto2020photonic}%
  \BibitemOpen
  \bibfield  {author} {\bibinfo {author} {\bibfnamefont {I.}~\bibnamefont {Carusotto}}, \bibinfo {author} {\bibfnamefont {A.~A.}\ \bibnamefont {Houck}}, \bibinfo {author} {\bibfnamefont {A.~J.}\ \bibnamefont {Koll{\'a}r}}, \bibinfo {author} {\bibfnamefont {P.}~\bibnamefont {Roushan}}, \bibinfo {author} {\bibfnamefont {D.~I.}\ \bibnamefont {Schuster}},\ and\ \bibinfo {author} {\bibfnamefont {J.}~\bibnamefont {Simon}},\ }\bibfield  {title} {\bibinfo {title} {Photonic materials in circuit quantum electrodynamics},\ }\href@noop {} {\bibfield  {journal} {\bibinfo  {journal} {Nature Physics}\ }\textbf {\bibinfo {volume} {16}},\ \bibinfo {pages} {268} (\bibinfo {year} {2020})}\BibitemShut {NoStop}%
\bibitem [{\citenamefont {Hatridge}\ \emph {et~al.}(2011)\citenamefont {Hatridge}, \citenamefont {Vijay}, \citenamefont {Slichter}, \citenamefont {Clarke},\ and\ \citenamefont {Siddiqi}}]{hatridge2011dispersive}%
  \BibitemOpen
  \bibfield  {author} {\bibinfo {author} {\bibfnamefont {M.}~\bibnamefont {Hatridge}}, \bibinfo {author} {\bibfnamefont {R.}~\bibnamefont {Vijay}}, \bibinfo {author} {\bibfnamefont {D.}~\bibnamefont {Slichter}}, \bibinfo {author} {\bibfnamefont {J.}~\bibnamefont {Clarke}},\ and\ \bibinfo {author} {\bibfnamefont {I.}~\bibnamefont {Siddiqi}},\ }\bibfield  {title} {\bibinfo {title} {Dispersive magnetometry with a quantum limited squid parametric amplifier},\ }\href@noop {} {\bibfield  {journal} {\bibinfo  {journal} {Physical Review B}\ }\textbf {\bibinfo {volume} {83}},\ \bibinfo {pages} {134501} (\bibinfo {year} {2011})}\BibitemShut {NoStop}%
\bibitem [{\citenamefont {Macklin}\ \emph {et~al.}(2015)\citenamefont {Macklin}, \citenamefont {O’brien}, \citenamefont {Hover}, \citenamefont {Schwartz}, \citenamefont {Bolkhovsky}, \citenamefont {Zhang}, \citenamefont {Oliver},\ and\ \citenamefont {Siddiqi}}]{macklin2015near}%
  \BibitemOpen
  \bibfield  {author} {\bibinfo {author} {\bibfnamefont {C.}~\bibnamefont {Macklin}}, \bibinfo {author} {\bibfnamefont {K.}~\bibnamefont {O’brien}}, \bibinfo {author} {\bibfnamefont {D.}~\bibnamefont {Hover}}, \bibinfo {author} {\bibfnamefont {M.}~\bibnamefont {Schwartz}}, \bibinfo {author} {\bibfnamefont {V.}~\bibnamefont {Bolkhovsky}}, \bibinfo {author} {\bibfnamefont {X.}~\bibnamefont {Zhang}}, \bibinfo {author} {\bibfnamefont {W.}~\bibnamefont {Oliver}},\ and\ \bibinfo {author} {\bibfnamefont {I.}~\bibnamefont {Siddiqi}},\ }\bibfield  {title} {\bibinfo {title} {A near--quantum-limited josephson traveling-wave parametric amplifier},\ }\href@noop {} {\bibfield  {journal} {\bibinfo  {journal} {Science}\ }\textbf {\bibinfo {volume} {350}},\ \bibinfo {pages} {307} (\bibinfo {year} {2015})}\BibitemShut {NoStop}%
\bibitem [{\citenamefont {Blais}\ \emph {et~al.}(2004)\citenamefont {Blais}, \citenamefont {Huang}, \citenamefont {Wallraff}, \citenamefont {Girvin},\ and\ \citenamefont {Schoelkopf}}]{blais2004cqed}%
  \BibitemOpen
  \bibfield  {author} {\bibinfo {author} {\bibfnamefont {A.}~\bibnamefont {Blais}}, \bibinfo {author} {\bibfnamefont {R.-S.}\ \bibnamefont {Huang}}, \bibinfo {author} {\bibfnamefont {A.}~\bibnamefont {Wallraff}}, \bibinfo {author} {\bibfnamefont {S.~M.}\ \bibnamefont {Girvin}},\ and\ \bibinfo {author} {\bibfnamefont {R.~J.}\ \bibnamefont {Schoelkopf}},\ }\bibfield  {title} {\bibinfo {title} {Cavity quantum electrodynamics for superconducting electrical circuits: An architecture for quantum computation},\ }\href@noop {} {\bibfield  {journal} {\bibinfo  {journal} {Physical Review A}\ }\textbf {\bibinfo {volume} {69}},\ \bibinfo {pages} {062320} (\bibinfo {year} {2004})}\BibitemShut {NoStop}%
\bibitem [{\citenamefont {Devoret}\ and\ \citenamefont {Schoelkopf}(2013)}]{devoret2013superconducting}%
  \BibitemOpen
  \bibfield  {author} {\bibinfo {author} {\bibfnamefont {M.~H.}\ \bibnamefont {Devoret}}\ and\ \bibinfo {author} {\bibfnamefont {R.~J.}\ \bibnamefont {Schoelkopf}},\ }\bibfield  {title} {\bibinfo {title} {Superconducting circuits for quantum information: an outlook},\ }\href@noop {} {\bibfield  {journal} {\bibinfo  {journal} {Science}\ }\textbf {\bibinfo {volume} {339}},\ \bibinfo {pages} {1169} (\bibinfo {year} {2013})}\BibitemShut {NoStop}%
\bibitem [{\citenamefont {Feynman}\ \emph {et~al.}(1965)\citenamefont {Feynman}, \citenamefont {Leighton},\ and\ \citenamefont {Sands}}]{feynman}%
  \BibitemOpen
  \bibfield  {author} {\bibinfo {author} {\bibfnamefont {R.~P.}\ \bibnamefont {Feynman}}, \bibinfo {author} {\bibfnamefont {R.~B.}\ \bibnamefont {Leighton}},\ and\ \bibinfo {author} {\bibfnamefont {M.}~\bibnamefont {Sands}},\ }\bibinfo {title} {The {S}chrödinger {E}quation in a {C}lassical {C}ontext: {A} {S}eminar on {S}uperconductivity},\ in\ \href@noop {} {\emph {\bibinfo {booktitle} {The Feynman Lectures on Physics}}},\ Vol.\ \bibinfo {volume} {III}\ (\bibinfo  {publisher} {Addison–Wesley},\ \bibinfo {year} {1965})\ Chap.~\bibinfo {chapter} {21}\BibitemShut {NoStop}%
\bibitem [{\citenamefont {Ao}\ \emph {et~al.}(1995)\citenamefont {Ao}, \citenamefont {Thouless},\ and\ \citenamefont {Zhu}}]{ao1995nonlinear}%
  \BibitemOpen
  \bibfield  {author} {\bibinfo {author} {\bibfnamefont {P.}~\bibnamefont {Ao}}, \bibinfo {author} {\bibfnamefont {D.~J.}\ \bibnamefont {Thouless}},\ and\ \bibinfo {author} {\bibfnamefont {X.-M.}\ \bibnamefont {Zhu}},\ }\bibfield  {title} {\bibinfo {title} {Nonlinear {S}chr{\"o}dinger equation for superconductors},\ }\href@noop {} {\bibfield  {journal} {\bibinfo  {journal} {Modern Physics Letters B}\ }\textbf {\bibinfo {volume} {9}},\ \bibinfo {pages} {755} (\bibinfo {year} {1995})}\BibitemShut {NoStop}%
\bibitem [{\citenamefont {Aitchison}\ \emph {et~al.}(1995)\citenamefont {Aitchison}, \citenamefont {Ao}, \citenamefont {Thouless},\ and\ \citenamefont {Zhu}}]{aitchison1995effective}%
  \BibitemOpen
  \bibfield  {author} {\bibinfo {author} {\bibfnamefont {I.~J.}\ \bibnamefont {Aitchison}}, \bibinfo {author} {\bibfnamefont {P.}~\bibnamefont {Ao}}, \bibinfo {author} {\bibfnamefont {D.~J.}\ \bibnamefont {Thouless}},\ and\ \bibinfo {author} {\bibfnamefont {X.-M.}\ \bibnamefont {Zhu}},\ }\bibfield  {title} {\bibinfo {title} {Effective {L}agrangians for {BCS} superconductors at {T}=0},\ }\href@noop {} {\bibfield  {journal} {\bibinfo  {journal} {Physical Review B}\ }\textbf {\bibinfo {volume} {51}},\ \bibinfo {pages} {6531} (\bibinfo {year} {1995})}\BibitemShut {NoStop}%
\bibitem [{\citenamefont {Pham}\ \emph {et~al.}(2023{\natexlab{a}})\citenamefont {Pham}, \citenamefont {Fan}, \citenamefont {Scheer},\ and\ \citenamefont {Tureci}}]{dec-qed}%
  \BibitemOpen
  \bibfield  {author} {\bibinfo {author} {\bibfnamefont {D.~N.}\ \bibnamefont {Pham}}, \bibinfo {author} {\bibfnamefont {W.}~\bibnamefont {Fan}}, \bibinfo {author} {\bibfnamefont {M.~G.}\ \bibnamefont {Scheer}},\ and\ \bibinfo {author} {\bibfnamefont {H.~E.}\ \bibnamefont {Tureci}},\ }\bibfield  {title} {\bibinfo {title} {Flux-based three-dimensional electrodynamic modeling approach to superconducting circuits and materials},\ }\href@noop {} {\bibfield  {journal} {\bibinfo  {journal} {Phys. Rev. A}\ }\textbf {\bibinfo {volume} {107}},\ \bibinfo {pages} {053704} (\bibinfo {year} {2023}{\natexlab{a}})}\BibitemShut {NoStop}%
\bibitem [{\citenamefont {Greiter}\ \emph {et~al.}(1989)\citenamefont {Greiter}, \citenamefont {Wilczek},\ and\ \citenamefont {Witten}}]{greiter1989hydrodynamic}%
  \BibitemOpen
  \bibfield  {author} {\bibinfo {author} {\bibfnamefont {M.}~\bibnamefont {Greiter}}, \bibinfo {author} {\bibfnamefont {F.}~\bibnamefont {Wilczek}},\ and\ \bibinfo {author} {\bibfnamefont {E.}~\bibnamefont {Witten}},\ }\bibfield  {title} {\bibinfo {title} {Hydrodynamic relations in superconductivity},\ }\href@noop {} {\bibfield  {journal} {\bibinfo  {journal} {Modern Physics Letters B}\ }\textbf {\bibinfo {volume} {3}},\ \bibinfo {pages} {903} (\bibinfo {year} {1989})}\BibitemShut {NoStop}%
\bibitem [{\citenamefont {Fisher}\ \emph {et~al.}(1990)\citenamefont {Fisher}, \citenamefont {Grinstein},\ and\ \citenamefont {Girvin}}]{PhysRevLett.64.587}%
  \BibitemOpen
  \bibfield  {author} {\bibinfo {author} {\bibfnamefont {M.~P.~A.}\ \bibnamefont {Fisher}}, \bibinfo {author} {\bibfnamefont {G.}~\bibnamefont {Grinstein}},\ and\ \bibinfo {author} {\bibfnamefont {S.~M.}\ \bibnamefont {Girvin}},\ }\bibfield  {title} {\bibinfo {title} {Presence of quantum diffusion in two dimensions: {U}niversal resistance at the superconductor-insulator transition},\ }\href {https://doi.org/10.1103/PhysRevLett.64.587} {\bibfield  {journal} {\bibinfo  {journal} {Phys. Rev. Lett.}\ }\textbf {\bibinfo {volume} {64}},\ \bibinfo {pages} {587} (\bibinfo {year} {1990})}\BibitemShut {NoStop}%
\bibitem [{\citenamefont {Salasnich}(2009)}]{salasnich2009hydrodynamics}%
  \BibitemOpen
  \bibfield  {author} {\bibinfo {author} {\bibfnamefont {L.}~\bibnamefont {Salasnich}},\ }\bibfield  {title} {\bibinfo {title} {Hydrodynamics of {B}ose and {F}ermi superfluids at zero temperature: the superfluid nonlinear {S}chr{\"o}dinger equation},\ }\href@noop {} {\bibfield  {journal} {\bibinfo  {journal} {Laser physics}\ }\textbf {\bibinfo {volume} {19}},\ \bibinfo {pages} {642} (\bibinfo {year} {2009})}\BibitemShut {NoStop}%
\bibitem [{\citenamefont {Blais}\ \emph {et~al.}(2021)\citenamefont {Blais}, \citenamefont {Grimsmo}, \citenamefont {Girvin},\ and\ \citenamefont {Wallraff}}]{blais_rmp}%
  \BibitemOpen
  \bibfield  {author} {\bibinfo {author} {\bibfnamefont {A.}~\bibnamefont {Blais}}, \bibinfo {author} {\bibfnamefont {A.~L.}\ \bibnamefont {Grimsmo}}, \bibinfo {author} {\bibfnamefont {S.~M.}\ \bibnamefont {Girvin}},\ and\ \bibinfo {author} {\bibfnamefont {A.}~\bibnamefont {Wallraff}},\ }\bibfield  {title} {\bibinfo {title} {Circuit quantum electrodynamics},\ }\href {https://doi.org/10.1103/RevModPhys.93.025005} {\bibfield  {journal} {\bibinfo  {journal} {Rev. Mod. Phys.}\ }\textbf {\bibinfo {volume} {93}},\ \bibinfo {pages} {025005} (\bibinfo {year} {2021})}\BibitemShut {NoStop}%
\bibitem [{\citenamefont {Nigg}\ \emph {et~al.}(2012)\citenamefont {Nigg}, \citenamefont {Paik}, \citenamefont {Vlastakis}, \citenamefont {Kirchmair}, \citenamefont {Shankar}, \citenamefont {Frunzio}, \citenamefont {Devoret}, \citenamefont {Schoelkopf},\ and\ \citenamefont {Girvin}}]{BBQ_2012}%
  \BibitemOpen
  \bibfield  {author} {\bibinfo {author} {\bibfnamefont {S.~E.}\ \bibnamefont {Nigg}}, \bibinfo {author} {\bibfnamefont {H.}~\bibnamefont {Paik}}, \bibinfo {author} {\bibfnamefont {B.}~\bibnamefont {Vlastakis}}, \bibinfo {author} {\bibfnamefont {G.}~\bibnamefont {Kirchmair}}, \bibinfo {author} {\bibfnamefont {S.}~\bibnamefont {Shankar}}, \bibinfo {author} {\bibfnamefont {L.}~\bibnamefont {Frunzio}}, \bibinfo {author} {\bibfnamefont {M.~H.}\ \bibnamefont {Devoret}}, \bibinfo {author} {\bibfnamefont {R.~J.}\ \bibnamefont {Schoelkopf}},\ and\ \bibinfo {author} {\bibfnamefont {S.~M.}\ \bibnamefont {Girvin}},\ }\bibfield  {title} {\bibinfo {title} {Black-box superconducting circuit quantization},\ }\href {https://doi.org/10.1103/PhysRevLett.108.240502} {\bibfield  {journal} {\bibinfo  {journal} {Phys. Revs. Letts.}\ }\textbf {\bibinfo {volume} {108}},\ \bibinfo {pages} {240502} (\bibinfo {year} {2012})}\BibitemShut {NoStop}%
\bibitem [{\citenamefont {Solgun}\ \emph {et~al.}(2014)\citenamefont {Solgun}, \citenamefont {Abraham},\ and\ \citenamefont {DiVincenzo}}]{solgun2014blackbox}%
  \BibitemOpen
  \bibfield  {author} {\bibinfo {author} {\bibfnamefont {F.}~\bibnamefont {Solgun}}, \bibinfo {author} {\bibfnamefont {D.~W.}\ \bibnamefont {Abraham}},\ and\ \bibinfo {author} {\bibfnamefont {D.~P.}\ \bibnamefont {DiVincenzo}},\ }\bibfield  {title} {\bibinfo {title} {Blackbox quantization of superconducting circuits using exact impedance synthesis},\ }\href@noop {} {\bibfield  {journal} {\bibinfo  {journal} {Physical Review B}\ }\textbf {\bibinfo {volume} {90}},\ \bibinfo {pages} {134504} (\bibinfo {year} {2014})}\BibitemShut {NoStop}%
\bibitem [{\citenamefont {Smith}\ \emph {et~al.}(2016)\citenamefont {Smith}, \citenamefont {Kou}, \citenamefont {Vool}, \citenamefont {Pop}, \citenamefont {Frunzio}, \citenamefont {Schoelkopf},\ and\ \citenamefont {Devoret}}]{quantize_shunted_sc_2016}%
  \BibitemOpen
  \bibfield  {author} {\bibinfo {author} {\bibfnamefont {W.~C.}\ \bibnamefont {Smith}}, \bibinfo {author} {\bibfnamefont {A.}~\bibnamefont {Kou}}, \bibinfo {author} {\bibfnamefont {U.}~\bibnamefont {Vool}}, \bibinfo {author} {\bibfnamefont {I.~M.}\ \bibnamefont {Pop}}, \bibinfo {author} {\bibfnamefont {L.}~\bibnamefont {Frunzio}}, \bibinfo {author} {\bibfnamefont {R.~J.}\ \bibnamefont {Schoelkopf}},\ and\ \bibinfo {author} {\bibfnamefont {M.~H.}\ \bibnamefont {Devoret}},\ }\bibfield  {title} {\bibinfo {title} {Quantization of inductively shunted superconducting circuits},\ }\href {https://doi.org/10.1103/PhysRevB.94.144507} {\bibfield  {journal} {\bibinfo  {journal} {Phys. Rev. B}\ }\textbf {\bibinfo {volume} {94}},\ \bibinfo {pages} {144507} (\bibinfo {year} {2016})}\BibitemShut {NoStop}%
\bibitem [{\citenamefont {Malekakhlagh}\ \emph {et~al.}(2017)\citenamefont {Malekakhlagh}, \citenamefont {Petrescu},\ and\ \citenamefont {T\"ureci}}]{cuttoff_free_cqed_2017}%
  \BibitemOpen
  \bibfield  {author} {\bibinfo {author} {\bibfnamefont {M.}~\bibnamefont {Malekakhlagh}}, \bibinfo {author} {\bibfnamefont {A.}~\bibnamefont {Petrescu}},\ and\ \bibinfo {author} {\bibfnamefont {H.~E.}\ \bibnamefont {T\"ureci}},\ }\bibfield  {title} {\bibinfo {title} {Cutoff-free circuit quantum electrodynamics},\ }\href {https://doi.org/10.1103/PhysRevLett.119.073601} {\bibfield  {journal} {\bibinfo  {journal} {Phys. Rev. Lett.}\ }\textbf {\bibinfo {volume} {119}},\ \bibinfo {pages} {073601} (\bibinfo {year} {2017})}\BibitemShut {NoStop}%
\bibitem [{\citenamefont {Parra-Rodriguez}\ \emph {et~al.}(2019)\citenamefont {Parra-Rodriguez}, \citenamefont {Egusquiza}, \citenamefont {DiVincenzo},\ and\ \citenamefont {Solano}}]{parra2019canonical}%
  \BibitemOpen
  \bibfield  {author} {\bibinfo {author} {\bibfnamefont {A.}~\bibnamefont {Parra-Rodriguez}}, \bibinfo {author} {\bibfnamefont {I.}~\bibnamefont {Egusquiza}}, \bibinfo {author} {\bibfnamefont {D.}~\bibnamefont {DiVincenzo}},\ and\ \bibinfo {author} {\bibfnamefont {E.}~\bibnamefont {Solano}},\ }\bibfield  {title} {\bibinfo {title} {Canonical circuit quantization with linear nonreciprocal devices},\ }\href@noop {} {\bibfield  {journal} {\bibinfo  {journal} {Physical Review B}\ }\textbf {\bibinfo {volume} {99}},\ \bibinfo {pages} {014514} (\bibinfo {year} {2019})}\BibitemShut {NoStop}%
\bibitem [{\citenamefont {Minev}\ \emph {et~al.}(2021)\citenamefont {Minev}, \citenamefont {Leghtas}, \citenamefont {Mundhada}, \citenamefont {Christakis}, \citenamefont {Pop},\ and\ \citenamefont {Devoret}}]{EPR}%
  \BibitemOpen
  \bibfield  {author} {\bibinfo {author} {\bibfnamefont {Z.~K.}\ \bibnamefont {Minev}}, \bibinfo {author} {\bibfnamefont {Z.}~\bibnamefont {Leghtas}}, \bibinfo {author} {\bibfnamefont {S.~O.}\ \bibnamefont {Mundhada}}, \bibinfo {author} {\bibfnamefont {L.}~\bibnamefont {Christakis}}, \bibinfo {author} {\bibfnamefont {I.~M.}\ \bibnamefont {Pop}},\ and\ \bibinfo {author} {\bibfnamefont {M.~H.}\ \bibnamefont {Devoret}},\ }\bibfield  {title} {\bibinfo {title} {Energy-participation quantization of josephson circuits},\ }\href {https://doi.org/10.1038/s41534-021-00461-8} {\bibfield  {journal} {\bibinfo  {journal} {npj Quantum Information}\ }\textbf {\bibinfo {volume} {7}},\ \bibinfo {pages} {131} (\bibinfo {year} {2021})}\BibitemShut {NoStop}%
\bibitem [{TCG()}]{TCGtoolbox}%
  \BibitemOpen
  \href@noop {} {\bibinfo {title} {The time-coarse graining toolbox}},\ \bibinfo {howpublished} {\url{https://github.com/leonbello/QuantumGraining.jl}}\BibitemShut {NoStop}%
\bibitem [{\citenamefont {Houck}\ \emph {et~al.}(2008)\citenamefont {Houck}, \citenamefont {Schreier}, \citenamefont {Johnson}, \citenamefont {Chow}, \citenamefont {Koch}, \citenamefont {Gambetta}, \citenamefont {Schuster}, \citenamefont {Frunzio}, \citenamefont {Devoret}, \citenamefont {Girvin} \emph {et~al.}}]{houck2008spontaneous}%
  \BibitemOpen
  \bibfield  {author} {\bibinfo {author} {\bibfnamefont {A.}~\bibnamefont {Houck}}, \bibinfo {author} {\bibfnamefont {J.}~\bibnamefont {Schreier}}, \bibinfo {author} {\bibfnamefont {B.}~\bibnamefont {Johnson}}, \bibinfo {author} {\bibfnamefont {J.}~\bibnamefont {Chow}}, \bibinfo {author} {\bibfnamefont {J.}~\bibnamefont {Koch}}, \bibinfo {author} {\bibfnamefont {J.}~\bibnamefont {Gambetta}}, \bibinfo {author} {\bibfnamefont {D.}~\bibnamefont {Schuster}}, \bibinfo {author} {\bibfnamefont {L.}~\bibnamefont {Frunzio}}, \bibinfo {author} {\bibfnamefont {M.}~\bibnamefont {Devoret}}, \bibinfo {author} {\bibfnamefont {S.}~\bibnamefont {Girvin}}, \emph {et~al.},\ }\bibfield  {title} {\bibinfo {title} {Controlling the spontaneous emission of a superconducting transmon qubit},\ }\href@noop {} {\bibfield  {journal} {\bibinfo  {journal} {Physical review letters}\ }\textbf {\bibinfo {volume} {101}},\ \bibinfo {pages} {080502} (\bibinfo {year} {2008})}\BibitemShut {NoStop}%
\bibitem [{\citenamefont {Reed}\ \emph {et~al.}(2010)\citenamefont {Reed}, \citenamefont {Johnson}, \citenamefont {Houck}, \citenamefont {DiCarlo}, \citenamefont {Chow}, \citenamefont {Schuster}, \citenamefont {Frunzio},\ and\ \citenamefont {Schoelkopf}}]{reed2010fast}%
  \BibitemOpen
  \bibfield  {author} {\bibinfo {author} {\bibfnamefont {M.~D.}\ \bibnamefont {Reed}}, \bibinfo {author} {\bibfnamefont {B.~R.}\ \bibnamefont {Johnson}}, \bibinfo {author} {\bibfnamefont {A.~A.}\ \bibnamefont {Houck}}, \bibinfo {author} {\bibfnamefont {L.}~\bibnamefont {DiCarlo}}, \bibinfo {author} {\bibfnamefont {J.~M.}\ \bibnamefont {Chow}}, \bibinfo {author} {\bibfnamefont {D.~I.}\ \bibnamefont {Schuster}}, \bibinfo {author} {\bibfnamefont {L.}~\bibnamefont {Frunzio}},\ and\ \bibinfo {author} {\bibfnamefont {R.~J.}\ \bibnamefont {Schoelkopf}},\ }\bibfield  {title} {\bibinfo {title} {Fast reset and suppressing spontaneous emission of a superconducting qubit},\ }\href@noop {} {\bibfield  {journal} {\bibinfo  {journal} {Applied Physics Letters}\ }\textbf {\bibinfo {volume} {96}} (\bibinfo {year} {2010})}\BibitemShut {NoStop}%
\bibitem [{\citenamefont {Jeffrey}\ \emph {et~al.}(2014)\citenamefont {Jeffrey}, \citenamefont {Sank}, \citenamefont {Mutus}, \citenamefont {White}, \citenamefont {Kelly}, \citenamefont {Barends}, \citenamefont {Chen}, \citenamefont {Chen}, \citenamefont {Chiaro}, \citenamefont {Dunsworth}, \citenamefont {Megrant}, \citenamefont {O'Malley}, \citenamefont {Neill}, \citenamefont {Roushan}, \citenamefont {Vainsencher}, \citenamefont {Wenner}, \citenamefont {Cleland},\ and\ \citenamefont {Martinis}}]{jeffrey2014fast}%
  \BibitemOpen
  \bibfield  {author} {\bibinfo {author} {\bibfnamefont {E.}~\bibnamefont {Jeffrey}}, \bibinfo {author} {\bibfnamefont {D.}~\bibnamefont {Sank}}, \bibinfo {author} {\bibfnamefont {J.~Y.}\ \bibnamefont {Mutus}}, \bibinfo {author} {\bibfnamefont {T.~C.}\ \bibnamefont {White}}, \bibinfo {author} {\bibfnamefont {J.}~\bibnamefont {Kelly}}, \bibinfo {author} {\bibfnamefont {R.}~\bibnamefont {Barends}}, \bibinfo {author} {\bibfnamefont {Y.}~\bibnamefont {Chen}}, \bibinfo {author} {\bibfnamefont {Z.}~\bibnamefont {Chen}}, \bibinfo {author} {\bibfnamefont {B.}~\bibnamefont {Chiaro}}, \bibinfo {author} {\bibfnamefont {A.}~\bibnamefont {Dunsworth}}, \bibinfo {author} {\bibfnamefont {A.}~\bibnamefont {Megrant}}, \bibinfo {author} {\bibfnamefont {P.~J.~J.}\ \bibnamefont {O'Malley}}, \bibinfo {author} {\bibfnamefont {C.}~\bibnamefont {Neill}}, \bibinfo {author} {\bibfnamefont {P.}~\bibnamefont {Roushan}}, \bibinfo {author} {\bibfnamefont {A.}~\bibnamefont {Vainsencher}}, \bibinfo {author} {\bibfnamefont {J.}~\bibnamefont
  {Wenner}}, \bibinfo {author} {\bibfnamefont {A.~N.}\ \bibnamefont {Cleland}},\ and\ \bibinfo {author} {\bibfnamefont {J.~M.}\ \bibnamefont {Martinis}},\ }\bibfield  {title} {\bibinfo {title} {Fast accurate state measurement with superconducting qubits},\ }\href {https://doi.org/10.1103/PhysRevLett.112.190504} {\bibfield  {journal} {\bibinfo  {journal} {Phys. Rev. Lett.}\ }\textbf {\bibinfo {volume} {112}},\ \bibinfo {pages} {190504} (\bibinfo {year} {2014})}\BibitemShut {NoStop}%
\bibitem [{\citenamefont {Bronn}\ \emph {et~al.}(2015)\citenamefont {Bronn}, \citenamefont {Liu}, \citenamefont {Hertzberg}, \citenamefont {C{\'o}rcoles}, \citenamefont {Houck}, \citenamefont {Gambetta},\ and\ \citenamefont {Chow}}]{bronn2015broadband}%
  \BibitemOpen
  \bibfield  {author} {\bibinfo {author} {\bibfnamefont {N.~T.}\ \bibnamefont {Bronn}}, \bibinfo {author} {\bibfnamefont {Y.}~\bibnamefont {Liu}}, \bibinfo {author} {\bibfnamefont {J.~B.}\ \bibnamefont {Hertzberg}}, \bibinfo {author} {\bibfnamefont {A.~D.}\ \bibnamefont {C{\'o}rcoles}}, \bibinfo {author} {\bibfnamefont {A.~A.}\ \bibnamefont {Houck}}, \bibinfo {author} {\bibfnamefont {J.~M.}\ \bibnamefont {Gambetta}},\ and\ \bibinfo {author} {\bibfnamefont {J.~M.}\ \bibnamefont {Chow}},\ }\bibfield  {title} {\bibinfo {title} {Broadband filters for abatement of spontaneous emission in circuit quantum electrodynamics},\ }\href@noop {} {\bibfield  {journal} {\bibinfo  {journal} {Applied Physics Letters}\ }\textbf {\bibinfo {volume} {107}} (\bibinfo {year} {2015})}\BibitemShut {NoStop}%
\bibitem [{\citenamefont {Boissonneault}\ \emph {et~al.}(2009)\citenamefont {Boissonneault}, \citenamefont {Gambetta},\ and\ \citenamefont {Blais}}]{Blais_etal_dispersive_cQED}%
  \BibitemOpen
  \bibfield  {author} {\bibinfo {author} {\bibfnamefont {M.}~\bibnamefont {Boissonneault}}, \bibinfo {author} {\bibfnamefont {J.~M.}\ \bibnamefont {Gambetta}},\ and\ \bibinfo {author} {\bibfnamefont {A.}~\bibnamefont {Blais}},\ }\bibfield  {title} {\bibinfo {title} {Dispersive regime of circuit qed: Photon-dependent qubit dephasing and relaxation rates},\ }\href {https://doi.org/10.1103/PhysRevA.79.013819} {\bibfield  {journal} {\bibinfo  {journal} {Phys. Rev. A}\ }\textbf {\bibinfo {volume} {79}},\ \bibinfo {pages} {013819} (\bibinfo {year} {2009})}\BibitemShut {NoStop}%
\bibitem [{\citenamefont {Slichter}\ \emph {et~al.}(2012)\citenamefont {Slichter}, \citenamefont {Vijay}, \citenamefont {Weber}, \citenamefont {Boutin}, \citenamefont {Boissonneault}, \citenamefont {Gambetta}, \citenamefont {Blais},\ and\ \citenamefont {Siddiqi}}]{Slichter2012StateMixing}%
  \BibitemOpen
  \bibfield  {author} {\bibinfo {author} {\bibfnamefont {D.~H.}\ \bibnamefont {Slichter}}, \bibinfo {author} {\bibfnamefont {R.}~\bibnamefont {Vijay}}, \bibinfo {author} {\bibfnamefont {S.~J.}\ \bibnamefont {Weber}}, \bibinfo {author} {\bibfnamefont {S.}~\bibnamefont {Boutin}}, \bibinfo {author} {\bibfnamefont {M.}~\bibnamefont {Boissonneault}}, \bibinfo {author} {\bibfnamefont {J.~M.}\ \bibnamefont {Gambetta}}, \bibinfo {author} {\bibfnamefont {A.}~\bibnamefont {Blais}},\ and\ \bibinfo {author} {\bibfnamefont {I.}~\bibnamefont {Siddiqi}},\ }\bibfield  {title} {\bibinfo {title} {Measurement-induced qubit state mixing in circuit qed from up-converted dephasing noise},\ }\href {https://doi.org/10.1103/PhysRevLett.109.153601} {\bibfield  {journal} {\bibinfo  {journal} {Phys. Rev. Lett.}\ }\textbf {\bibinfo {volume} {109}},\ \bibinfo {pages} {153601} (\bibinfo {year} {2012})}\BibitemShut {NoStop}%
\bibitem [{\citenamefont {Mundhada}\ \emph {et~al.}(2016)\citenamefont {Mundhada}, \citenamefont {Shankar}, \citenamefont {Narla}, \citenamefont {Zalys-Geller}, \citenamefont {Girvin},\ and\ \citenamefont {Devoret}}]{Devoret_March_Meeting}%
  \BibitemOpen
  \bibfield  {author} {\bibinfo {author} {\bibfnamefont {S.}~\bibnamefont {Mundhada}}, \bibinfo {author} {\bibfnamefont {S.}~\bibnamefont {Shankar}}, \bibinfo {author} {\bibfnamefont {A.}~\bibnamefont {Narla}}, \bibinfo {author} {\bibfnamefont {E.}~\bibnamefont {Zalys-Geller}}, \bibinfo {author} {\bibfnamefont {S.}~\bibnamefont {Girvin}},\ and\ \bibinfo {author} {\bibfnamefont {M.}~\bibnamefont {Devoret}},\ }\bibfield  {title} {\bibinfo {title} {Dependence of transmon qubit relaxation rate on readout drive power},\ }\href@noop {} {\bibfield  {journal} {\bibinfo  {journal} {APS March Meeting}\ }\textbf {\bibinfo {volume} {V48}} (\bibinfo {year} {2016})}\BibitemShut {NoStop}%
\bibitem [{\citenamefont {Minev}\ \emph {et~al.}(2019)\citenamefont {Minev}, \citenamefont {Mundhada}, \citenamefont {Shankar}, \citenamefont {Reinhold}, \citenamefont {Gutiérrez-Jáuregui}, \citenamefont {Schoelkopf}, \citenamefont {Mirrahimi}, \citenamefont {Carmichael},\ and\ \citenamefont {Devoret}}]{Minev_Nature}%
  \BibitemOpen
  \bibfield  {author} {\bibinfo {author} {\bibfnamefont {Z.}~\bibnamefont {Minev}}, \bibinfo {author} {\bibfnamefont {S.}~\bibnamefont {Mundhada}}, \bibinfo {author} {\bibfnamefont {S.}~\bibnamefont {Shankar}}, \bibinfo {author} {\bibfnamefont {P.}~\bibnamefont {Reinhold}}, \bibinfo {author} {\bibfnamefont {R.}~\bibnamefont {Gutiérrez-Jáuregui}}, \bibinfo {author} {\bibfnamefont {R.}~\bibnamefont {Schoelkopf}}, \bibinfo {author} {\bibfnamefont {M.}~\bibnamefont {Mirrahimi}}, \bibinfo {author} {\bibfnamefont {H.}~\bibnamefont {Carmichael}},\ and\ \bibinfo {author} {\bibfnamefont {M.}~\bibnamefont {Devoret}},\ }\bibfield  {title} {\bibinfo {title} {To catch and reverse a quantum jump mid-flight},\ }\bibfield  {journal} {\bibinfo  {journal} {Nature}\ }\textbf {\bibinfo {volume} {570}},\ \href {https://doi.org/10.1038/s41586-019-1287-z} {10.1038/s41586-019-1287-z} (\bibinfo {year} {2019})\BibitemShut {NoStop}%
\bibitem [{\citenamefont {Petrescu}\ \emph {et~al.}(2020)\citenamefont {Petrescu}, \citenamefont {Malekakhlagh},\ and\ \citenamefont {T\"ureci}}]{PetrescuReadout2020}%
  \BibitemOpen
  \bibfield  {author} {\bibinfo {author} {\bibfnamefont {A.}~\bibnamefont {Petrescu}}, \bibinfo {author} {\bibfnamefont {M.}~\bibnamefont {Malekakhlagh}},\ and\ \bibinfo {author} {\bibfnamefont {H.~E.}\ \bibnamefont {T\"ureci}},\ }\bibfield  {title} {\bibinfo {title} {Lifetime renormalization of driven weakly anharmonic superconducting qubits. ii. the readout problem},\ }\href {https://doi.org/10.1103/PhysRevB.101.134510} {\bibfield  {journal} {\bibinfo  {journal} {Phys. Rev. B}\ }\textbf {\bibinfo {volume} {101}},\ \bibinfo {pages} {134510} (\bibinfo {year} {2020})}\BibitemShut {NoStop}%
\bibitem [{\citenamefont {Hanai}\ \emph {et~al.}(2021)\citenamefont {Hanai}, \citenamefont {McDonald},\ and\ \citenamefont {Clerk}}]{Hanai2021}%
  \BibitemOpen
  \bibfield  {author} {\bibinfo {author} {\bibfnamefont {R.}~\bibnamefont {Hanai}}, \bibinfo {author} {\bibfnamefont {A.}~\bibnamefont {McDonald}},\ and\ \bibinfo {author} {\bibfnamefont {A.}~\bibnamefont {Clerk}},\ }\bibfield  {title} {\bibinfo {title} {Intrinsic mechanisms for drive-dependent purcell decay in superconducting quantum circuits},\ }\href {https://doi.org/10.1103/PhysRevResearch.3.043228} {\bibfield  {journal} {\bibinfo  {journal} {Phys. Rev. Res.}\ }\textbf {\bibinfo {volume} {3}},\ \bibinfo {pages} {043228} (\bibinfo {year} {2021})}\BibitemShut {NoStop}%
\bibitem [{\citenamefont {Gusenkova}\ \emph {et~al.}(2021)\citenamefont {Gusenkova}, \citenamefont {Spiecker}, \citenamefont {Gebauer}, \citenamefont {Willsch}, \citenamefont {Willsch}, \citenamefont {Valenti}, \citenamefont {Karcher}, \citenamefont {Gr\"unhaupt}, \citenamefont {Takmakov}, \citenamefont {Winkel}, \citenamefont {Rieger}, \citenamefont {Ustinov}, \citenamefont {Roch}, \citenamefont {Wernsdorfer}, \citenamefont {Michielsen}, \citenamefont {Sander},\ and\ \citenamefont {Pop}}]{Pop_et_al_readout_with_large_photon_number}%
  \BibitemOpen
  \bibfield  {author} {\bibinfo {author} {\bibfnamefont {D.}~\bibnamefont {Gusenkova}}, \bibinfo {author} {\bibfnamefont {M.}~\bibnamefont {Spiecker}}, \bibinfo {author} {\bibfnamefont {R.}~\bibnamefont {Gebauer}}, \bibinfo {author} {\bibfnamefont {M.}~\bibnamefont {Willsch}}, \bibinfo {author} {\bibfnamefont {D.}~\bibnamefont {Willsch}}, \bibinfo {author} {\bibfnamefont {F.}~\bibnamefont {Valenti}}, \bibinfo {author} {\bibfnamefont {N.}~\bibnamefont {Karcher}}, \bibinfo {author} {\bibfnamefont {L.}~\bibnamefont {Gr\"unhaupt}}, \bibinfo {author} {\bibfnamefont {I.}~\bibnamefont {Takmakov}}, \bibinfo {author} {\bibfnamefont {P.}~\bibnamefont {Winkel}}, \bibinfo {author} {\bibfnamefont {D.}~\bibnamefont {Rieger}}, \bibinfo {author} {\bibfnamefont {A.~V.}\ \bibnamefont {Ustinov}}, \bibinfo {author} {\bibfnamefont {N.}~\bibnamefont {Roch}}, \bibinfo {author} {\bibfnamefont {W.}~\bibnamefont {Wernsdorfer}}, \bibinfo {author} {\bibfnamefont {K.}~\bibnamefont {Michielsen}}, \bibinfo {author} {\bibfnamefont
  {O.}~\bibnamefont {Sander}},\ and\ \bibinfo {author} {\bibfnamefont {I.~M.}\ \bibnamefont {Pop}},\ }\bibfield  {title} {\bibinfo {title} {Quantum nondemolition dispersive readout of a superconducting artificial atom using large photon numbers},\ }\href {https://doi.org/10.1103/PhysRevApplied.15.064030} {\bibfield  {journal} {\bibinfo  {journal} {Phys. Rev. Appl.}\ }\textbf {\bibinfo {volume} {15}},\ \bibinfo {pages} {064030} (\bibinfo {year} {2021})}\BibitemShut {NoStop}%
\bibitem [{\citenamefont {Shillito}\ \emph {et~al.}(2022)\citenamefont {Shillito}, \citenamefont {Petrescu}, \citenamefont {Cohen}, \citenamefont {Beall}, \citenamefont {Hauru}, \citenamefont {Ganahl}, \citenamefont {Lewis}, \citenamefont {Vidal},\ and\ \citenamefont {Blais}}]{Blais_etal_transmon_ionization}%
  \BibitemOpen
  \bibfield  {author} {\bibinfo {author} {\bibfnamefont {R.}~\bibnamefont {Shillito}}, \bibinfo {author} {\bibfnamefont {A.}~\bibnamefont {Petrescu}}, \bibinfo {author} {\bibfnamefont {J.}~\bibnamefont {Cohen}}, \bibinfo {author} {\bibfnamefont {J.}~\bibnamefont {Beall}}, \bibinfo {author} {\bibfnamefont {M.}~\bibnamefont {Hauru}}, \bibinfo {author} {\bibfnamefont {M.}~\bibnamefont {Ganahl}}, \bibinfo {author} {\bibfnamefont {A.~G.}\ \bibnamefont {Lewis}}, \bibinfo {author} {\bibfnamefont {G.}~\bibnamefont {Vidal}},\ and\ \bibinfo {author} {\bibfnamefont {A.}~\bibnamefont {Blais}},\ }\bibfield  {title} {\bibinfo {title} {Dynamics of transmon ionization},\ }\href {https://doi.org/10.1103/PhysRevApplied.18.034031} {\bibfield  {journal} {\bibinfo  {journal} {Phys. Rev. Appl.}\ }\textbf {\bibinfo {volume} {18}},\ \bibinfo {pages} {034031} (\bibinfo {year} {2022})}\BibitemShut {NoStop}%
\bibitem [{\citenamefont {Cohen}\ \emph {et~al.}(2023)\citenamefont {Cohen}, \citenamefont {Petrescu}, \citenamefont {Shillito},\ and\ \citenamefont {Blais}}]{Blais_et_al_chaos_in_transmon}%
  \BibitemOpen
  \bibfield  {author} {\bibinfo {author} {\bibfnamefont {J.}~\bibnamefont {Cohen}}, \bibinfo {author} {\bibfnamefont {A.}~\bibnamefont {Petrescu}}, \bibinfo {author} {\bibfnamefont {R.}~\bibnamefont {Shillito}},\ and\ \bibinfo {author} {\bibfnamefont {A.}~\bibnamefont {Blais}},\ }\bibfield  {title} {\bibinfo {title} {Reminiscence of classical chaos in driven transmons},\ }\href {https://doi.org/10.1103/PRXQuantum.4.020312} {\bibfield  {journal} {\bibinfo  {journal} {PRX Quantum}\ }\textbf {\bibinfo {volume} {4}},\ \bibinfo {pages} {020312} (\bibinfo {year} {2023})}\BibitemShut {NoStop}%
\bibitem [{\citenamefont {Caldeira}\ and\ \citenamefont {Leggett}(1983)}]{Caldeira_Leggett}%
  \BibitemOpen
  \bibfield  {author} {\bibinfo {author} {\bibfnamefont {A.}~\bibnamefont {Caldeira}}\ and\ \bibinfo {author} {\bibfnamefont {A.}~\bibnamefont {Leggett}},\ }\bibfield  {title} {\bibinfo {title} {Quantum tunnelling in a dissipative system},\ }\href {https://doi.org/https://doi.org/10.1016/0003-4916(83)90202-6} {\bibfield  {journal} {\bibinfo  {journal} {Annals of Physics}\ }\textbf {\bibinfo {volume} {149}},\ \bibinfo {pages} {374} (\bibinfo {year} {1983})}\BibitemShut {NoStop}%
\bibitem [{\citenamefont {Filipp}\ \emph {et~al.}(2011)\citenamefont {Filipp}, \citenamefont {G\"oppl}, \citenamefont {Fink}, \citenamefont {Baur}, \citenamefont {Bianchetti}, \citenamefont {Steffen},\ and\ \citenamefont {Wallraff}}]{Filipp2011}%
  \BibitemOpen
  \bibfield  {author} {\bibinfo {author} {\bibfnamefont {S.}~\bibnamefont {Filipp}}, \bibinfo {author} {\bibfnamefont {M.}~\bibnamefont {G\"oppl}}, \bibinfo {author} {\bibfnamefont {J.~M.}\ \bibnamefont {Fink}}, \bibinfo {author} {\bibfnamefont {M.}~\bibnamefont {Baur}}, \bibinfo {author} {\bibfnamefont {R.}~\bibnamefont {Bianchetti}}, \bibinfo {author} {\bibfnamefont {L.}~\bibnamefont {Steffen}},\ and\ \bibinfo {author} {\bibfnamefont {A.}~\bibnamefont {Wallraff}},\ }\bibfield  {title} {\bibinfo {title} {Multimode mediated qubit-qubit coupling and dark-state symmetries in circuit quantum electrodynamics},\ }\href {https://doi.org/10.1103/PhysRevA.83.063827} {\bibfield  {journal} {\bibinfo  {journal} {Phys. Rev. A}\ }\textbf {\bibinfo {volume} {83}},\ \bibinfo {pages} {063827} (\bibinfo {year} {2011})}\BibitemShut {NoStop}%
\bibitem [{\citenamefont {Gely}\ \emph {et~al.}(2017)\citenamefont {Gely}, \citenamefont {Parra-Rodriguez}, \citenamefont {Bothner}, \citenamefont {Blanter}, \citenamefont {Bosman}, \citenamefont {Solano},\ and\ \citenamefont {Steele}}]{Gely2017multimode}%
  \BibitemOpen
  \bibfield  {author} {\bibinfo {author} {\bibfnamefont {M.~F.}\ \bibnamefont {Gely}}, \bibinfo {author} {\bibfnamefont {A.}~\bibnamefont {Parra-Rodriguez}}, \bibinfo {author} {\bibfnamefont {D.}~\bibnamefont {Bothner}}, \bibinfo {author} {\bibfnamefont {Y.~M.}\ \bibnamefont {Blanter}}, \bibinfo {author} {\bibfnamefont {S.~J.}\ \bibnamefont {Bosman}}, \bibinfo {author} {\bibfnamefont {E.}~\bibnamefont {Solano}},\ and\ \bibinfo {author} {\bibfnamefont {G.~A.}\ \bibnamefont {Steele}},\ }\bibfield  {title} {\bibinfo {title} {Convergence of the multimode quantum rabi model of circuit quantum electrodynamics},\ }\href {https://doi.org/10.1103/PhysRevB.95.245115} {\bibfield  {journal} {\bibinfo  {journal} {Phys. Rev. B}\ }\textbf {\bibinfo {volume} {95}},\ \bibinfo {pages} {245115} (\bibinfo {year} {2017})}\BibitemShut {NoStop}%
\bibitem [{\citenamefont {Sinha}\ \emph {et~al.}(2022)\citenamefont {Sinha}, \citenamefont {Khan}, \citenamefont {C{\"u}ce},\ and\ \citenamefont {T{\"u}reci}}]{kanupaper}%
  \BibitemOpen
  \bibfield  {author} {\bibinfo {author} {\bibfnamefont {K.}~\bibnamefont {Sinha}}, \bibinfo {author} {\bibfnamefont {S.~A.}\ \bibnamefont {Khan}}, \bibinfo {author} {\bibfnamefont {E.}~\bibnamefont {C{\"u}ce}},\ and\ \bibinfo {author} {\bibfnamefont {H.~E.}\ \bibnamefont {T{\"u}reci}},\ }\bibfield  {title} {\bibinfo {title} {Radiative properties of an artificial atom coupled to a josephson-junction array},\ }\href {https://doi.org/10.1103/PhysRevA.106.033714} {\bibfield  {journal} {\bibinfo  {journal} {Phys. Rev. A}\ }\textbf {\bibinfo {volume} {106}},\ \bibinfo {pages} {033714} (\bibinfo {year} {2022})}\BibitemShut {NoStop}%
\bibitem [{\citenamefont {Yoshihara}\ \emph {et~al.}(2014)\citenamefont {Yoshihara}, \citenamefont {Nakamura}, \citenamefont {Yan}, \citenamefont {Gustavsson}, \citenamefont {Bylander}, \citenamefont {Oliver},\ and\ \citenamefont {Tsai}}]{yoshihara2014flux}%
  \BibitemOpen
  \bibfield  {author} {\bibinfo {author} {\bibfnamefont {F.}~\bibnamefont {Yoshihara}}, \bibinfo {author} {\bibfnamefont {Y.}~\bibnamefont {Nakamura}}, \bibinfo {author} {\bibfnamefont {F.}~\bibnamefont {Yan}}, \bibinfo {author} {\bibfnamefont {S.}~\bibnamefont {Gustavsson}}, \bibinfo {author} {\bibfnamefont {J.}~\bibnamefont {Bylander}}, \bibinfo {author} {\bibfnamefont {W.~D.}\ \bibnamefont {Oliver}},\ and\ \bibinfo {author} {\bibfnamefont {J.-S.}\ \bibnamefont {Tsai}},\ }\bibfield  {title} {\bibinfo {title} {Flux qubit noise spectroscopy using rabi oscillations under strong driving conditions},\ }\href@noop {} {\bibfield  {journal} {\bibinfo  {journal} {Physical Review B}\ }\textbf {\bibinfo {volume} {89}},\ \bibinfo {pages} {020503} (\bibinfo {year} {2014})}\BibitemShut {NoStop}%
\bibitem [{\citenamefont {Kumar}\ \emph {et~al.}(2016)\citenamefont {Kumar}, \citenamefont {Sendelbach}, \citenamefont {Beck}, \citenamefont {Freeland}, \citenamefont {Wang}, \citenamefont {Wang}, \citenamefont {Clare}, \citenamefont {Wu}, \citenamefont {Pappas},\ and\ \citenamefont {McDermott}}]{kumar2016origin}%
  \BibitemOpen
  \bibfield  {author} {\bibinfo {author} {\bibfnamefont {P.}~\bibnamefont {Kumar}}, \bibinfo {author} {\bibfnamefont {S.}~\bibnamefont {Sendelbach}}, \bibinfo {author} {\bibfnamefont {M.}~\bibnamefont {Beck}}, \bibinfo {author} {\bibfnamefont {J.}~\bibnamefont {Freeland}}, \bibinfo {author} {\bibfnamefont {Z.}~\bibnamefont {Wang}}, \bibinfo {author} {\bibfnamefont {H.}~\bibnamefont {Wang}}, \bibinfo {author} {\bibfnamefont {C.~Y.}\ \bibnamefont {Clare}}, \bibinfo {author} {\bibfnamefont {R.}~\bibnamefont {Wu}}, \bibinfo {author} {\bibfnamefont {D.}~\bibnamefont {Pappas}},\ and\ \bibinfo {author} {\bibfnamefont {R.}~\bibnamefont {McDermott}},\ }\bibfield  {title} {\bibinfo {title} {Origin and reduction of 1/f magnetic flux noise in superconducting devices},\ }\href@noop {} {\bibfield  {journal} {\bibinfo  {journal} {Physical Review Applied}\ }\textbf {\bibinfo {volume} {6}},\ \bibinfo {pages} {041001} (\bibinfo {year} {2016})}\BibitemShut {NoStop}%
\bibitem [{\citenamefont {Wilen}\ \emph {et~al.}(2021)\citenamefont {Wilen}, \citenamefont {Abdullah}, \citenamefont {Kurinsky}, \citenamefont {Stanford}, \citenamefont {Cardani}, \citenamefont {d’Imperio}, \citenamefont {Tomei}, \citenamefont {Faoro}, \citenamefont {Ioffe}, \citenamefont {Liu} \emph {et~al.}}]{wilen2021correlated}%
  \BibitemOpen
  \bibfield  {author} {\bibinfo {author} {\bibfnamefont {C.~D.}\ \bibnamefont {Wilen}}, \bibinfo {author} {\bibfnamefont {S.}~\bibnamefont {Abdullah}}, \bibinfo {author} {\bibfnamefont {N.}~\bibnamefont {Kurinsky}}, \bibinfo {author} {\bibfnamefont {C.}~\bibnamefont {Stanford}}, \bibinfo {author} {\bibfnamefont {L.}~\bibnamefont {Cardani}}, \bibinfo {author} {\bibfnamefont {G.}~\bibnamefont {d’Imperio}}, \bibinfo {author} {\bibfnamefont {C.}~\bibnamefont {Tomei}}, \bibinfo {author} {\bibfnamefont {L.}~\bibnamefont {Faoro}}, \bibinfo {author} {\bibfnamefont {L.}~\bibnamefont {Ioffe}}, \bibinfo {author} {\bibfnamefont {C.}~\bibnamefont {Liu}}, \emph {et~al.},\ }\bibfield  {title} {\bibinfo {title} {Correlated charge noise and relaxation errors in superconducting qubits},\ }\href@noop {} {\bibfield  {journal} {\bibinfo  {journal} {Nature}\ }\textbf {\bibinfo {volume} {594}},\ \bibinfo {pages} {369} (\bibinfo {year} {2021})}\BibitemShut {NoStop}%
\bibitem [{\citenamefont {Pop}\ \emph {et~al.}(2014)\citenamefont {Pop}, \citenamefont {Geerlings}, \citenamefont {Catelani}, \citenamefont {Schoelkopf}, \citenamefont {Glazman},\ and\ \citenamefont {Devoret}}]{pop2014}%
  \BibitemOpen
  \bibfield  {author} {\bibinfo {author} {\bibfnamefont {I.~M.}\ \bibnamefont {Pop}}, \bibinfo {author} {\bibfnamefont {K.}~\bibnamefont {Geerlings}}, \bibinfo {author} {\bibfnamefont {G.}~\bibnamefont {Catelani}}, \bibinfo {author} {\bibfnamefont {R.~J.}\ \bibnamefont {Schoelkopf}}, \bibinfo {author} {\bibfnamefont {L.~I.}\ \bibnamefont {Glazman}},\ and\ \bibinfo {author} {\bibfnamefont {M.~H.}\ \bibnamefont {Devoret}},\ }\bibfield  {title} {\bibinfo {title} {Coherent suppression of electromagnetic dissipation due to superconducting quasiparticles},\ }\href@noop {} {\bibfield  {journal} {\bibinfo  {journal} {Nature}\ }\textbf {\bibinfo {volume} {508}},\ \bibinfo {pages} {369} (\bibinfo {year} {2014})}\BibitemShut {NoStop}%
\bibitem [{\citenamefont {Frattini}\ \emph {et~al.}(2018)\citenamefont {Frattini}, \citenamefont {Sivak}, \citenamefont {Lingenfelter}, \citenamefont {Shankar},\ and\ \citenamefont {Devoret}}]{frattini2018optimizing}%
  \BibitemOpen
  \bibfield  {author} {\bibinfo {author} {\bibfnamefont {N.}~\bibnamefont {Frattini}}, \bibinfo {author} {\bibfnamefont {V.}~\bibnamefont {Sivak}}, \bibinfo {author} {\bibfnamefont {A.}~\bibnamefont {Lingenfelter}}, \bibinfo {author} {\bibfnamefont {S.}~\bibnamefont {Shankar}},\ and\ \bibinfo {author} {\bibfnamefont {M.}~\bibnamefont {Devoret}},\ }\bibfield  {title} {\bibinfo {title} {Optimizing the nonlinearity and dissipation of a snail parametric amplifier for dynamic range},\ }\href@noop {} {\bibfield  {journal} {\bibinfo  {journal} {Physical Review Applied}\ }\textbf {\bibinfo {volume} {10}},\ \bibinfo {pages} {054020} (\bibinfo {year} {2018})}\BibitemShut {NoStop}%
\bibitem [{\citenamefont {Wenner}\ \emph {et~al.}(2011)\citenamefont {Wenner}, \citenamefont {Neeley}, \citenamefont {Bialczak}, \citenamefont {Lenander}, \citenamefont {Lucero}, \citenamefont {O’Connell}, \citenamefont {Sank}, \citenamefont {Wang}, \citenamefont {Weides}, \citenamefont {Cleland},\ and\ \citenamefont {Martinis}}]{Crosstalk1_2011}%
  \BibitemOpen
  \bibfield  {author} {\bibinfo {author} {\bibfnamefont {J.}~\bibnamefont {Wenner}}, \bibinfo {author} {\bibfnamefont {M.}~\bibnamefont {Neeley}}, \bibinfo {author} {\bibfnamefont {R.~C.}\ \bibnamefont {Bialczak}}, \bibinfo {author} {\bibfnamefont {M.}~\bibnamefont {Lenander}}, \bibinfo {author} {\bibfnamefont {E.}~\bibnamefont {Lucero}}, \bibinfo {author} {\bibfnamefont {A.~D.}\ \bibnamefont {O’Connell}}, \bibinfo {author} {\bibfnamefont {D.}~\bibnamefont {Sank}}, \bibinfo {author} {\bibfnamefont {H.}~\bibnamefont {Wang}}, \bibinfo {author} {\bibfnamefont {M.}~\bibnamefont {Weides}}, \bibinfo {author} {\bibfnamefont {A.~N.}\ \bibnamefont {Cleland}},\ and\ \bibinfo {author} {\bibfnamefont {J.~M.}\ \bibnamefont {Martinis}},\ }\bibfield  {title} {\bibinfo {title} {Wirebond crosstalk and cavity modes in large chip mounts for superconducting qubits},\ }\href {https://doi.org/10.1088/0953-2048/24/6/065001} {\bibfield  {journal} {\bibinfo  {journal} {Supercond. Sci. Technol.}\ }\textbf {\bibinfo {volume} {24}},\
  \bibinfo {pages} {065001} (\bibinfo {year} {2011})}\BibitemShut {NoStop}%
\bibitem [{\citenamefont {Huang}\ \emph {et~al.}(2021)\citenamefont {Huang}, \citenamefont {Lienhard}, \citenamefont {Calusine}, \citenamefont {Veps{\"a}l{\"a}inen}, \citenamefont {Braum{\"u}ller}, \citenamefont {Kim}, \citenamefont {Melville}, \citenamefont {Niedzielski}, \citenamefont {Yoder}, \citenamefont {Kannan} \emph {et~al.}}]{huang2021microwavepackage}%
  \BibitemOpen
  \bibfield  {author} {\bibinfo {author} {\bibfnamefont {S.}~\bibnamefont {Huang}}, \bibinfo {author} {\bibfnamefont {B.}~\bibnamefont {Lienhard}}, \bibinfo {author} {\bibfnamefont {G.}~\bibnamefont {Calusine}}, \bibinfo {author} {\bibfnamefont {A.}~\bibnamefont {Veps{\"a}l{\"a}inen}}, \bibinfo {author} {\bibfnamefont {J.}~\bibnamefont {Braum{\"u}ller}}, \bibinfo {author} {\bibfnamefont {D.~K.}\ \bibnamefont {Kim}}, \bibinfo {author} {\bibfnamefont {A.~J.}\ \bibnamefont {Melville}}, \bibinfo {author} {\bibfnamefont {B.~M.}\ \bibnamefont {Niedzielski}}, \bibinfo {author} {\bibfnamefont {J.~L.}\ \bibnamefont {Yoder}}, \bibinfo {author} {\bibfnamefont {B.}~\bibnamefont {Kannan}}, \emph {et~al.},\ }\bibfield  {title} {\bibinfo {title} {Microwave package design for superconducting quantum processors},\ }\href@noop {} {\bibfield  {journal} {\bibinfo  {journal} {PRX Quantum}\ }\textbf {\bibinfo {volume} {2}},\ \bibinfo {pages} {020306} (\bibinfo {year} {2021})}\BibitemShut {NoStop}%
\bibitem [{\citenamefont {Berenger}(1994)}]{berenger1994pml}%
  \BibitemOpen
  \bibfield  {author} {\bibinfo {author} {\bibfnamefont {J.-P.}\ \bibnamefont {Berenger}},\ }\bibfield  {title} {\bibinfo {title} {A perfectly matched layer for the absorption of electromagnetic waves},\ }\href@noop {} {\bibfield  {journal} {\bibinfo  {journal} {Journal of computational physics}\ }\textbf {\bibinfo {volume} {114}},\ \bibinfo {pages} {185} (\bibinfo {year} {1994})}\BibitemShut {NoStop}%
\bibitem [{\citenamefont {Berenger}(1996)}]{berenger1996_3Dpml}%
  \BibitemOpen
  \bibfield  {author} {\bibinfo {author} {\bibfnamefont {J.-P.}\ \bibnamefont {Berenger}},\ }\bibfield  {title} {\bibinfo {title} {Three-dimensional perfectly matched layer for the absorption of electromagnetic waves},\ }\href@noop {} {\bibfield  {journal} {\bibinfo  {journal} {Journal of computational physics}\ }\textbf {\bibinfo {volume} {127}},\ \bibinfo {pages} {363} (\bibinfo {year} {1996})}\BibitemShut {NoStop}%
\bibitem [{\citenamefont {Clayton}\ and\ \citenamefont {Engquist}(1977)}]{clayton1977absorbing}%
  \BibitemOpen
  \bibfield  {author} {\bibinfo {author} {\bibfnamefont {R.}~\bibnamefont {Clayton}}\ and\ \bibinfo {author} {\bibfnamefont {B.}~\bibnamefont {Engquist}},\ }\bibfield  {title} {\bibinfo {title} {Absorbing boundary conditions for acoustic and elastic wave equations},\ }\href@noop {} {\bibfield  {journal} {\bibinfo  {journal} {Bulletin of the seismological society of America}\ }\textbf {\bibinfo {volume} {67}},\ \bibinfo {pages} {1529} (\bibinfo {year} {1977})}\BibitemShut {NoStop}%
\bibitem [{\citenamefont {Higdon}(1987)}]{absorbingBCs}%
  \BibitemOpen
  \bibfield  {author} {\bibinfo {author} {\bibfnamefont {R.~L.}\ \bibnamefont {Higdon}},\ }\bibfield  {title} {\bibinfo {title} {Numerical absorbing boundary conditions for the wave equation},\ }\href {http://www.jstor.org/stable/2008250} {\bibfield  {journal} {\bibinfo  {journal} {Mathematics of Computation}\ }\textbf {\bibinfo {volume} {49}},\ \bibinfo {pages} {65} (\bibinfo {year} {1987})}\BibitemShut {NoStop}%
\bibitem [{\citenamefont {Gell-Mann}\ and\ \citenamefont {E.}(1954)}]{Gell-Mann_qed}%
  \BibitemOpen
  \bibfield  {author} {\bibinfo {author} {\bibfnamefont {M.}~\bibnamefont {Gell-Mann}}\ and\ \bibinfo {author} {\bibfnamefont {L.~F.}\ \bibnamefont {E.}},\ }\bibfield  {title} {\bibinfo {title} {Quantum electrodynamics at small distances},\ }\href {https://doi.org/10.1103/PhysRev.95.1300} {\bibfield  {journal} {\bibinfo  {journal} {Phys. Rev.}\ }\textbf {\bibinfo {volume} {95}},\ \bibinfo {pages} {1300} (\bibinfo {year} {1954})}\BibitemShut {NoStop}%
\bibitem [{\citenamefont {Qi}\ \emph {et~al.}(2018)\citenamefont {Qi}, \citenamefont {Zhang},\ and\ \citenamefont {Ge}}]{qi2018defect}%
  \BibitemOpen
  \bibfield  {author} {\bibinfo {author} {\bibfnamefont {B.}~\bibnamefont {Qi}}, \bibinfo {author} {\bibfnamefont {L.}~\bibnamefont {Zhang}},\ and\ \bibinfo {author} {\bibfnamefont {L.}~\bibnamefont {Ge}},\ }\bibfield  {title} {\bibinfo {title} {Defect states emerging from a non-hermitian flatband of photonic zero modes},\ }\href@noop {} {\bibfield  {journal} {\bibinfo  {journal} {Physical review letters}\ }\textbf {\bibinfo {volume} {120}},\ \bibinfo {pages} {093901} (\bibinfo {year} {2018})}\BibitemShut {NoStop}%
\bibitem [{\citenamefont {El-Ganainy}\ \emph {et~al.}(2018)\citenamefont {El-Ganainy}, \citenamefont {Makris}, \citenamefont {Khajavikhan}, \citenamefont {Musslimani}, \citenamefont {Rotter},\ and\ \citenamefont {Christodoulides}}]{el2018non}%
  \BibitemOpen
  \bibfield  {author} {\bibinfo {author} {\bibfnamefont {R.}~\bibnamefont {El-Ganainy}}, \bibinfo {author} {\bibfnamefont {K.~G.}\ \bibnamefont {Makris}}, \bibinfo {author} {\bibfnamefont {M.}~\bibnamefont {Khajavikhan}}, \bibinfo {author} {\bibfnamefont {Z.~H.}\ \bibnamefont {Musslimani}}, \bibinfo {author} {\bibfnamefont {S.}~\bibnamefont {Rotter}},\ and\ \bibinfo {author} {\bibfnamefont {D.~N.}\ \bibnamefont {Christodoulides}},\ }\bibfield  {title} {\bibinfo {title} {Non-hermitian physics and pt symmetry},\ }\href@noop {} {\bibfield  {journal} {\bibinfo  {journal} {Nature Physics}\ }\textbf {\bibinfo {volume} {14}},\ \bibinfo {pages} {11} (\bibinfo {year} {2018})}\BibitemShut {NoStop}%
\bibitem [{\citenamefont {Gigli}\ \emph {et~al.}(2020)\citenamefont {Gigli}, \citenamefont {Wu}, \citenamefont {Marino}, \citenamefont {Borne}, \citenamefont {Leo},\ and\ \citenamefont {Lalanne}}]{gigli2020quasinormal}%
  \BibitemOpen
  \bibfield  {author} {\bibinfo {author} {\bibfnamefont {C.}~\bibnamefont {Gigli}}, \bibinfo {author} {\bibfnamefont {T.}~\bibnamefont {Wu}}, \bibinfo {author} {\bibfnamefont {G.}~\bibnamefont {Marino}}, \bibinfo {author} {\bibfnamefont {A.}~\bibnamefont {Borne}}, \bibinfo {author} {\bibfnamefont {G.}~\bibnamefont {Leo}},\ and\ \bibinfo {author} {\bibfnamefont {P.}~\bibnamefont {Lalanne}},\ }\bibfield  {title} {\bibinfo {title} {Quasinormal-mode non-hermitian modeling and design in nonlinear nano-optics},\ }\href@noop {} {\bibfield  {journal} {\bibinfo  {journal} {ACS photonics}\ }\textbf {\bibinfo {volume} {7}},\ \bibinfo {pages} {1197} (\bibinfo {year} {2020})}\BibitemShut {NoStop}%
\bibitem [{\citenamefont {Purcell}\ \emph {et~al.}(1946)\citenamefont {Purcell}, \citenamefont {Torrey},\ and\ \citenamefont {Pound}}]{purcell1946resonance}%
  \BibitemOpen
  \bibfield  {author} {\bibinfo {author} {\bibfnamefont {E.~M.}\ \bibnamefont {Purcell}}, \bibinfo {author} {\bibfnamefont {H.~C.}\ \bibnamefont {Torrey}},\ and\ \bibinfo {author} {\bibfnamefont {R.~V.}\ \bibnamefont {Pound}},\ }\bibfield  {title} {\bibinfo {title} {Resonance absorption by nuclear magnetic moments in a solid},\ }\href@noop {} {\bibfield  {journal} {\bibinfo  {journal} {Physical review}\ }\textbf {\bibinfo {volume} {69}},\ \bibinfo {pages} {37} (\bibinfo {year} {1946})}\BibitemShut {NoStop}%
\bibitem [{\citenamefont {Malekakhlagh}\ \emph {et~al.}(2016)\citenamefont {Malekakhlagh}, \citenamefont {Petrescu},\ and\ \citenamefont {T{\"u}reci}}]{malekakhlagh2016non}%
  \BibitemOpen
  \bibfield  {author} {\bibinfo {author} {\bibfnamefont {M.}~\bibnamefont {Malekakhlagh}}, \bibinfo {author} {\bibfnamefont {A.}~\bibnamefont {Petrescu}},\ and\ \bibinfo {author} {\bibfnamefont {H.~E.}\ \bibnamefont {T{\"u}reci}},\ }\bibfield  {title} {\bibinfo {title} {Non-markovian dynamics of a superconducting qubit in an open multimode resonator},\ }\href@noop {} {\bibfield  {journal} {\bibinfo  {journal} {Physical Review A}\ }\textbf {\bibinfo {volume} {94}},\ \bibinfo {pages} {063848} (\bibinfo {year} {2016})}\BibitemShut {NoStop}%
\bibitem [{\citenamefont {Bosman}\ \emph {et~al.}(2017)\citenamefont {Bosman}, \citenamefont {Gely}, \citenamefont {Singh}, \citenamefont {Bruno}, \citenamefont {Bothner},\ and\ \citenamefont {Steele}}]{bosman2017multi}%
  \BibitemOpen
  \bibfield  {author} {\bibinfo {author} {\bibfnamefont {S.~J.}\ \bibnamefont {Bosman}}, \bibinfo {author} {\bibfnamefont {M.~F.}\ \bibnamefont {Gely}}, \bibinfo {author} {\bibfnamefont {V.}~\bibnamefont {Singh}}, \bibinfo {author} {\bibfnamefont {A.}~\bibnamefont {Bruno}}, \bibinfo {author} {\bibfnamefont {D.}~\bibnamefont {Bothner}},\ and\ \bibinfo {author} {\bibfnamefont {G.~A.}\ \bibnamefont {Steele}},\ }\bibfield  {title} {\bibinfo {title} {Multi-mode ultra-strong coupling in circuit quantum electrodynamics},\ }\href@noop {} {\bibfield  {journal} {\bibinfo  {journal} {npj Quantum Information}\ }\textbf {\bibinfo {volume} {3}},\ \bibinfo {pages} {46} (\bibinfo {year} {2017})}\BibitemShut {NoStop}%
\bibitem [{\citenamefont {Puertas~Mart{\'\i}nez}\ \emph {et~al.}(2019)\citenamefont {Puertas~Mart{\'\i}nez}, \citenamefont {L{\'e}ger}, \citenamefont {Gheeraert}, \citenamefont {Dassonneville}, \citenamefont {Planat}, \citenamefont {Foroughi}, \citenamefont {Krupko}, \citenamefont {Buisson}, \citenamefont {Naud}, \citenamefont {Hasch-Guichard} \emph {et~al.}}]{puertas2019tunable}%
  \BibitemOpen
  \bibfield  {author} {\bibinfo {author} {\bibfnamefont {J.}~\bibnamefont {Puertas~Mart{\'\i}nez}}, \bibinfo {author} {\bibfnamefont {S.}~\bibnamefont {L{\'e}ger}}, \bibinfo {author} {\bibfnamefont {N.}~\bibnamefont {Gheeraert}}, \bibinfo {author} {\bibfnamefont {R.}~\bibnamefont {Dassonneville}}, \bibinfo {author} {\bibfnamefont {L.}~\bibnamefont {Planat}}, \bibinfo {author} {\bibfnamefont {F.}~\bibnamefont {Foroughi}}, \bibinfo {author} {\bibfnamefont {Y.}~\bibnamefont {Krupko}}, \bibinfo {author} {\bibfnamefont {O.}~\bibnamefont {Buisson}}, \bibinfo {author} {\bibfnamefont {C.}~\bibnamefont {Naud}}, \bibinfo {author} {\bibfnamefont {W.}~\bibnamefont {Hasch-Guichard}}, \emph {et~al.},\ }\bibfield  {title} {\bibinfo {title} {A tunable josephson platform to explore many-body quantum optics in circuit-qed},\ }\href@noop {} {\bibfield  {journal} {\bibinfo  {journal} {npj Quantum Information}\ }\textbf {\bibinfo {volume} {5}},\ \bibinfo {pages} {19} (\bibinfo {year} {2019})}\BibitemShut {NoStop}%
\bibitem [{\citenamefont {Roth}\ and\ \citenamefont {Chew}(2021)}]{roth2021_fullwaveCQED}%
  \BibitemOpen
  \bibfield  {author} {\bibinfo {author} {\bibfnamefont {T.~E.}\ \bibnamefont {Roth}}\ and\ \bibinfo {author} {\bibfnamefont {W.~C.}\ \bibnamefont {Chew}},\ }\bibfield  {title} {\bibinfo {title} {Macroscopic circuit quantum electrodynamics: A new look toward developing full-wave numerical models},\ }\href@noop {} {\bibfield  {journal} {\bibinfo  {journal} {IEEE Journal on Multiscale and Multiphysics Computational Techniques}\ }\textbf {\bibinfo {volume} {6}},\ \bibinfo {pages} {109} (\bibinfo {year} {2021})}\BibitemShut {NoStop}%
\bibitem [{\citenamefont {Chew}\ \emph {et~al.}(2016)\citenamefont {Chew}, \citenamefont {Liu}, \citenamefont {Salazar-Lazaro},\ and\ \citenamefont {Sha}}]{chew2016quantum}%
  \BibitemOpen
  \bibfield  {author} {\bibinfo {author} {\bibfnamefont {W.~C.}\ \bibnamefont {Chew}}, \bibinfo {author} {\bibfnamefont {A.~Y.}\ \bibnamefont {Liu}}, \bibinfo {author} {\bibfnamefont {C.}~\bibnamefont {Salazar-Lazaro}},\ and\ \bibinfo {author} {\bibfnamefont {W.~E.}\ \bibnamefont {Sha}},\ }\bibfield  {title} {\bibinfo {title} {Quantum electromagnetics: A new look—part ii},\ }\href@noop {} {\bibfield  {journal} {\bibinfo  {journal} {IEEE Journal on Multiscale and Multiphysics Computational Techniques}\ }\textbf {\bibinfo {volume} {1}},\ \bibinfo {pages} {85} (\bibinfo {year} {2016})}\BibitemShut {NoStop}%
\bibitem [{\citenamefont {T{\"u}reci}\ \emph {et~al.}(2008)\citenamefont {T{\"u}reci}, \citenamefont {Ge}, \citenamefont {Rotter},\ and\ \citenamefont {Stone}}]{tureci2008randomlaser}%
  \BibitemOpen
  \bibfield  {author} {\bibinfo {author} {\bibfnamefont {H.~E.}\ \bibnamefont {T{\"u}reci}}, \bibinfo {author} {\bibfnamefont {L.}~\bibnamefont {Ge}}, \bibinfo {author} {\bibfnamefont {S.}~\bibnamefont {Rotter}},\ and\ \bibinfo {author} {\bibfnamefont {A.~D.}\ \bibnamefont {Stone}},\ }\bibfield  {title} {\bibinfo {title} {Strong interactions in multimode random lasers},\ }\href@noop {} {\bibfield  {journal} {\bibinfo  {journal} {Science}\ }\textbf {\bibinfo {volume} {320}},\ \bibinfo {pages} {643} (\bibinfo {year} {2008})}\BibitemShut {NoStop}%
\bibitem [{\citenamefont {Wiersig}(2002)}]{wiersig2002boundary}%
  \BibitemOpen
  \bibfield  {author} {\bibinfo {author} {\bibfnamefont {J.}~\bibnamefont {Wiersig}},\ }\bibfield  {title} {\bibinfo {title} {Boundary element method for resonances in dielectric microcavities},\ }\href@noop {} {\bibfield  {journal} {\bibinfo  {journal} {Journal of Optics A: Pure and Applied Optics}\ }\textbf {\bibinfo {volume} {5}},\ \bibinfo {pages} {53} (\bibinfo {year} {2002})}\BibitemShut {NoStop}%
\bibitem [{dec()}]{decqed_repo}%
  \BibitemOpen
  \href@noop {} {\bibinfo {title} {{DEC-QED} toolbox}},\ \bibinfo {howpublished} {\url{https://github.com/dnpham23/DEC-QED}}\BibitemShut {NoStop}%
\bibitem [{\citenamefont {London}\ and\ \citenamefont {London}(1935)}]{londontheory}%
  \BibitemOpen
  \bibfield  {author} {\bibinfo {author} {\bibfnamefont {F.}~\bibnamefont {London}}\ and\ \bibinfo {author} {\bibfnamefont {H.}~\bibnamefont {London}},\ }\bibfield  {title} {\bibinfo {title} {The electromagnetic equations of the supraconductor},\ }\href {https://doi.org/10.1098/rspa.1935.0048} {\bibfield  {journal} {\bibinfo  {journal} {Proceedings of the Royal Society of London. Series A - Mathematical and Physical Sciences}\ }\textbf {\bibinfo {volume} {149}},\ \bibinfo {pages} {71} (\bibinfo {year} {1935})}\BibitemShut {NoStop}%
\bibitem [{\citenamefont {Shaw}(1974)}]{Shaw_1974}%
  \BibitemOpen
  \bibfield  {author} {\bibinfo {author} {\bibfnamefont {G.~B.}\ \bibnamefont {Shaw}},\ }\bibfield  {title} {\bibinfo {title} {Degeneracy in the particle-in-a-box problem},\ }\href {https://doi.org/10.1088/0305-4470/7/13/008} {\bibfield  {journal} {\bibinfo  {journal} {Journal of Physics A: Mathematical, Nuclear and General}\ }\textbf {\bibinfo {volume} {7}},\ \bibinfo {pages} {1537} (\bibinfo {year} {1974})}\BibitemShut {NoStop}%
\bibitem [{\citenamefont {Josephson}(1965)}]{JosephsonReview_1965}%
  \BibitemOpen
  \bibfield  {author} {\bibinfo {author} {\bibfnamefont {B.~D.}\ \bibnamefont {Josephson}},\ }\bibfield  {title} {\bibinfo {title} {Supercurrents through barriers},\ }\href {https://doi.org/10.1080/00018736500101091} {\bibfield  {journal} {\bibinfo  {journal} {Advances in Physics}\ }\textbf {\bibinfo {volume} {14}},\ \bibinfo {pages} {419} (\bibinfo {year} {1965})}\BibitemShut {NoStop}%
\bibitem [{\citenamefont {Kosztin}\ and\ \citenamefont {Schulten}(1997)}]{kosztin1997boundary}%
  \BibitemOpen
  \bibfield  {author} {\bibinfo {author} {\bibfnamefont {I.}~\bibnamefont {Kosztin}}\ and\ \bibinfo {author} {\bibfnamefont {K.}~\bibnamefont {Schulten}},\ }\bibfield  {title} {\bibinfo {title} {Boundary integral method for stationary states of two-dimensional quantum systems},\ }\href@noop {} {\bibfield  {journal} {\bibinfo  {journal} {International Journal of Modern Physics C}\ }\textbf {\bibinfo {volume} {8}},\ \bibinfo {pages} {293} (\bibinfo {year} {1997})}\BibitemShut {NoStop}%
\bibitem [{\citenamefont {Pham}\ \emph {et~al.}(2023{\natexlab{b}})\citenamefont {Pham}, \citenamefont {Bharadwaj}, \citenamefont {Rodriguez}, \citenamefont {Rodriguez},\ and\ \citenamefont {Ram-Mohan}}]{pham2023singularfields}%
  \BibitemOpen
  \bibfield  {author} {\bibinfo {author} {\bibfnamefont {D.~N.}\ \bibnamefont {Pham}}, \bibinfo {author} {\bibfnamefont {S.}~\bibnamefont {Bharadwaj}}, \bibinfo {author} {\bibfnamefont {S.}~\bibnamefont {Rodriguez}}, \bibinfo {author} {\bibfnamefont {L.}~\bibnamefont {Rodriguez}},\ and\ \bibinfo {author} {\bibfnamefont {L.~R.}\ \bibnamefont {Ram-Mohan}},\ }\bibfield  {title} {\bibinfo {title} {High-accuracy calculation of singular electromagnetic fields in regions with re-entrant peripheries},\ }\href@noop {} {\bibfield  {journal} {\bibinfo  {journal} {Journal of Applied Physics}\ }\textbf {\bibinfo {volume} {134}},\ \bibinfo {pages} {153102} (\bibinfo {year} {2023}{\natexlab{b}})}\BibitemShut {NoStop}%
\bibitem [{\citenamefont {B{\"a}cker}(2003)}]{Backer_2003}%
  \BibitemOpen
  \bibfield  {author} {\bibinfo {author} {\bibfnamefont {A.}~\bibnamefont {B{\"a}cker}},\ }\bibfield  {title} {\bibinfo {title} {Numerical aspects of eigenvalue and eigenfunction computations for chaotic quantum systems},\ }in\ \href@noop {} {\emph {\bibinfo {booktitle} {The Mathematical Aspects of Quantum Maps}}},\ \bibinfo {editor} {edited by\ \bibinfo {editor} {\bibfnamefont {M.~D.}\ \bibnamefont {Esposti}}\ and\ \bibinfo {editor} {\bibfnamefont {S.}~\bibnamefont {Graffi}}}\ (\bibinfo  {publisher} {Springer Berlin Heidelberg},\ \bibinfo {address} {Berlin, Heidelberg},\ \bibinfo {year} {2003})\ pp.\ \bibinfo {pages} {91--144}\BibitemShut {NoStop}%
\bibitem [{\citenamefont {Barrera}\ \emph {et~al.}(1985)\citenamefont {Barrera}, \citenamefont {Estevez},\ and\ \citenamefont {Giraldo}}]{barrera1985vector}%
  \BibitemOpen
  \bibfield  {author} {\bibinfo {author} {\bibfnamefont {R.~G.}\ \bibnamefont {Barrera}}, \bibinfo {author} {\bibfnamefont {G.}~\bibnamefont {Estevez}},\ and\ \bibinfo {author} {\bibfnamefont {J.}~\bibnamefont {Giraldo}},\ }\bibfield  {title} {\bibinfo {title} {Vector spherical harmonics and their application to magnetostatics},\ }\href@noop {} {\bibfield  {journal} {\bibinfo  {journal} {European Journal of Physics}\ }\textbf {\bibinfo {volume} {6}},\ \bibinfo {pages} {287} (\bibinfo {year} {1985})}\BibitemShut {NoStop}%
\bibitem [{\citenamefont {Hirani}(2003)}]{DEC_HiraniThesis}%
  \BibitemOpen
  \bibfield  {author} {\bibinfo {author} {\bibfnamefont {A.~N.}\ \bibnamefont {Hirani}},\ }\emph {\bibinfo {title} {Discrete Exterior Calculus}},\ \href@noop {} {Ph.D. thesis},\ \bibinfo  {school} {California Institute of Technology} (\bibinfo {year} {2003})\BibitemShut {NoStop}%
\bibitem [{\citenamefont {Morse}\ and\ \citenamefont {Feshbach}(1954)}]{morse1954methods}%
  \BibitemOpen
  \bibfield  {author} {\bibinfo {author} {\bibfnamefont {P.~M.}\ \bibnamefont {Morse}}\ and\ \bibinfo {author} {\bibfnamefont {H.}~\bibnamefont {Feshbach}},\ }\bibfield  {title} {\bibinfo {title} {Methods of theoretical physics},\ }\href@noop {} {\bibfield  {journal} {\bibinfo  {journal} {American Journal of Physics}\ }\textbf {\bibinfo {volume} {22}},\ \bibinfo {pages} {410} (\bibinfo {year} {1954})}\BibitemShut {NoStop}%
\end{thebibliography}%
\end{document}